\documentclass[useAMS,usenatbib]{mnras}

\usepackage{graphicx}
\usepackage{times}
\usepackage[usenames, dvipsnames]{xcolor}
\usepackage{rotating}
\usepackage{txfonts}
\usepackage{longtable,lscape}
\usepackage{anysize}
\usepackage{amssymb}
\usepackage{natbib}
\usepackage{booktabs}
\usepackage{bm}
\usepackage{pdflscape}

\newcommand{\mgc} {\emph{mGC3}}
\newcommand{\ngc} {\emph{nGC3}}
\newcommand{\xgc} {\emph{xGC3}}
\newcommand{\gc} {\emph{GC3}}

\newcommand{\FeH} {[\mathrm{Fe}/\mathrm{H}]}

\newcommand{\Lsun}{{\rm L}_\odot}

\newcommand{\Rhel}{R_{\rm hel}}
\newcommand{\Rgal}{R_{\rm gal}}

\newcommand{\rrab} {\mbox{RR\emph{ab}}}
\newcommand{\rrc} {\mbox{RR\emph{c}}}
\newcommand{\typeab} {\mbox{\emph{ab}}}
\newcommand{\typec} {\mbox{\emph{c}}}

\title[RR Lyrae star streams in the Catalina survey]{Fourteen candidate RR Lyrae star streams in the inner Galaxy}
\author[C. Mateu et. al.]{Cecilia~Mateu$^{1}$\thanks{E-mail:cmateu@cida.gob.ve, cmateu.cida@gmail.com}, 
Justin~I.~Read$^{2}$, Daisuke~Kawata$^{3}$ \\
$^{1}${{Centro de Investigaciones de Astronom\'{\i}a, AP 264, M\'erida 5101--A, Venezuela}}\\
$^{2}${{Department of Physics, University of Surrey, Guildford, GU2 7XH, UK}}\\
$^{3}${{Mullard Space Science Laboratory, University College London, Holmbury St Mary, Dorking, Surrey RH5 6NT, UK}}\\
}

\begin{document}

\date{Accepted. Received ; in original form }
\pagerange{\pageref{firstpage}--\pageref{lastpage}} \pubyear{2017}
\maketitle

\label{firstpage}

\begin{abstract}
We apply the GC3 stream-finding method to RR Lyrae stars (RRLS) in the Catalina survey. We find two RRLS stream candidates at $>4\sigma$ confidence and another 12 at $>3.5\sigma$ confidence over the Galactocentric distance range $4 < D/{\rm kpc} < 26$. Of these, only two are associated with known globular clusters (NGC 1261 and Arp2). The remainder are candidate `orphan' streams, consistent with the idea that globular cluster streams are most visible close to dissolution. Our detections are likely a lower bound on the total number of dissolving globulars in the inner galaxy, since many globulars have few RRLS while only the brightest streams are visible over the Galactic RRLS background, particularly given the current lack of kinematical information. We make all of our candidate streams publicly available 
and provide a new  {{\tt galstreams}} Python library for the footprints of all known streams and overdensities in the Milky Way.
\end{abstract}


\section{Introduction}\label{s:intro}

In the standard $\Lambda$ Cold Dark Matter ($\Lambda$CDM) cosmological model, structures grow through the successive mergers of smaller structures \citep[e.g.][]{1978MNRAS.183..341W}. This model gives a remarkable match to the cosmic microwave background radiation \citep{1992ApJ...396L...1S,2013arXiv1303.5076P}, the growth of large scale structure in the Universe \citep[e.g.][]{2006Natur.440.1137S,2015arXiv151201981B}, and the abundance of isolated gas rich dwarf galaxies \citep[e.g.][]{2017MNRAS.467.2019R}. However, on smaller scales inside galaxies and groups there have been long standing tensions \citep[e.g][]{2017ARA&A..55..343B}. Key amongst these is the `missing satellites' problem (MSP). This is a discrepancy between the number of visible satellites orbiting the Milky Way (MW) and M31 and the expected number of bound dark matter halos in $\Lambda$CDM \citep{1999ApJ...522...82K,1999ApJ...524L..19M}.

To date, understanding the origin of the MSP has focussed on the {\it surviving} population of dwarf galaxies \citep[e.g.][]{2007ApJ...667..859D,2010arXiv1001.1731L,2013JCAP...03..014A,2017arXiv170804247N} and globular clusters (GCs; e.g. \citealt{2005astro.ph.10370M}). However, many dwarfs and GCs are expected to be tidally disrupted on infall to the MW, a process made more efficient by the presence of the Milky Way stellar disc \citep[e.g.][]{2010ApJ...709.1138D} and by any process that can lower the central density of the MW satellites \citep[e.g.][]{ReadEtal2006:B,2010MNRAS.406.1290P}. Thus, some solutions to the MSP posit a significant depletion of satellites \citep[see e.g. the discussion in][]{2016MNRAS.459.2573R}, while others primarily make satellites dark \citep[e.g.][]{2016MNRAS.457.1931S}. These lead to detectable differences, however, in the number and properties of dissolving satellites and their stellar streams \citep[e.g.][]{2001ApJ...548...33B}.

The above motivates building a complete census of stellar streams in the Galaxy. Such streams can also be used to directly probe the mass distribution and shape of the MW dark matter halo \citep[e.g.][]{2001ApJ...547L.133I,2005ApJ...619..800J,2009MNRAS.400..548E,2010ApJ...712..260K,2011MNRAS.417..198V,2013MNRAS.436.2386L,2015ApJ...803...80K}, to test alternative gravity models \citep[e.g.][]{2005MNRAS.361..971R,2017A&A...603A..65T}, to hunt for ghostly dark matter `mini-halos' \citep[e.g.][]{2002ApJ...570..656J,2002MNRAS.332..915I,2012ApJ...748...20C,2017MNRAS.470...60E}, and -- for GC streams -- to constrain GC formation and evolution models \citep[e.g.][]{Balbinot2017}.

With the advent of large surveys like the Sloan Digital Sky Survey, PanSTARRS and the Dark Energy Survey, the number of known stellar streams in the MW has grown dramatically \citep[e.g.][]{2006ApJ...642L.137B,Grillmair2014,2016ApJ...817..135L,Bernard2016,Torrealba2015,2016ApJ...820...58B,Grillmair2016}. Yet, due to incomplete sky coverage, crowding in the Galactic centre, and dust obscuration, the full census of GC and dwarf streams remains far from complete. In this paper, we apply the `GC3' stream-finding method \citep{LyndenBell1995,Johnston1996,Mateu2011} to RR Lyrae stars (RRLSs) in the Catalina Sky Survey \citep{Drake2013a,Drake2013b,Torrealba2015} to hunt for stellar streams. RRLSs are well known standard candles for which precise distances can be estimated with errors of order $\sim5\%$ out to a distance beyond $D \sim 100$~kpc, or even down to $\sim2\%$ when metallicity information and infrared data are available \citep{Sesar2017b,Neeley2017}. They are also present in all known dwarf galaxies \citep{Vivas2006,Mateu2009}, many GCs \citep{Clement2001}, as well as in all Galactic components except the thin disc \citep{Martin1998}, which keeps foreground contamination relatively low at low Galactic latitude compared to other tracers.
These properties allow us to detect even relatively faint streams (with just a dozen RRLS of type \typeab, corresponding to a typical luminosity for a GC stream of few $\times10^4\Lsun$), even towards the Galactic centre ($D > 4\,{\rm kpc}$). Due to crowding, this region has been relatively unexplored to date, yet it is where most surviving GCs are found in the MW today \citep[e.g.][]{2006ARA&A..44..193B}. As such, we may expect to find many GC streams over the region $4 < D/{\rm kpc} < 26$ that we are sensitive to (Section~\ref{s:pcms_and_peak_det}).

This paper is organised as follows. In \S\ref{s:catalina_hsoy}, we present the survey data and describe our analysis pipeline. In \S\ref{s:stream_search} we describe the GC3 method used to search for tidal streams, in the context of the larger \xgc~family of great-circle cell methods. In \S\ref{s:stream_candidate_results} we present the candidate tidal streams found and summarise their properties. In \S\ref{s:comparison_with_known_streams} we discuss our stream candidates in comparison with known streams, clouds and GCs in the MW and in \S\ref{s:comparison_with_predictions} we contrast our findings with predictions for GC tidal tails.

\section{The Catalina+HSOY \rrab~catalogue}\label{s:catalina_hsoy}

We made a compilation of the RRLSs of type \emph{ab} from the Catalina Sky Survey (CSS) in the northern hemisphere ferom \citep{Drake2013a,Drake2013b} and its southern hemisphere extension, the Siding Springs Survey (SSS) from \citet{Torrealba2015}. The joint CSS+SSS comprises $>$20,000 \rrab~stars, covering $>$34,000 sq. deg in the magnitude range $14<V<20$ across the whole sky, except an avoidance zone at low galactic latitude $|b|\leqslant 15\degr$ \citep{Drake2013a,Drake2013b,Torrealba2015}. 

The CSS+SSS \rrab~catalogue constitutes a deep, clean and homogeneous sample, with consistent distances computed following the same methods in the northern and southern parts of the survey. The CSS and SSS surveys have an average completeness of $70\%$ and are fairly uncontaminated as they focus on RRLSs of type~\typeab, easy to discriminate against other types of variables with their well sampled light curves. 

In the SSS catalogue, \citet{Torrealba2015} report photometric metallicities derived following \citet{Jurcsik1996} for their 10,540 \rrab~stars. In the CSS catalogue, \citet{Drake2013a} report spectroscopic SDSS metallicities for 1416 \rrab~stars. For the remaining stars, we used the publicly available CSS time series data\footnote{Available at \href{http://nesssi.cacr.caltech.edu/DataRelease/RRL.html}{http://nesssi.cacr.caltech.edu/DataRelease/RRL.html}.} to compute photometric metallicities using the {\sc tff} code from \citet{Kovacs2007} to perform the Fourier light curve decomposition and compute $\phi_{31}$. We used Eq. 7 from \citet{Torrealba2015} to obtain $\FeH$ which, as in \citet{Mateu2012}, we report only for stars with $D_m>3$. The resulting combined CSS+SSS catalogue contains 21,920 \rrab~stars.

We also matched the combined CSS+SSS \rrab~catalogue with the Hot Stuff for One Year (HSOY) catalogue from \citet{Altmann2017} to obtain supplementary proper motion information. HSOY is a proper motion catalogue compiled using PPMXL \citep{Roeser2010} and Gaia DR1 \citep{Gaia2016a,Gaia2016b} as first and second epoch catalogues respectively, and contains a total of $\sim500$ million stars down to $G=20$. 
A total of 20,610 matches were found, out of which 2,665 stars (12\%) have relative proper motion errors smaller than 30\% at a median distance of $\sim13$ kpc. 
The Catalina+HSOY \rrab~catalogue, including the photometric metallicities computed here for CSS stars, is publicly available here\footnote{\url{https://cmateu.github.io/Cecilia_Mateu_WebPage/CatalinaGC3_Streams.html}}


\subsection{Removing stars around known globular clusters}\label{s:removing_globulars}

Stars in the main body of known GCs and dwarf galaxies will produce a strong great circle signature in the pole count maps when kinematic data are unavailable or scarce. To avoid these contaminating signatures, we flag and discard from our analysis all RRLSs that fulfill these two criteria: i) lie within the angular tidal radius $r_t$ of each GC and ii) lie within an interval $[R_{min},R_{max}]$ in heliocentric distance $\Rhel$ around the cluster.  

For the first criterion, we compute the angular tidal radius as $r_t=r_c10^{c}$ \citep{Navin2016}, where the concentration $c$ and (angular) core radius $r_c$ are taken from the GC compilation of \citet{Harris1996} (in its 2010 edition). For dwarf galaxies, we remove stars within ten times the half-light radius of each galaxy, taken from the compilation of \citet{McConnachie2012}.

For the second criterion, in principle, we simply should have been able to remove the stars in the volume within the selected physical threshold radius. However, when looking at the heliocentric distance distribution of RRLSs around a few known GCs, a large number of the RRLS --within the tidal radius-- appear to be much closer than the nominal distance of the cluster.
For example, for M3 which is located at 10 kpc \citep{Harris1996}, RRLS are found down to $\sim9$ kpc, clearly beyond its tidal radius of $\sim 85$pc.

This suggests a problem with the absolute magnitude $M_V$ assumed to compute the RRLS distances, since the distribution appears more extended only in heliocentric distance. This is most likely due to the presence of over-luminous RRLS, noted by several authors in their GC studies  \citep[e.g.][]{Cacciari2005,VandenBerg2016}. \citet{Cacciari2005} proposes these are  RRLS that have evolved off the Zero Age Horizontal Branch (ZAHB).  These over-luminous RRLS are observed to be $\sim$0.2 to 0.25 mag brighter, see e.g. \citet{VandenBerg2016},  meaning their $\Rhel$ can be underestimated by as much as $\sim$10\%, consistent with what we see in the $\Rhel$ distribution of M3 RRLS in CSS.  

This illustrates the need for using a distance scale that deals with post-ZAHB evolution which, for field RRLS, is a highly non-trivial issue. A proper treatment of this is beyond the scope of this paper, so for our purposes we assume maximum offsets of $-0.3$ and $+0.05$ mag due to post-ZAHB evolution and photometric errors respectively, and remove stars in the interval $[mM-0.3, mM+0.05]$ mag, where $mM$ corresponds to the cluster's distance modulus. For future works, however, a possible route has been suggested by \citet{Kunder2010} who, in an Oosterhoff analysis of the absolute magnitude of bulge RRLS, propose using an $M_V$-period-shift relation instead of the traditionally used $M_V$-metallicity. This may offer a way to properly estimate the absolute magnitude of  these over-luminous RRLS, as \citeauthor{Cacciari2005}'s  findings support a connection between the Oosterhoff dichotomy -- and hence, the period shift -- and post-ZAHB evolution.

\section{The stream search}\label{s:stream_search}

\subsection{The \xgc~family}\label{s:xgc3}

The \xgc~family encompasses a suite of methods to search for tidal streams by looking for overdensities in great circle bands in the sky, as seen from the Galactic centre, an idea introduced originally by \citet{LyndenBell1995} and \citet{Johnston1996} and expanded in \citet{Mateu2011}.  The different methods in the family are defined by their use of the different layers of information available:

\begin{itemize}
\item \gc~(3D): 3D position
\item \ngc~(5D): 3D position + proper motions
\item \mgc~(6D): 3D position + 3D velocity
\end{itemize}

Briefly, the \xgc~methods in general consist in counting, as seen from the Galactic centre, how many stars lie along each great circle and, in the case of \ngc~and \mgc, counting how many of these also have their velocity vectors along that great circle. Labelling each great circle by its normal vector or `pole' and going over all possible poles in the celestial sphere, a pole count map (PCM) is produced which displays the number of stars associated to each pole. A tidal tail will show up as a maximum or localised peak in any \xgc~PCM, whereas a completely bound cluster or galaxy (i.e. a localised clump of comoving stars) will show up either as a great circle in an \gc~PCM, or as a localised peak in an \ngc/\mgc~PCM thanks to the addition of kinematic information. For various examples of the PCM signatures produced by different stellar structures and the effects of observational errors in them, we refer the reader to \citet{Mateu2017} (their Fig. 3 and Sec. 5.2), and for a detailed explanation of each of the methods in the family, to \citet{Mateu2011} (\gc, \mgc) and \citet{Abedi2014} (\ngc).

The \xgc~methods offer several advantages in the search for tidal streams: i) they are based in the simple principle that tidal streams are approximately planar if produced in a potential that is approximately symmetric, and no knowledge or assumption of the underlying potential is required; ii) even though in theory the methods work best for streams produced in symmetric static potentials,  \citet{Mateu2017} have shown streams produced in realistic cosmologically evolving potentials can be recovered with \ngc; iii) in a PCM a tidal stream, however complicated its distribution in radius might be, produces a peak which can simply be more or less stretched or deformed depending on the effect of precession and of the observational errors \citep[see Figs. 3 and 4 in][]{Mateu2017}; iv) the \xgc~methods are implemented directly in terms observable quantities \citep[see][]{Mateu2011} in a way that minimises error propagation, in particular when using the parallax by avoiding the computation of its reciprocal; v) the \xgc~methods are linear, hence, depending on the information available for different stars, PCMs can be produced with the different methods and later  combined into a single composite PCM by simple addition.

\subsection{Computation of PCMs and peak detection}\label{s:pcms_and_peak_det}

The Catalina+HSOY RRLS catalogue produced contains HSOY proper motions, however, for the majority of the RRLS the relative errors are quite large, as mentioned in Sec.~\ref{s:catalina_hsoy}. In what follows, we take as acceptable proper motions those with relative errors $<30\%$ and combine \ngc~PCMs for those stars (2,665) with \gc~PCMs for the rest (19,255). Although these stars are few, the use of their kinematic information can help reduce foreground contamination in the combined \gc+\ngc~PCM.

To compute the \gc~and \ngc~PCMs we use the {\sc PyMGC3} toolkit, a Python implementation of the \xgc~methods publicly available at \href{https://github.com/cmateu/PyMGC3}{this GitHub \underline{repository}} \citep{Mateu2014}. The \gc+\ngc~PCMs were computed with a great circle tolerance of $1\degr$ (both in position and velocity) and in Galactocentric bins 1 kpc wide with offsets of 0.5 kpc to cover the full range of Galactocentric distance from 4 to 25 kpc. We assume a reference system with the Sun is located at $X=-8.5$~kpc. The grid spacing used was $0\fdg5$, therefore our PCMs have in total 82,958 pixels, out of which half are independent.

The first step before peak detection is to produce unsharp masked PCMs, as described in detail in Sec.~6.2 of \citet{Mateu2017}. For this, the smooth background of each PCM is estimated with a median filter by computing a pixel-by-pixel estimate of the median in a $20\degr$ radius, i.e. in a neighbourhood much larger than the features we want to identify.  The smooth background is subtracted from the PCM resulting in an unsharp-masked PCM, in which sharp features are highlighted. This unsharp-masked PCM is expressed in $N\sigma$ units dividing it pixel-by-pixel by the typical standard deviation. In \citet{Mateu2017} the pixel-by-pixel standard deviation was computed as the square root of the smoothed PCM, assuming pole counts follow a Poissonian distribution. We have improved this procedure by now computing the standard deviation of the smoothed counts in an annulus around each pixel, with an inner radius equal to the size of the box used for the median estimate and an annulus width of $5\degr$. This allows for a much better estimation of the significance of pixel counts, especially near areas with a sharply varying background which, as in this case, can appear when the input catalogue has a sharply defined avoidance zone.

The detection of peaks is made in the unsharp-masked \gc+\ngc~PCMs, using the Fellwalker\footnote{FellWalker is publicly available as part of the Starlink Software Distribution at \url{http://starlink.eao.hawaii.edu/starlink}} algorithm from \citet{Berry2015} \citep[see][ for a detailed description]{Mateu2017}. The peak detections were made in the combined \gc+\ngc~PCMs with a tolerance of $1\degr$.

To search for GC tidal streams, one would ideally use a smaller tolerance, around $0\fdg25$ to $0\fdg5$ \citep[see e.g.][]{Ibata2001b}, as these are dynamically colder than dwarf galaxy streams. However, we have chosen to perform overall detections on PCMs with a larger tolerance to reduce random noise in the PCMs as there are relatively few RRLS per Galactocentric distance bin ($\sim700$ to $1000$ stars per 1 kpc bin up to $\sim25$ kpc). In the next section, Table~\ref{t:stream_detections} summarises all detected stream candidates.

\section{Stream Candidates}\label{s:stream_candidate_results}

The geometric and detection properties of the stream candidates are summarised in Table~\ref{t:stream_detections}.

Table~\ref{t:stream_detections} indicates, for each stream candidate, an ID, the pole's Galactocentric and heliocentric coordinates $(\phi,\theta)$ and $(l,b)$ respectively and the minimum and maximum heliocentric distance spanned by the candidates RRLS. The `central' galactic coordinates ($l_\circ,b_\circ$), mean Galactocentric distance and standard deviation, and Galactocentric stream width $\Delta\theta$ (perpendicular to the stream's plane)  reported correspond to those of the densest part of each stream candidate. This  was found by computing the mode of a gaussian kernel density estimation (KDE) in the azimuth distribution along the stream's plane, after subtracting the KDE for background stars in neighbouring poles selected in an annulus around the pole detection\footnote{The annulus used was the same as for the computation of the smoothed PCM standard deviation, see Sec.~\ref{s:pcms_and_peak_det}} and renormalised to match the total number of stars in each detection. The detection significance is reported in $N\sigma$ units and indicated in parenthesis is the detection significance for those poles with a counterpart detection in the \gc+\ngc~PCM with a $0\fdg5$ tolerance. The average bootstrap significance (Bts) (see following subsection) is also reported. The last column indicates whether the candidate might be associated to a known GC, based on the PCMs shown in Figure~\ref{f:pcms_detections} and discussed in Sec.~\ref{s:cand_pcms}.

\subsection{Candidate Classification: Bootstrap and Artefact tests}\label{s:bootstrap}

The sparsity of the RRLS sample in each radial bin, combined with the catalogue's avoidance zone ($|b|\geqslant 20 \degr$), might lead to significant stochastic noise and the presence of artefacts due to abrupt changes in the PCM background caused by the catalogue edges. These effects are mitigated by the unsharp masking and by the new procedure to estimate the standard deviation locally (Sec.~\ref{s:pcms_and_peak_det}), but it is reasonable to expect them not to be completely eliminated.

To obtain a more robust estimate of the detection significance we performed bootstrap tests: 100 bootstrap  realisations of the RRLS catalogue were produced and the corresponding unsharp-masked \gc+\ngc~PCMs computed in N$\sigma$ units (as in Sec.~\ref{s:pcms_and_peak_det}), in the same radial bins 1~kpc wide. For each of our initial candidates, we find the maximum pole counts within $2\degr$ of the candidate's pole and store it for each bootstrapped catalogue. Finally, we compute the bootstrapped mean detection significance which is summarised in column Bts-Av in Table~\ref{t:stream_detections}. 

A high mean significance in the bootstrap tests supports the robustness of a detection against stochastic fluctuations. The bootstrap test results show two of our initial candidates have a mean significance $>$4$\sigma$: 11.0-1 and 20.0-1, we consider these as our \emph{high-confidence} candidates. Two more detections are just below the $4\sigma$ threshold, 17.0-1 and 23.5-1, which have 3.9$\sigma$, but the remaining ones -- a clear majority --  have lower mean significance closer to $\sim$3.5$\sigma$, which means overall our initial significance estimates are slightly overestimated. We classify the candidates with bootstrap mean significance $<$4$\sigma$ as \emph{tentative} candidates if they pass the artefact test described below. 

To check whether a given detection is a likely artefact caused by possible edge-effects of the survey in the PCMs, we create a perturbed realisation of the RRLS catalogue and 100 bootstrap realisations of this perturbed  catalogue to analyse the detection significance as described above. The perturbed catalogue was produced by adding a random step to the equatorial coordinates of each RRLS. The steps were drawn at random from a gaussian distribution with a standard deviation of $3\degr$, a value chosen to be higher than the width of our \gc~great-circle cells ($2\degr$) to ensure the stars have a non-negligible probability of being perturbed out of their great-circle cell. Repeating the procedure used for the bootstrap analysis we compute the mean bootstrap significance for the 100 bootstrap realisations of the perturbed catalogue.

In the perturbed catalogue, any real stream within our selected tolerance should have been erased by the random perturbation, so if the bootstrap significance threshold chosen above is appropriate, no detections with a larger significance should be found in the perturbed catalogue.  The artefact test results show that no detections have bootstrap significance above $4\sigma$ in the high-confidence candidates radial bins, which confirms these are not likely to be artefacts. For the tentative candidates, we mark as artefacts those with a mean bootstrap significance lower than the maximum bootstrap significance of any detections found in the corresponding radial bin in the perturbed catalogue. These are thus reported as \emph{possible artefacts}\footnote{In further experiments with perturbed catalogues we estimate that for bins with $\Rgal<20$~kpc no detections with mean bootstrap significance  $>3\sigma$ are expected; while for $\Rgal>20$~kpc we find on average one per $\Rgal$-bin with mean bootstrap significance above $3\sigma$, and none at $>4\sigma$.}.

In some cases, e.g. 17.5-1 and 19.5-1, there are multiple detections of the same candidate, due to the 0.5 kpc overlap in our $\Rgal$~bins. When the pole detections in adjacent radial bins coincide within $\lesssim5\degr$, we report that with the highest mean bootstrap significance as the main one, and the rest as \emph{repeated detections}.

Finally, we report detections 19.5-2, 20.0-2 and 20.5-1 as unambiguous detections of the Sagittarius stream (Sgr). These detection's poles coincide within $\sim4\degr$ with the galactocentric pole $(93\fdg8,+13\fdg5)$ reported by \citet{Majewski2003}\footnote{\citet{Majewski2003} reports the antipodal pole $(273\fdg8,-13\fdg5)$ since it coincides with the direction of Sgr's angular momentum. Nevertheless, antipodal poles are equivalent since both define the same plane under the \xgc~criteria.}.

\subsection{Candidate PCMs and Sky Distribution}\label{s:cand_pcms}

Figure~\ref{f:pcms_detections} shows the \gc+\ngc~PCMs in the distance bins where significant ($>$4$\sigma$) peak detections were found. 

Figure~\ref{f:pcms_detections}, in its left panels, shows the unsharp-masked \gc+\ngc~PCMs in an equidistant north-polar azimuthal projection, with a colour scale proportional to the detection significance expressed in $N\sigma$ units (see Sec.\ref{s:pcms_and_peak_det}). Detected peaks are marked with a circle and labelled in each distance bin with an integer number. The Galactocentric distance bin is indicated at the top of each PCM. Each stream candidate is given a unique ID constructed as the integer label shown in Fig.~\ref{f:pcms_detections} and the lower limit of the distance bin in which the detection was made (e.g. stream candidates 04.5-1 and 04.5-2 are shown in the top panel of Fig.~\ref{f:pcms_detections}). This ID is used to identify each stream candidate throughout this paper. The right panels of Figure~\ref{f:pcms_detections} show the distribution of RRLS in the current $\Rgal$ bin, in galactocentric spherical coordinates $(\phi,\theta)$\footnote{In this reference system the Galactic Disc is located at $\theta=0\degr$ and $\phi=0\degr$ points \emph{away} from the Sun.}. The stars associated to each of the peaks marked as detected in the corresponding PCM are plotted with different colours, as shown in each panel's legend.

The left panels of Figure~\ref{f:pcms_detections} also show the expected \ngc~signature for all known GCs and dwarf galaxies present in each distance bin, calculated using positions and distances from \citet{Harris1996} and proper motions from \citet{Balbinot2017} for GCs; and using all data from \citet{McConnachie2012} for dwarf galaxies. The greyscale represents the signature expected within $3\sigma$ of each cluster's proper motion errors, with the darkest parts corresponding to the highest probability areas. Clusters (or galaxies) produce a signature along a great circle in the PCM, the length of which is inversely proportional to the proper motion errors: good proper motion data constrains the orbital plane well and produces a localised peak, while bad or no proper motion data does not, and so, produces a peak that stretches along a great circle. 

These plots allow us to see very quickly which of our stream candidates might be associated to a known GC or dwarf galaxy. The pole detection of candidate 07.0-2 lies right on top of the signature expected from GC Pal11 on a very high probability region. The pole detection of candidate 17.5-3 lies just off NGC1261's great-circle PCM signature and that of 20.0-1 lies along the great-circle signature due to Arp 2.  The remaining candidates do not seem to be associated with any known GCs. 

\subsection{Candidate RRL Properties}\label{s:cand_rrls_props}

The information related to or inferred from the RRLS in each candidate is summarised in Table~\ref{t:stream_rrprops}, shown only for the high confidence and tentative candidates. The list of RRLS associated to each candidate is given in Table~\ref{t:cand_rrls_list}.

Table~\ref{t:stream_rrprops} summarises $N_{RR}^{all}$ the total number of RRLS associated to each pole detection, the expected purity of each detection and $N_{RR}^{exp}$ the expected number of RRLS that would truly belong to the stream candidate, i.e. excluding contaminants. The expected purity is the fraction of stars expected to actually belong to a stream in each detection, after accounting for the contribution of background contaminants, estimated by integrating the pole counts in the smoothed PCM. The fractions of RRLS of Oosterhoff type I (OoI), Int (OoInt) and II (OoII) are also reported. Stars were classified as OoI, OoInt and OoII if $\Delta\log P\geqslant-0.005$, $-0.005\leqslant \Delta\log P<-0.04$ and $\Delta\log P\leqslant-0.04$ respectively, with $\Delta\log P=-0.14V_{amp}-0.12-\log P$, where  $V_{amp}$ and $P$ are the light curve V-band amplitude and period, following \citet{Kunder2009} and \citet{Clement1999}.

To report an `expected' Oosterhoff type for each candidate we cannot simply look at the relative fractions of stars of each Oosterhoff type, as this will always be dominated by the more abundant OoI stars. Nevertheless, we can use our knowledge of the typical fractions expected for each Oosterhoff type from the full CSS+SSS catalogue, which are respectively 71\%, 14\% and 15\% for OoI, OoInt and OoII RRLS. We use these fractions, and the total number of RRLS observed in each candidate $N_{RR}^{all}$, to compute the expected number of stars of each Oosterhoff type and compare these with the actual observed number of RRLS of each Oosterhoff type. We report as the expected Oosterhoff type that for which the observed number gives the smallest Poisson probability, when this probability is smaller than $0.05$. The expected Oosterhoff type and Poisson probability are reported in Table~\ref{t:stream_rrprops}. Note that, although the CSS catalogue reports spectroscopic and photometric metallicities \citep{Drake2013b,Torrealba2015}, we chose not to report a mean metallicity for the candidate streams as this will be dominated by the metallicity of `normal' halo contaminant stars. Lacking kinematical information, these contaminant RRLS cannot be distinguished from those that belong to the identified streams, leading to an erroneous metallicity estimate.

The absolute magnitude and total luminosity of the stream candidates, inferred from $N_{RR}^{exp}$, the expected number of RRLS in each stream, are also reported along with their respective 1$\sigma$ confidence intervals. The absolute magnitude $M_V$ was estimated using Bayesian Inference and the well-known linear relationship between the absolute magnitude and the (log) number of RRLS in a stellar population \citep{Suntzeff1991,Vivas2006,Mateu2009}. This inference thus corresponds to the absolute magnitude of a system given an observed number of \rrab~stars, $N_{ab}$, so it will correspond to the stream's absolute magnitude in most cases; however, if the candidate contains the majority of satellite progenitor stars this estimate reflects the total $M_V$ of the progenitor plus the stream.

We write the posterior probability density  $P(M_V|\log N_{ab})$ as the product of a Gaussian likelihood $L=P(\log N_{ab}|M_V)=\exp{(\log N_{ab} - \log N_{ab}^T(M_V))^2/(2\sigma_N^2)}$, which assumes a (constant) Gaussian uncertainty in the $M_V$--$\log N_{ab}$ relationship; and a power-law prior $P(M_V)=10^{0.1(M_V+5.)}$, given by the luminosity function of Milky Way satellite galaxies obtained by \citet{Koposov2008}. We express the $M_V$--$\log N_{ab}$ relationship as $\log N_{ab}=aM_V+b$, with $a=-0.2402$ and $b=-0.6167$, and assume a standard deviation $\sigma_N=0.328$ about this relationship. These values were estimated from a least squares fit using data from a compilation of Milky Way classical and ultra-faint dwarf galaxies, as detailed in Appendix~A\footnote{Note that this relationship is valid for $M_V>-5$ and $N_{ab}\geqslant3$}. The posterior probability density for the luminosity is expressed in terms of that for the absolute magnitude as $P(L_V|\log N_{ab}) = P(M_V(L_V)|\log N_{ab})/L_V$. In Table~\ref{t:stream_rrprops} we report the posterior mode and 1$\sigma$ confidence intervals for $M_V$ and $L_V$. 

\begin{figure*}
\begin{center}
 \includegraphics[width=0.8\columnwidth]{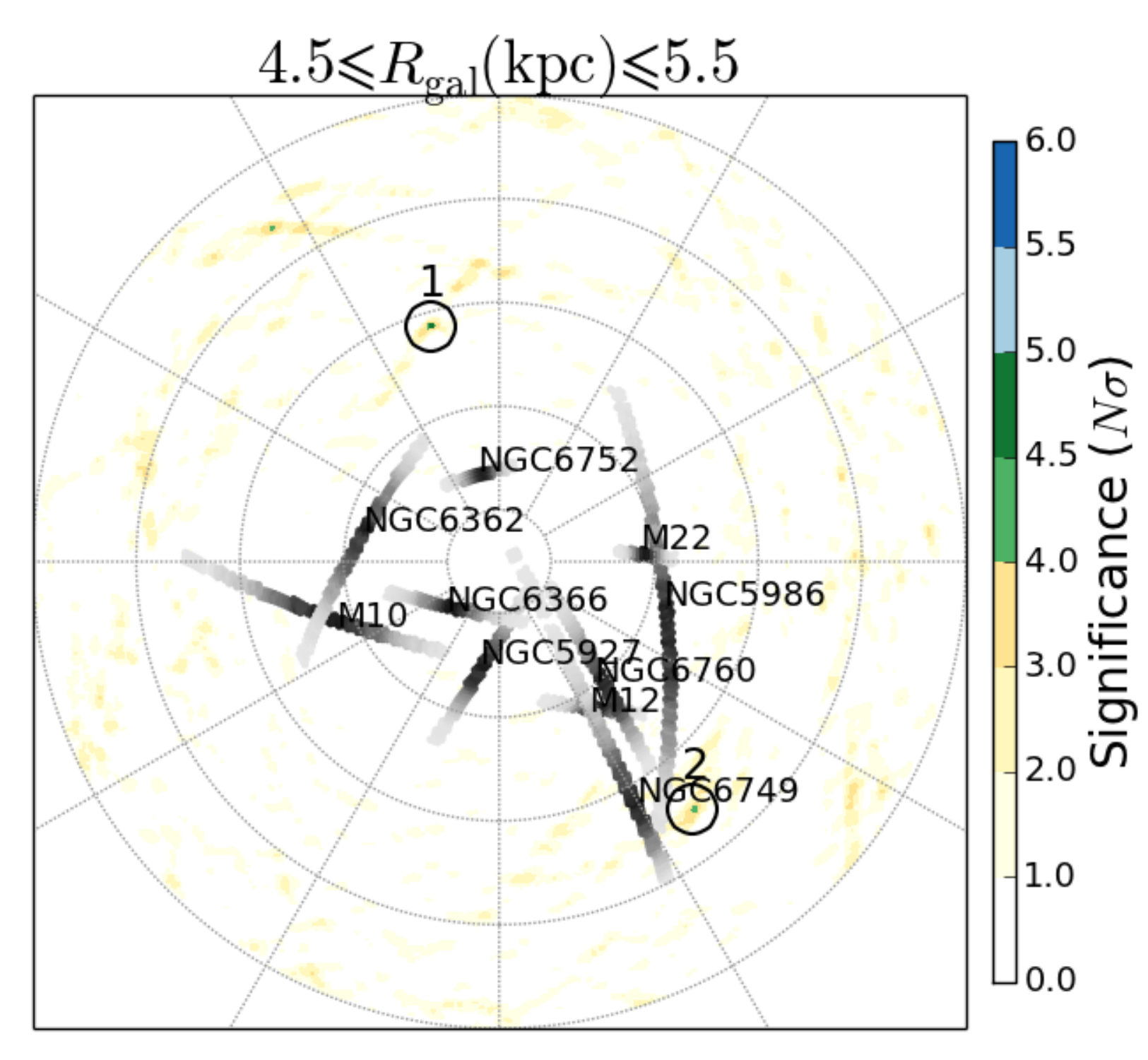} 
  \includegraphics[width=1.1\columnwidth]{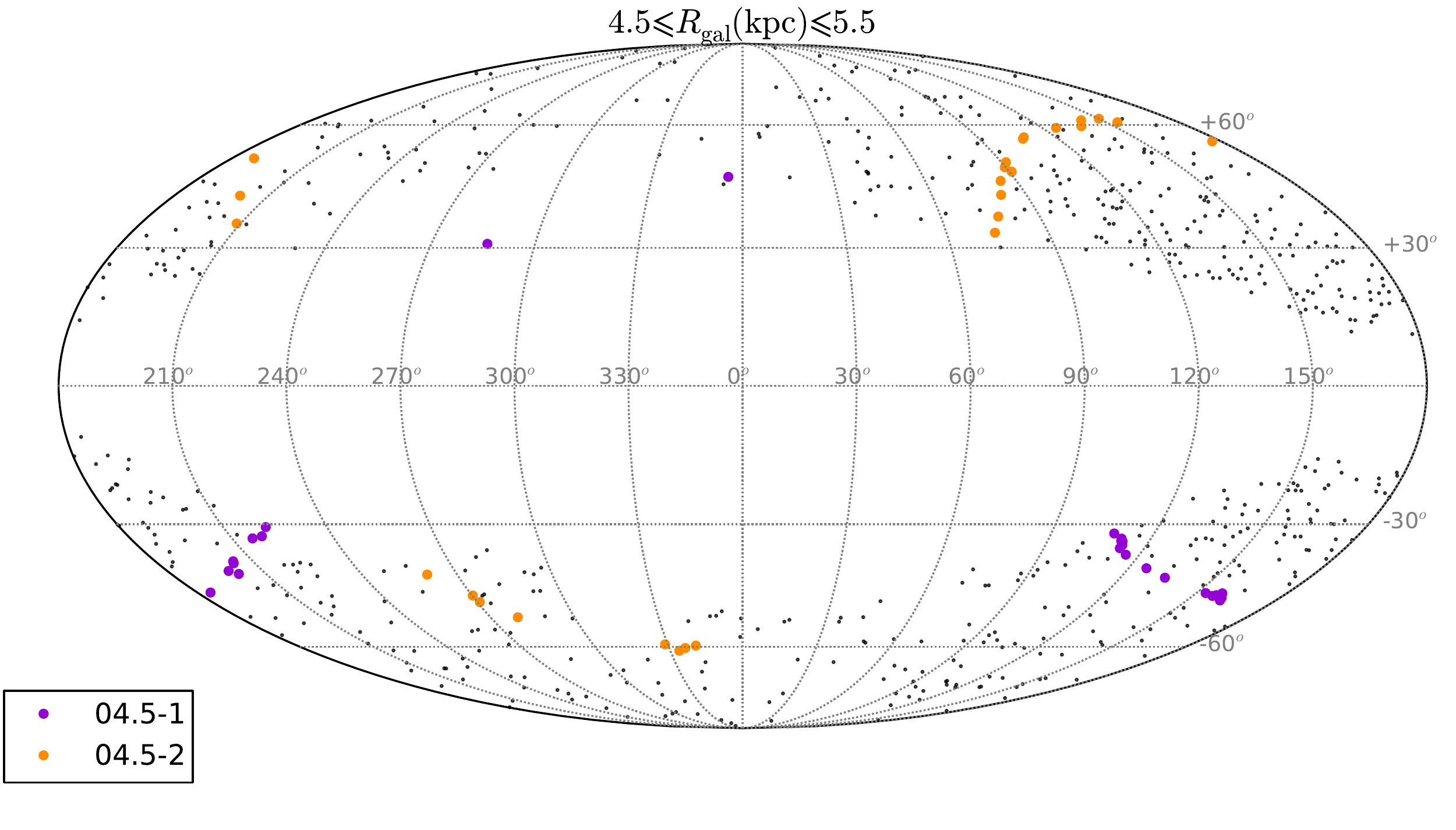}   
 \includegraphics[width=0.8\columnwidth]{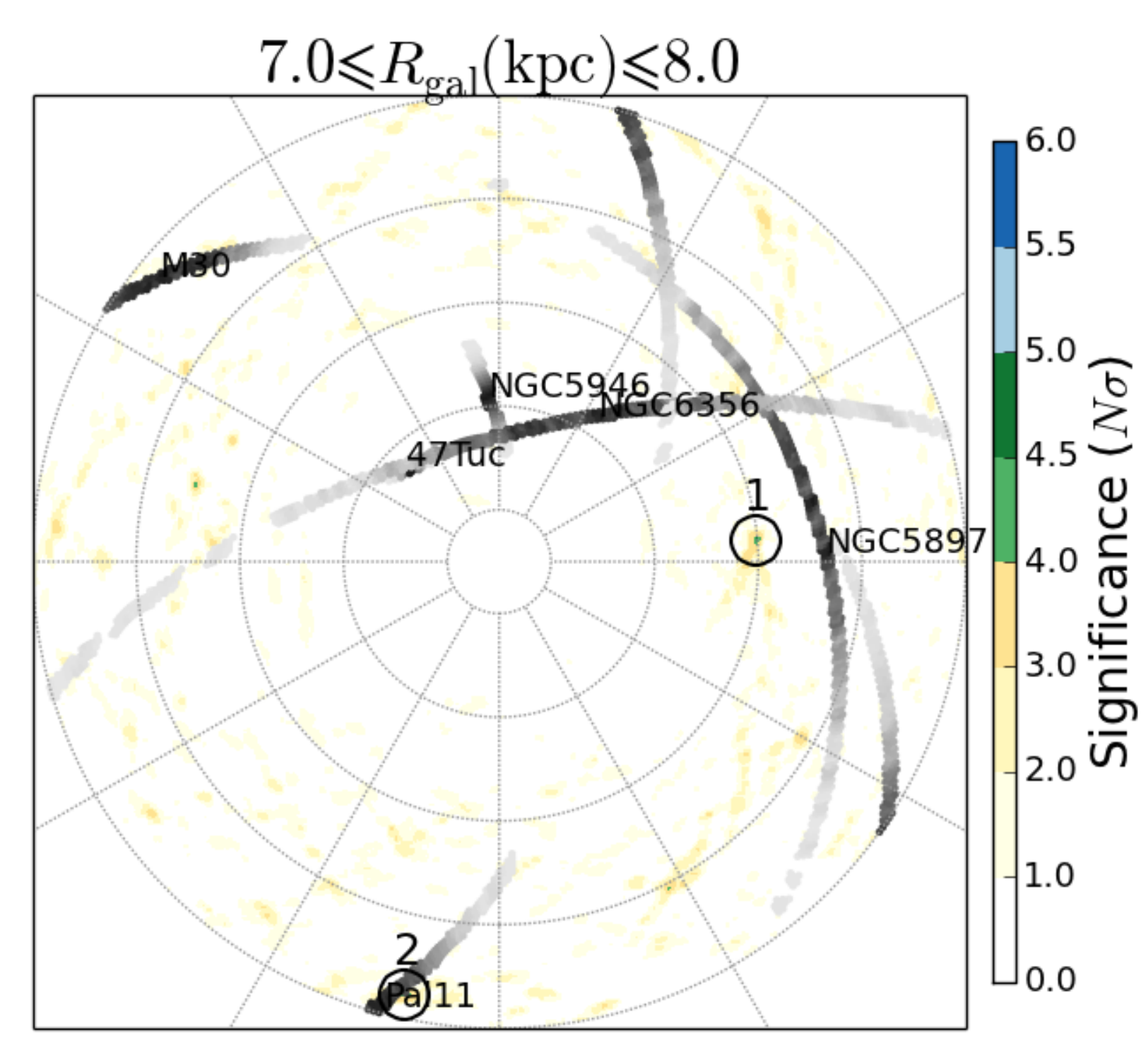} 
  \includegraphics[width=1.1\columnwidth]{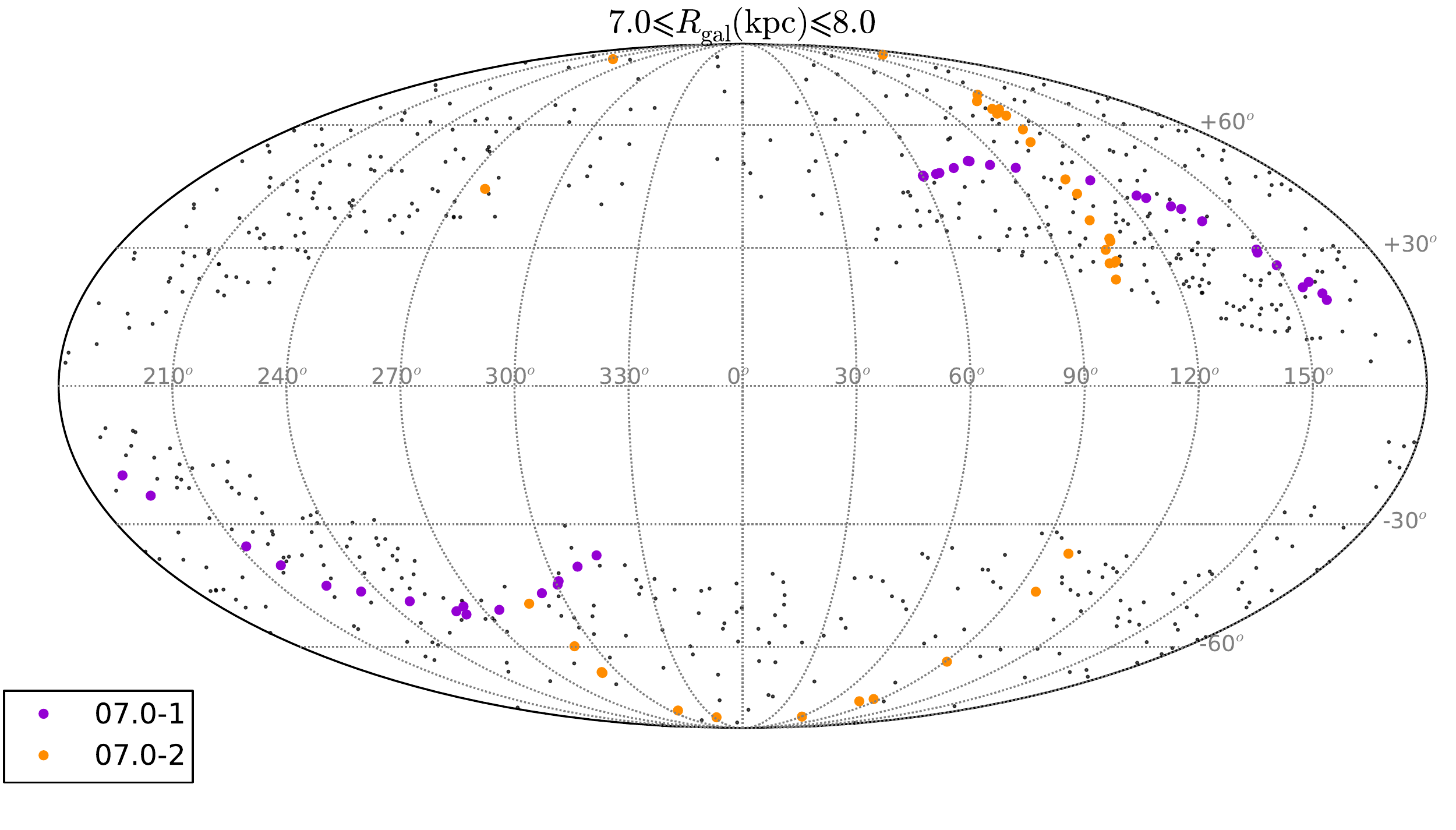}  
 \includegraphics[width=0.8\columnwidth]{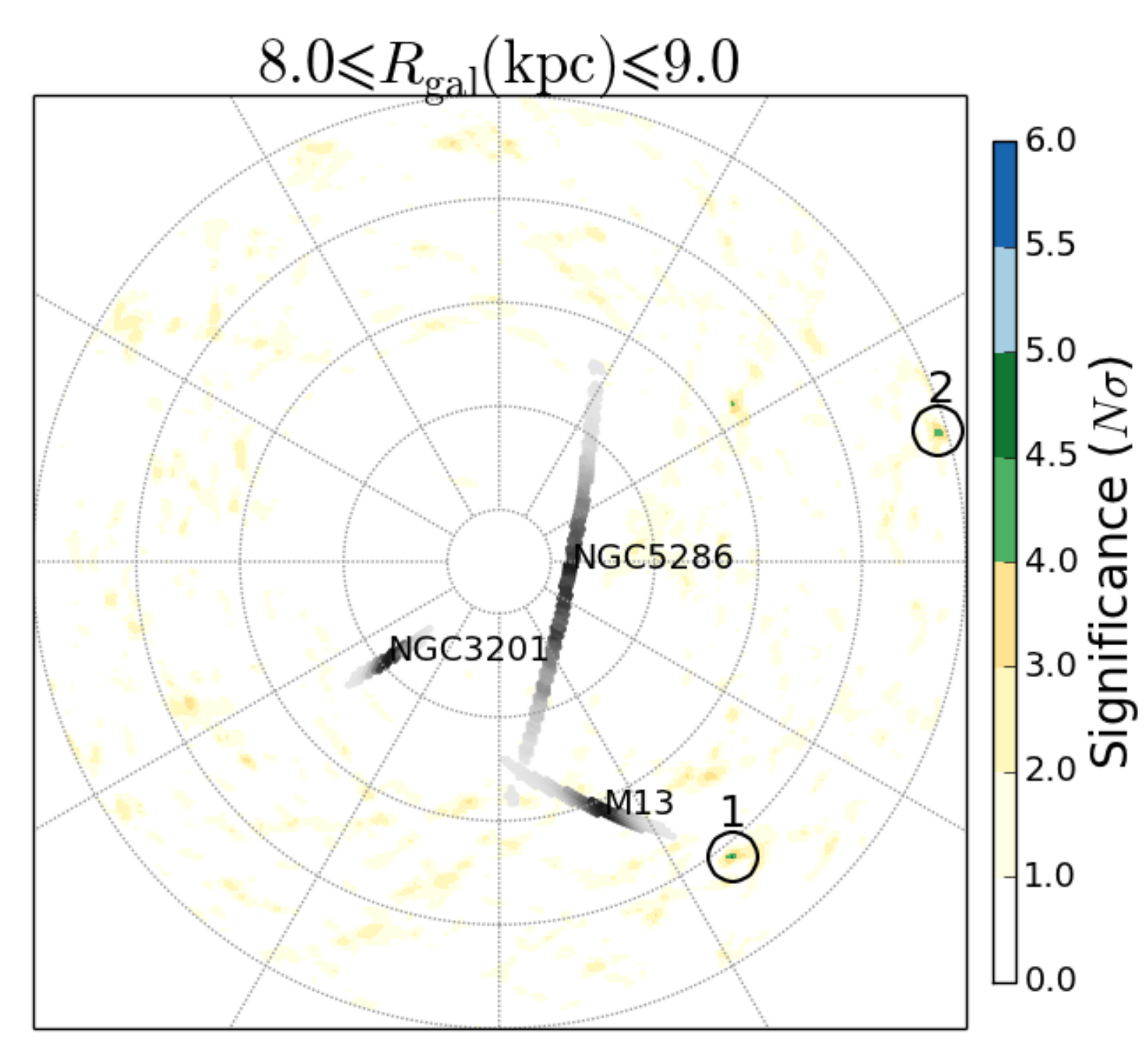}
  \includegraphics[width=1.1\columnwidth]{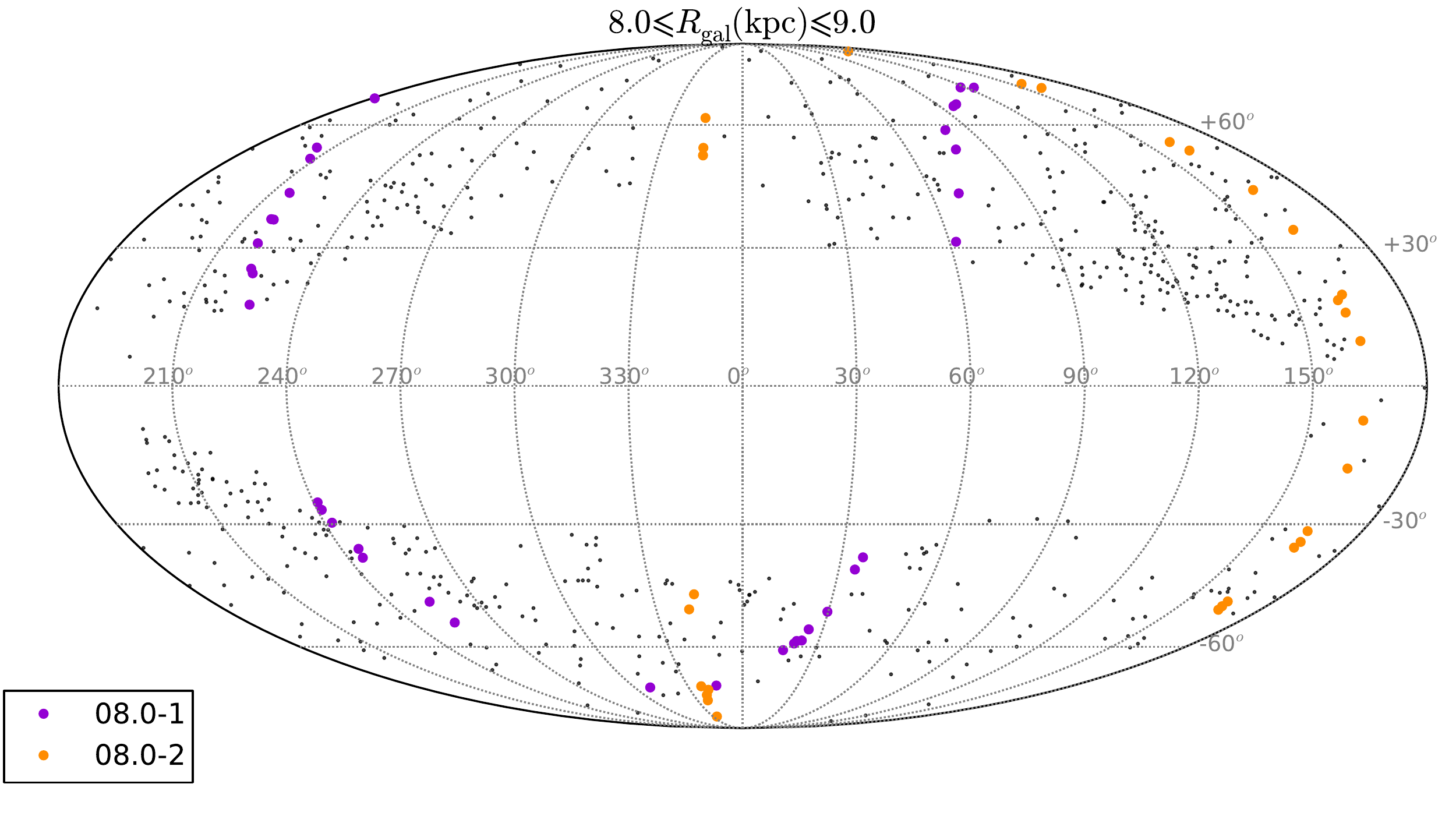}  
  \includegraphics[width=0.8\columnwidth]{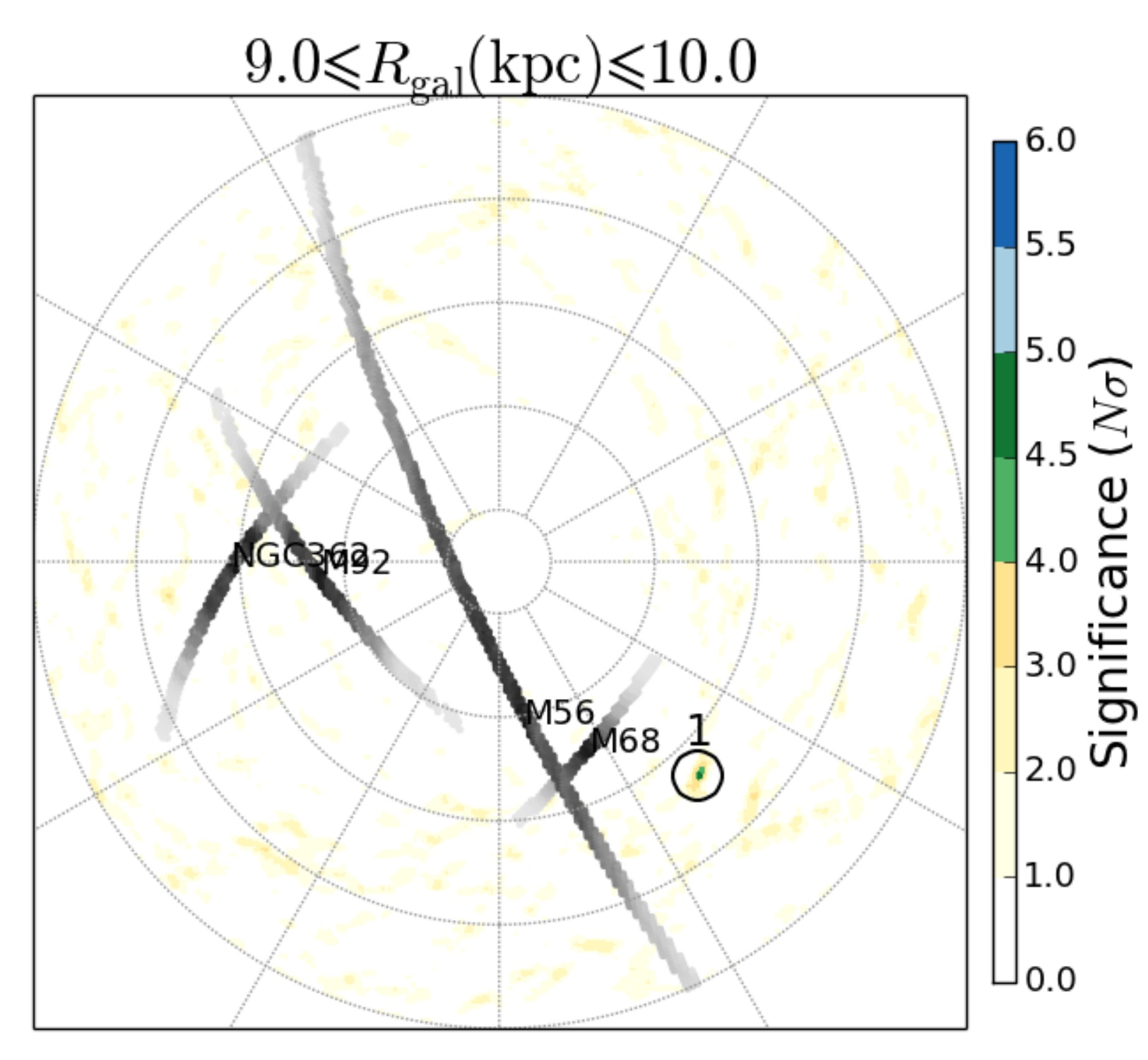}  
   \includegraphics[width=1.1\columnwidth]{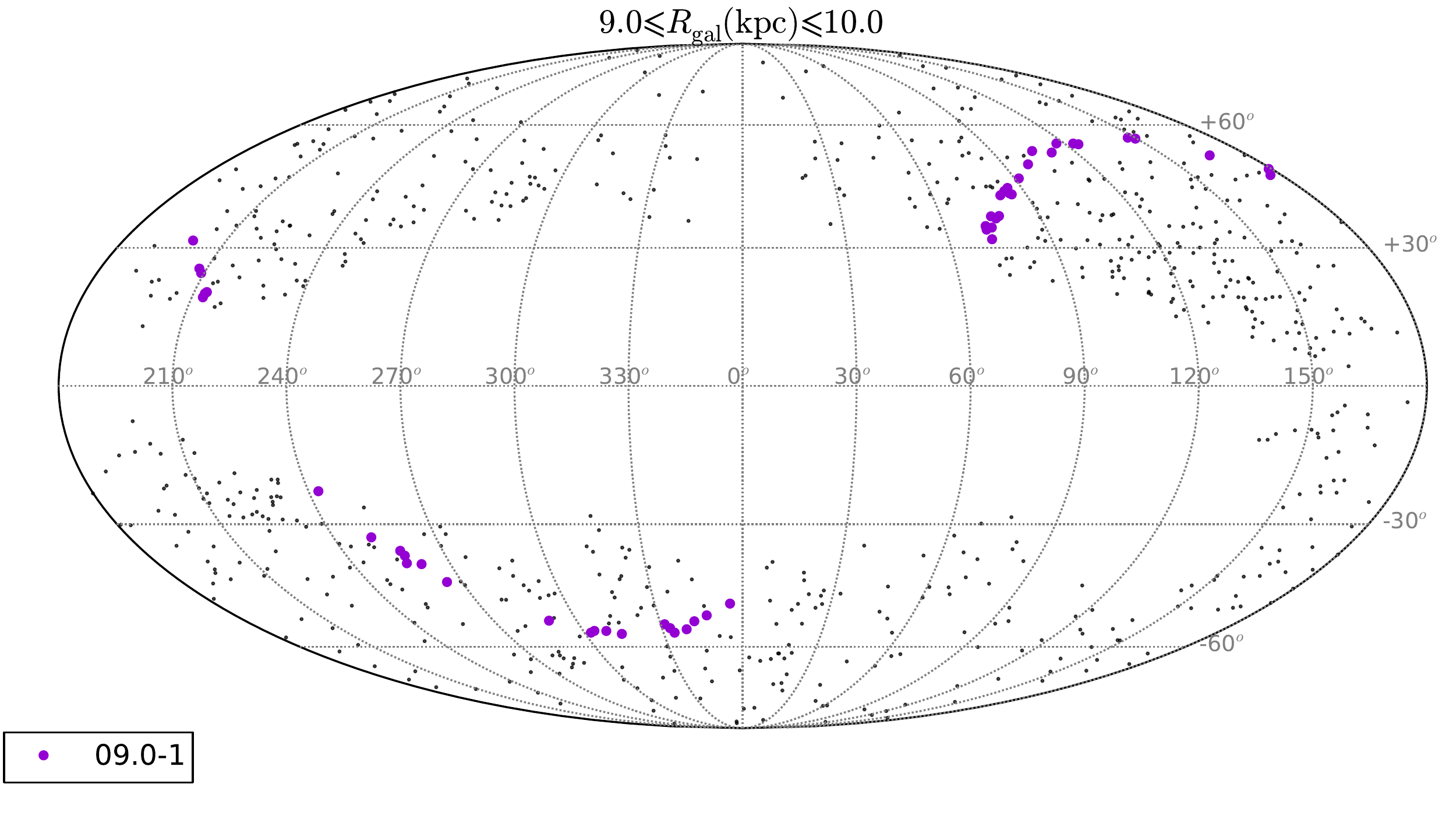}  
 \caption{\emph{Left:} Combined unsharp-masked \gc+\ngc~PCMs. The colour scale is proportional to the detection significance in $N\sigma$ units and significant detections ($>4\sigma$) are labelled and marked with an empty circle. The PCM signatures from known GCs and dwarf galaxies in each distance bin is shown as greyscale,  representing the probability density within $3\sigma$ of each cluster's proper motion errors. \emph{Right:} Mollweide projection map in \emph{Galactocentric} coordinates. The black dots indicate all the CSS+HSOY RRLSs in the given Galactocentric distance bin, the coloured dots represent the RRLSs associated to the detections made in the corresponding \gc+\ngc~PCM as indicated in each map's legend.} 
 \label{f:pcms_detections}
\end{center}
\end{figure*}

\begin{figure*}
\begin{center}
\centering
\includegraphics[width=0.8\columnwidth]{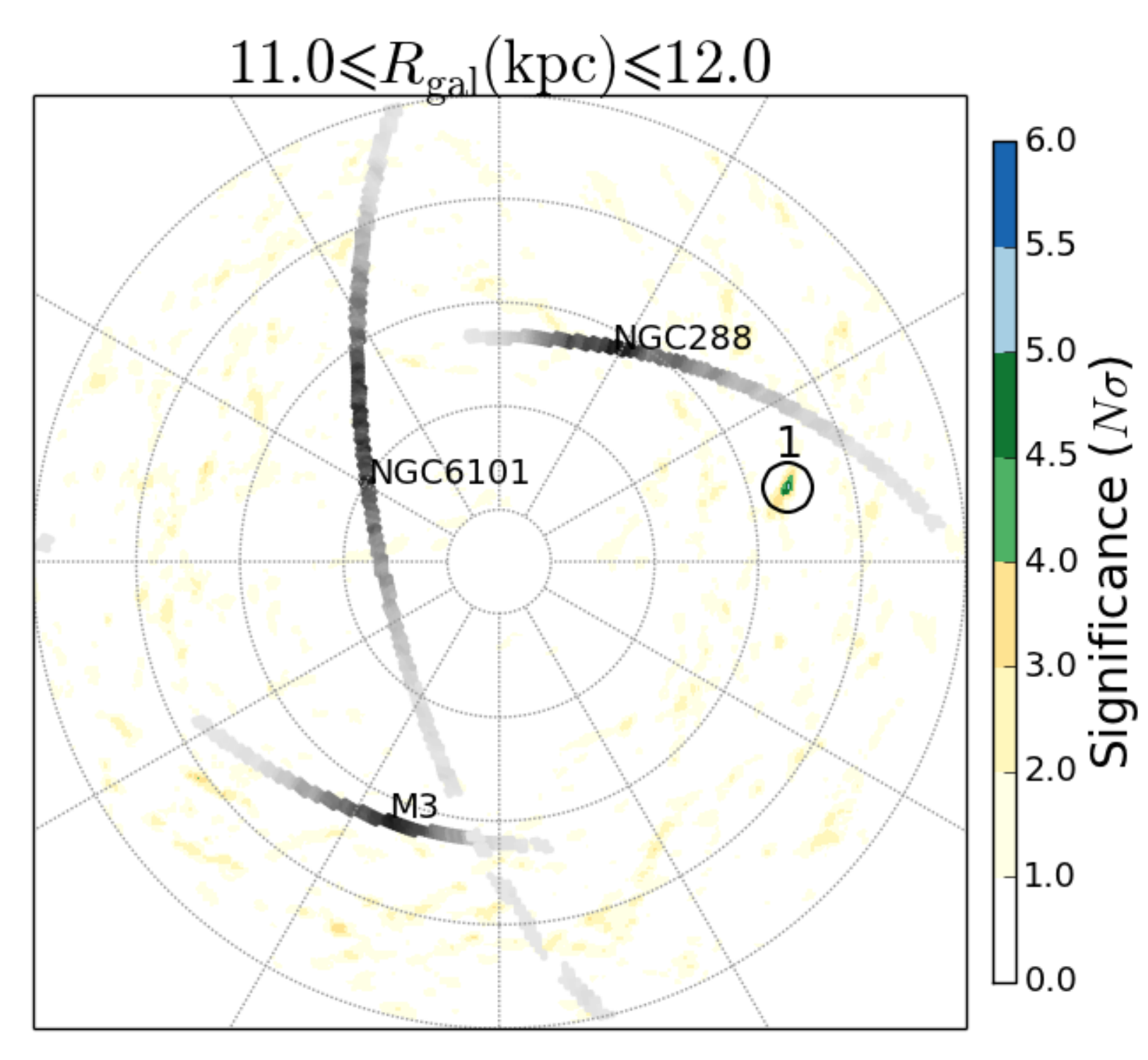} 
\includegraphics[width=1.1\columnwidth]{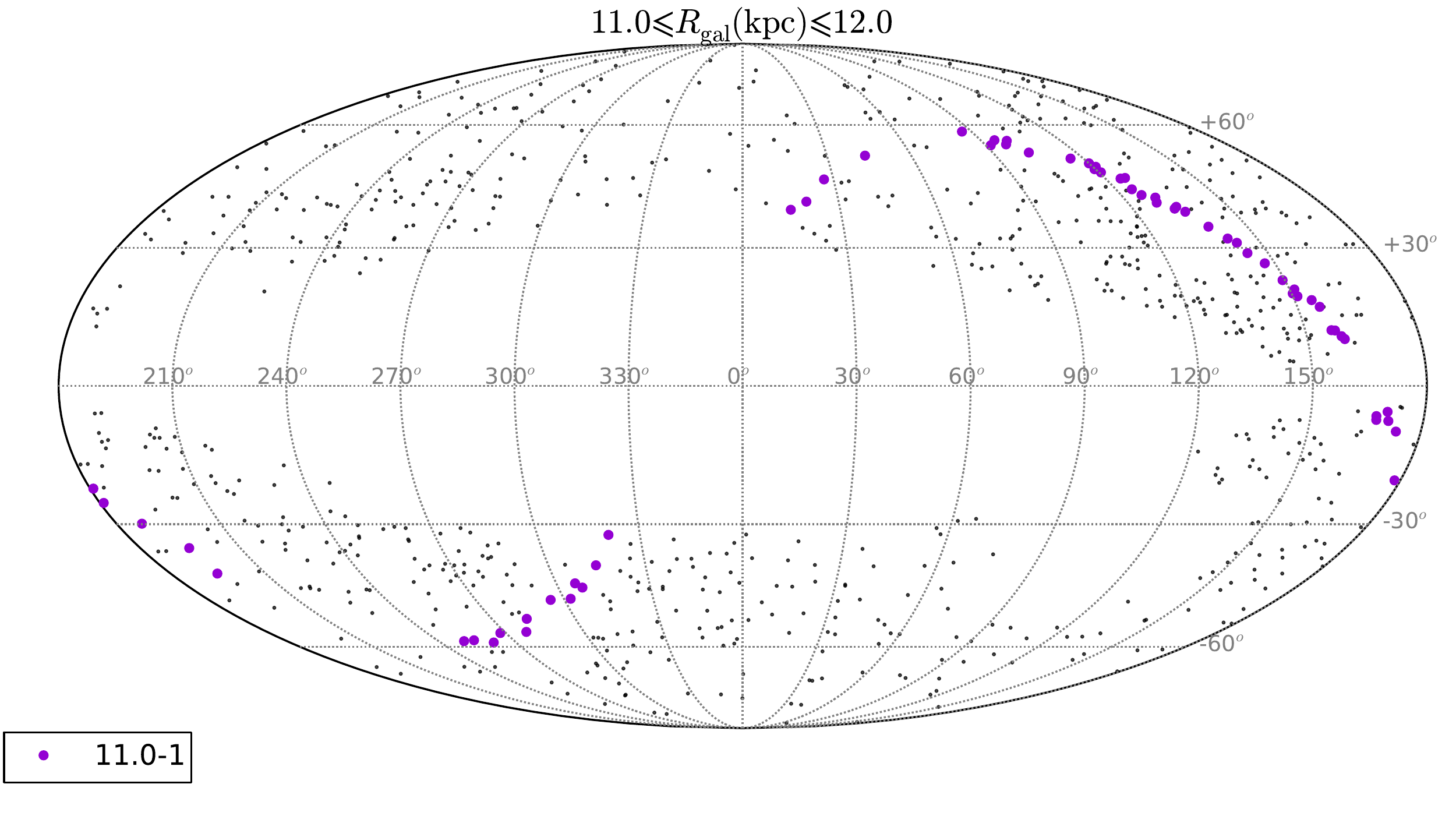}   
\includegraphics[width=0.8\columnwidth]{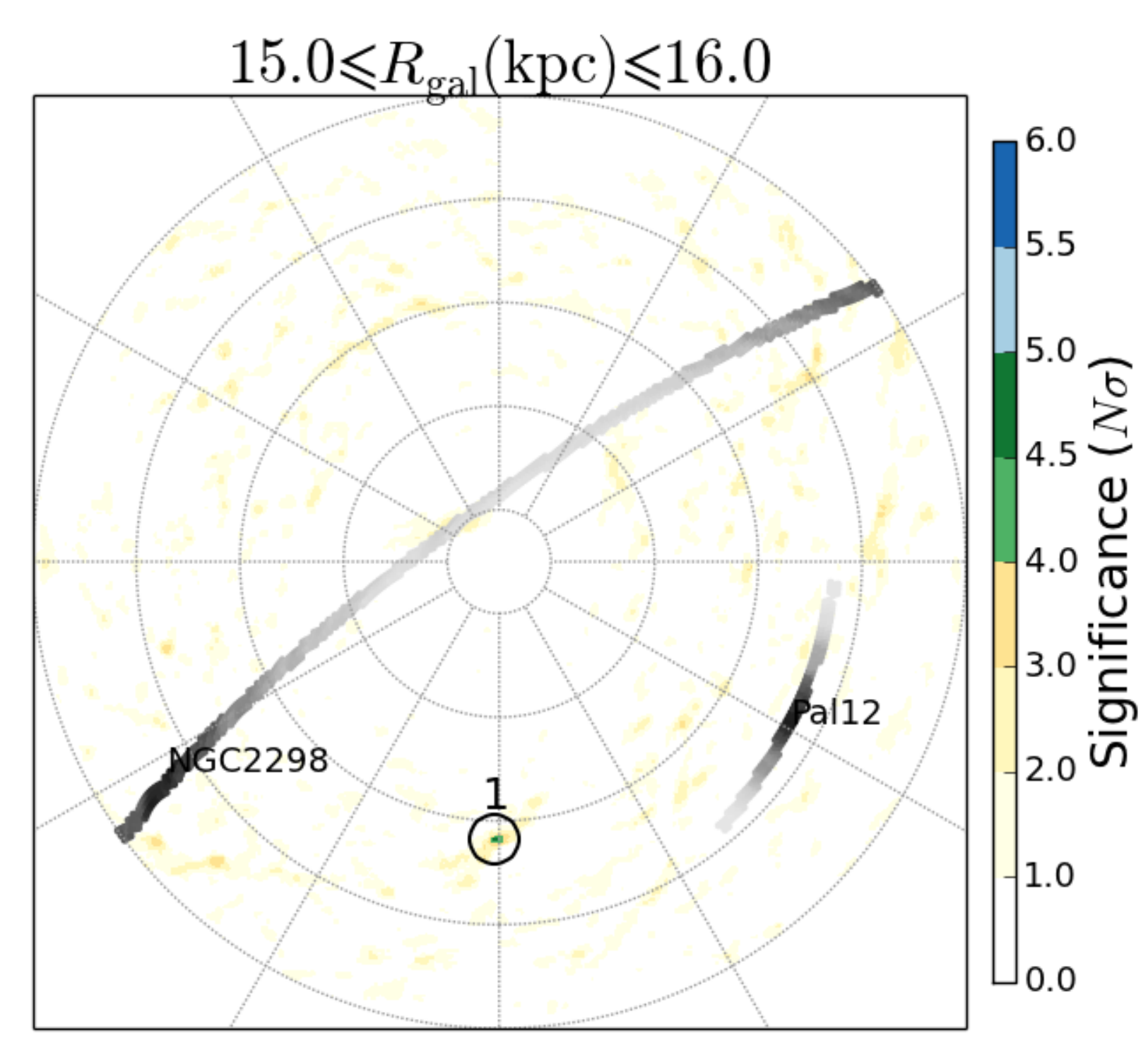} 
\includegraphics[width=1.1\columnwidth]{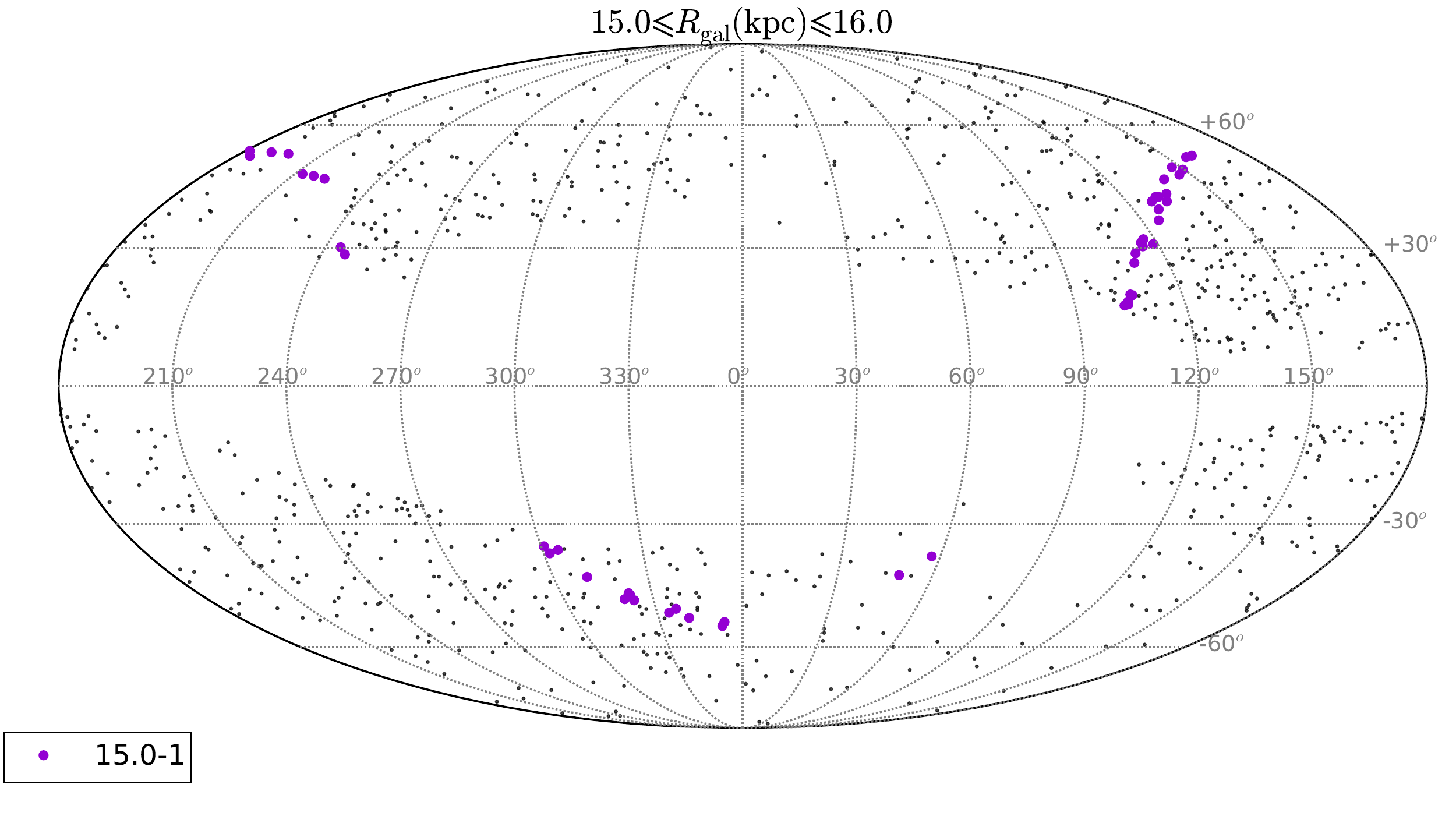}  
\includegraphics[width=0.8\columnwidth]{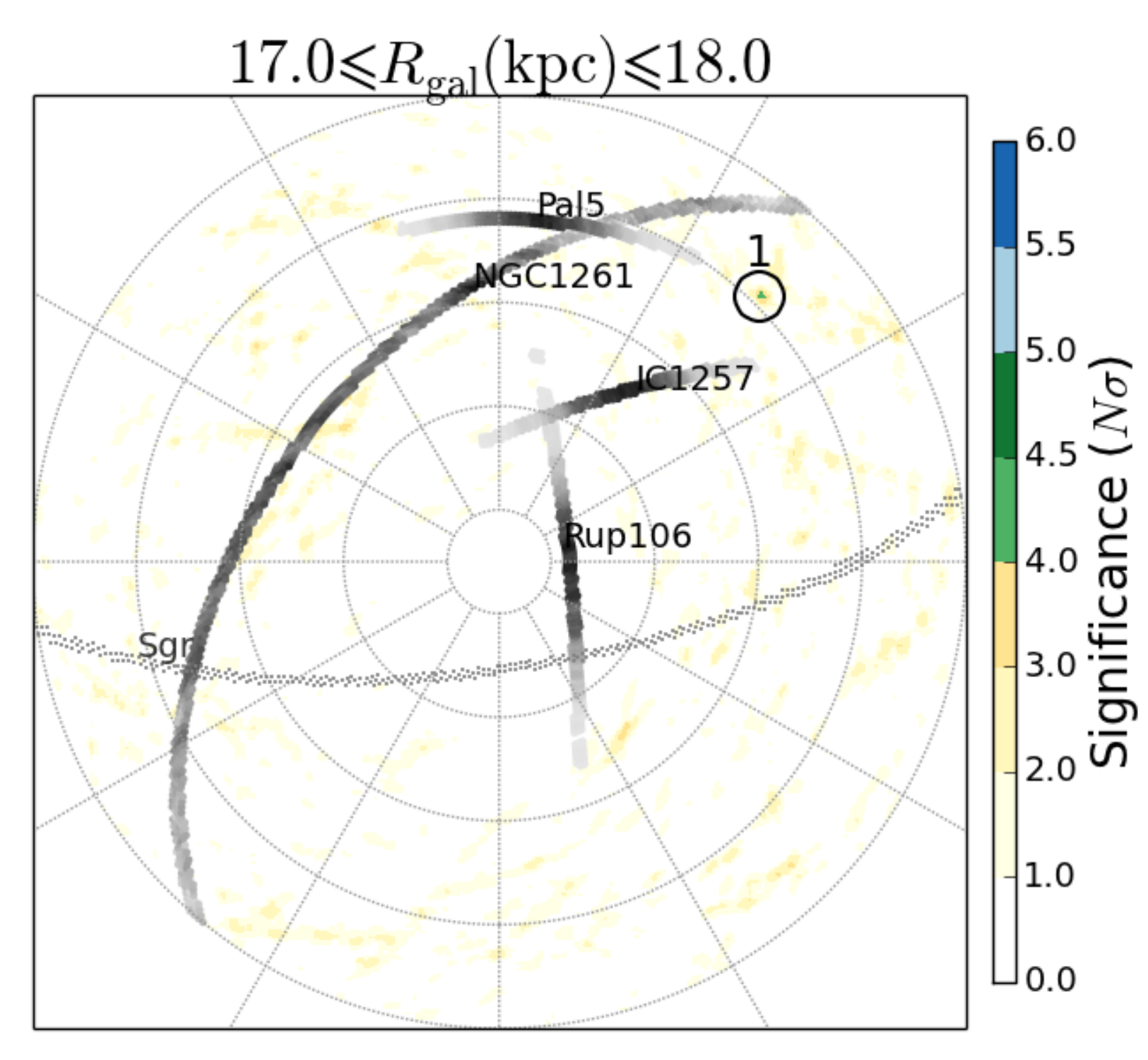}
\includegraphics[width=1.1\columnwidth]{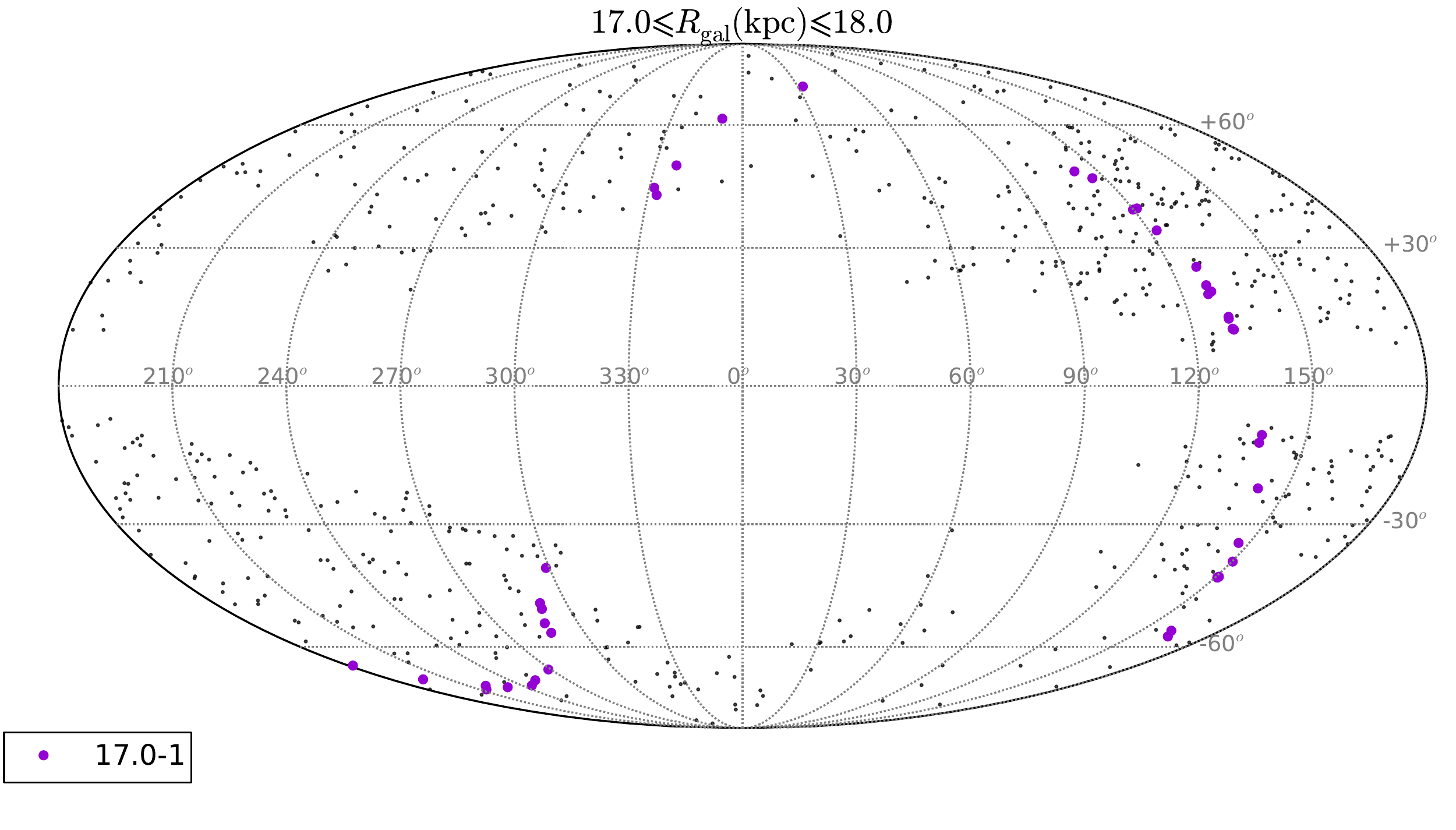}   
\includegraphics[width=0.8\columnwidth]{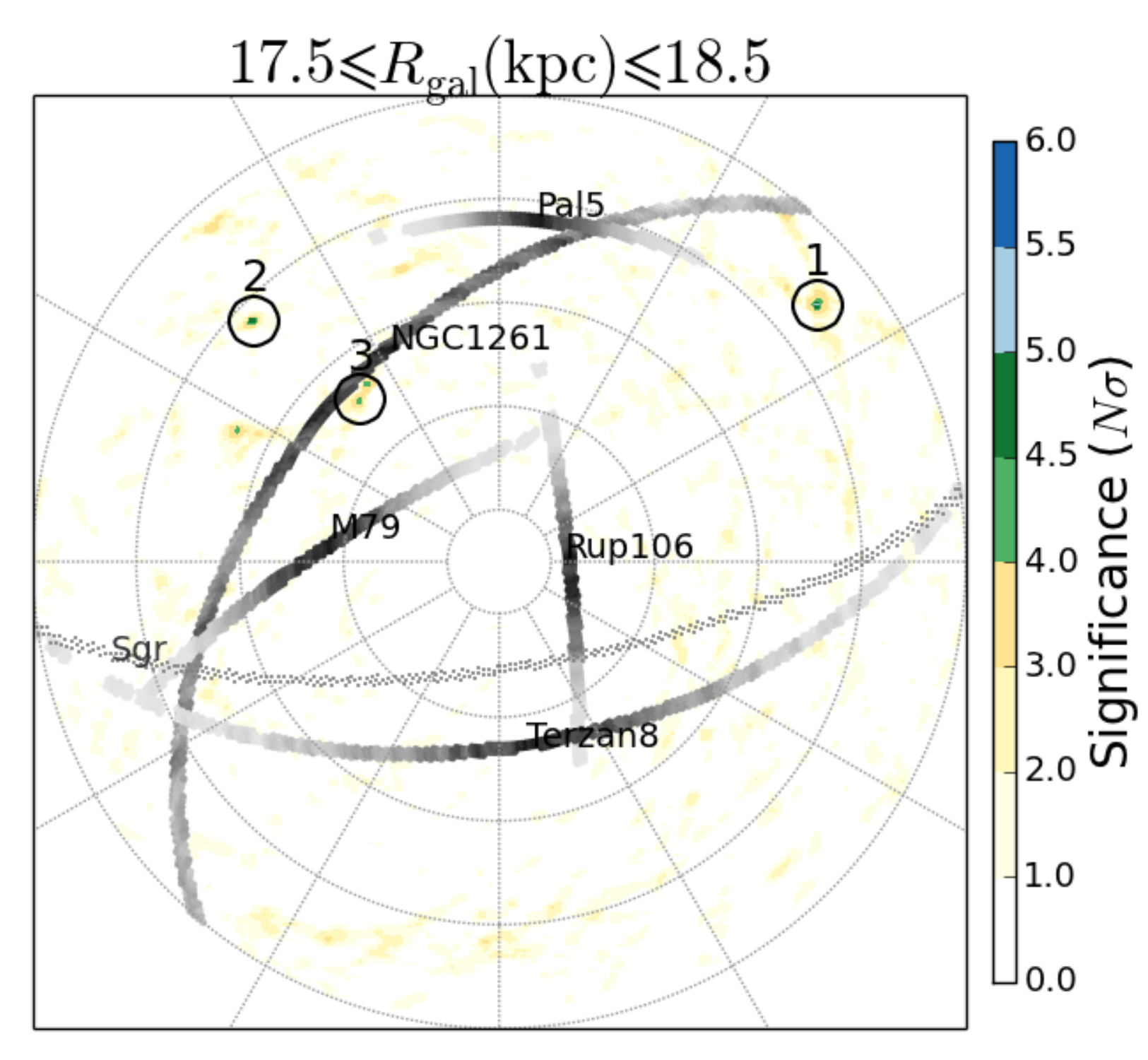}  
\includegraphics[width=1.1\columnwidth]{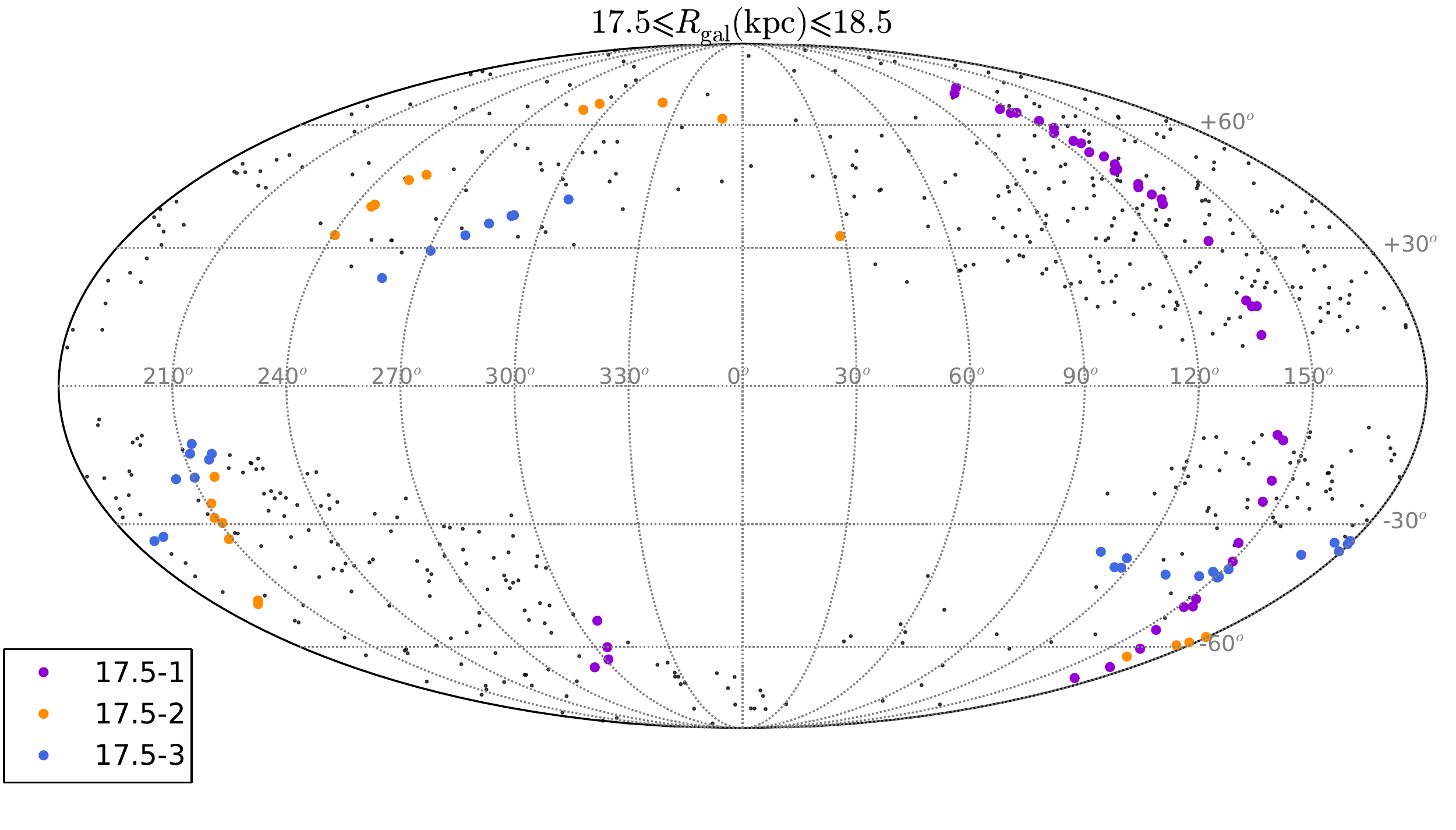} 
   \end{center}
\contcaption{ -- \emph{Left:} Combined unsharp-masked \gc+\ngc~PCMs.  \emph{Right:} \emph{Galactocentric} coordinates map. }
\end{figure*}

\begin{figure*}
\begin{center}
\includegraphics[width=0.8\columnwidth]{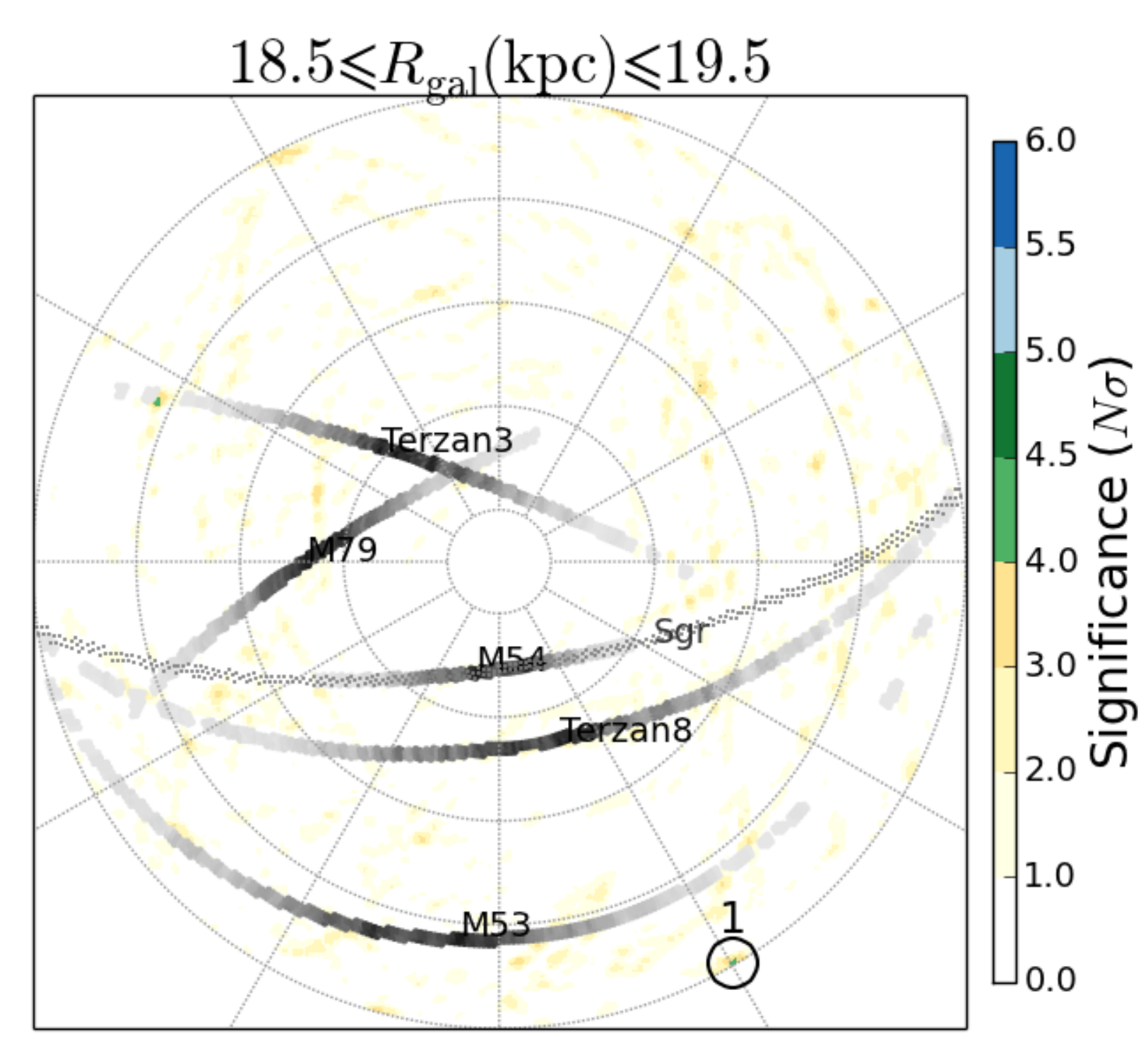} 
\includegraphics[width=1.1\columnwidth]{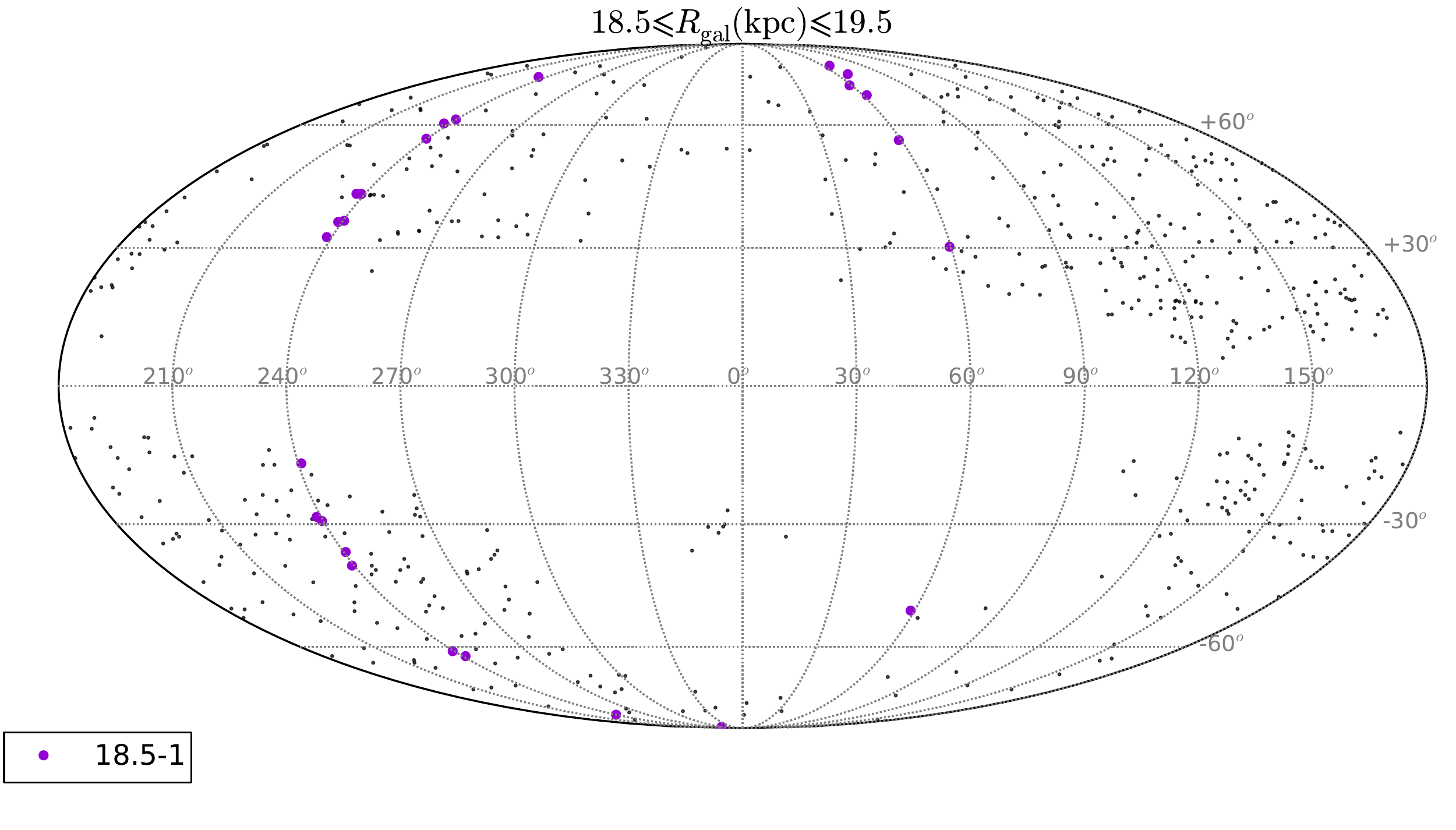}   
\includegraphics[width=0.8\columnwidth]{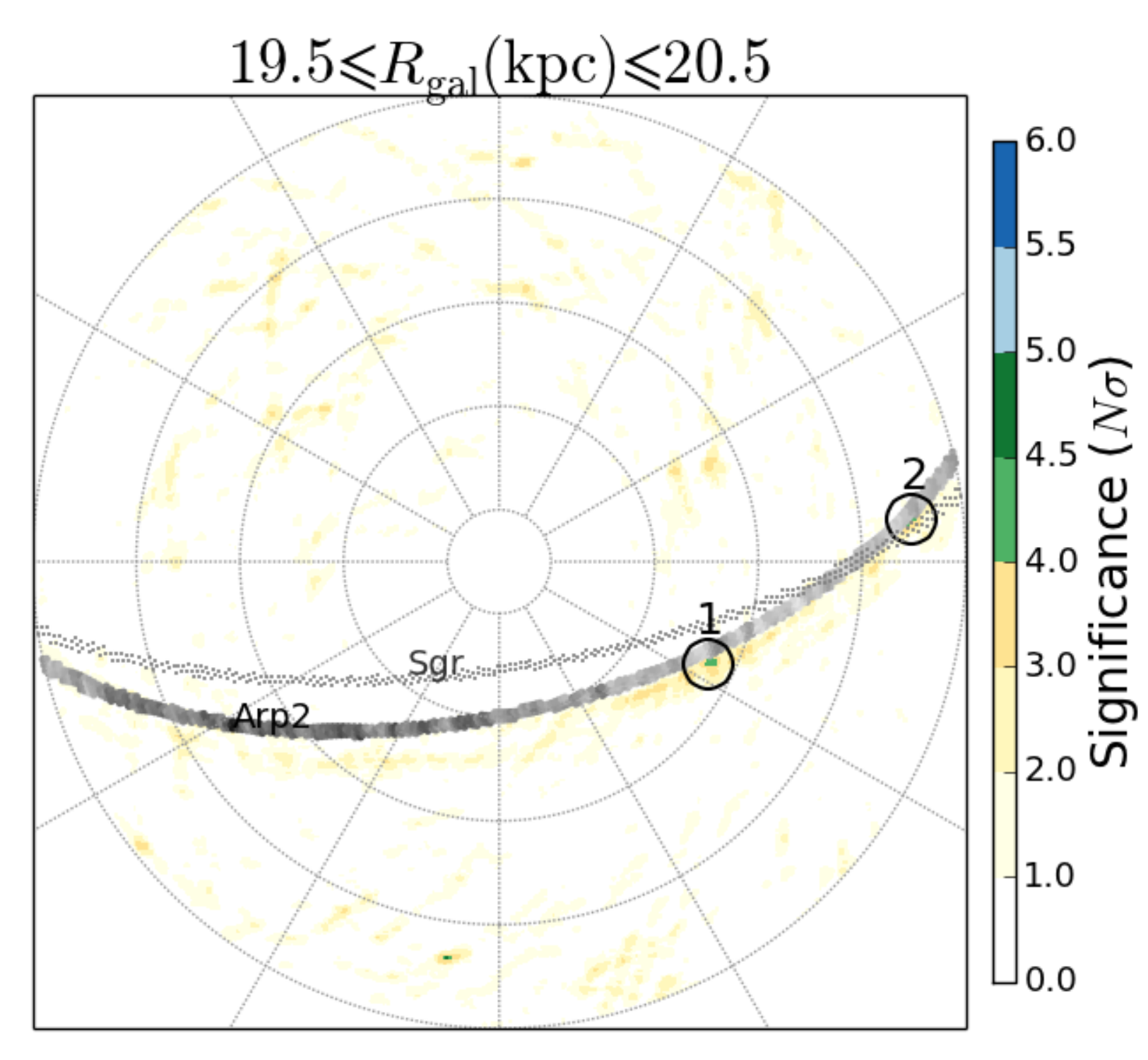} 
\includegraphics[width=1.1\columnwidth]{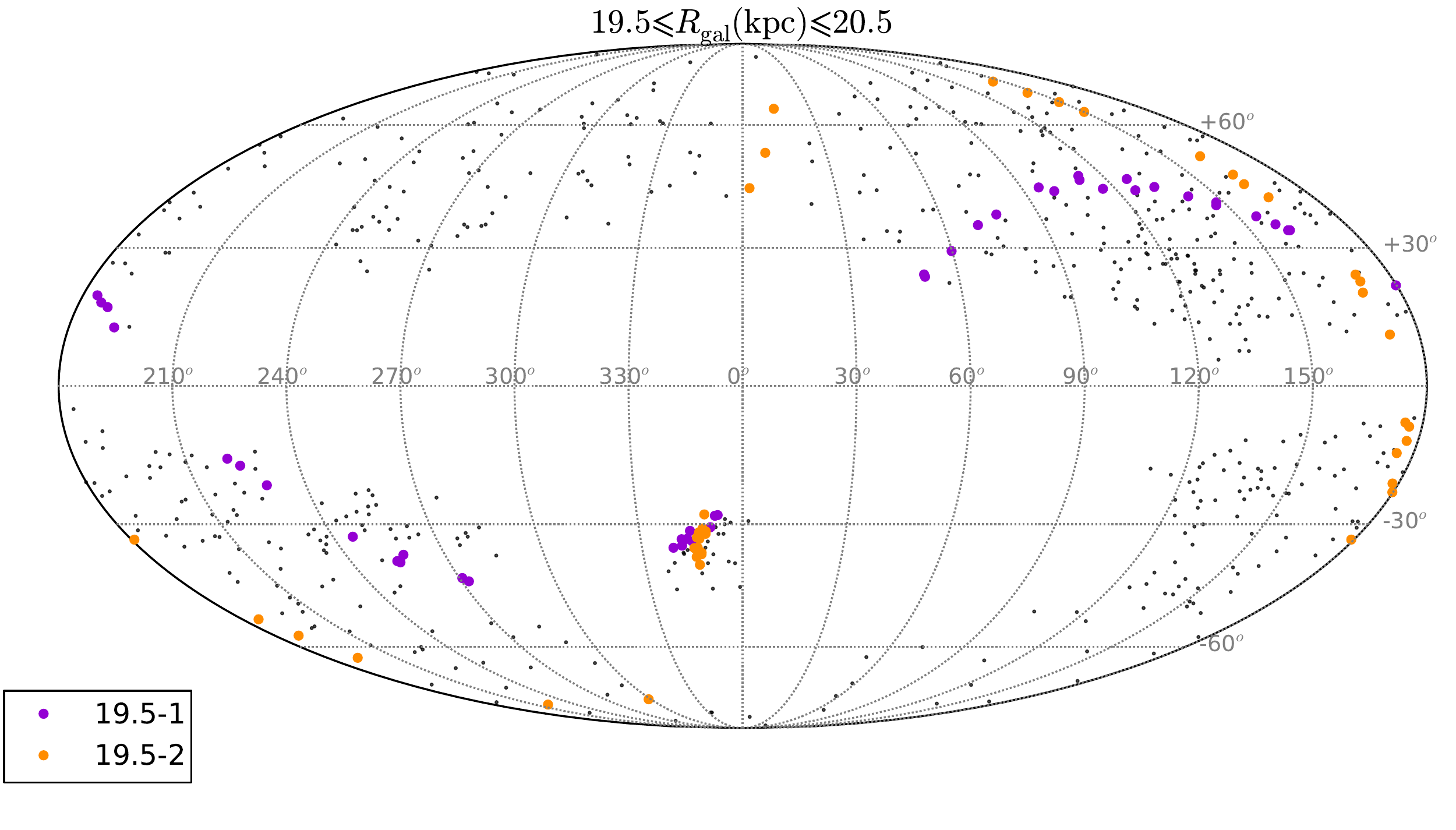}  
\includegraphics[width=0.8\columnwidth]{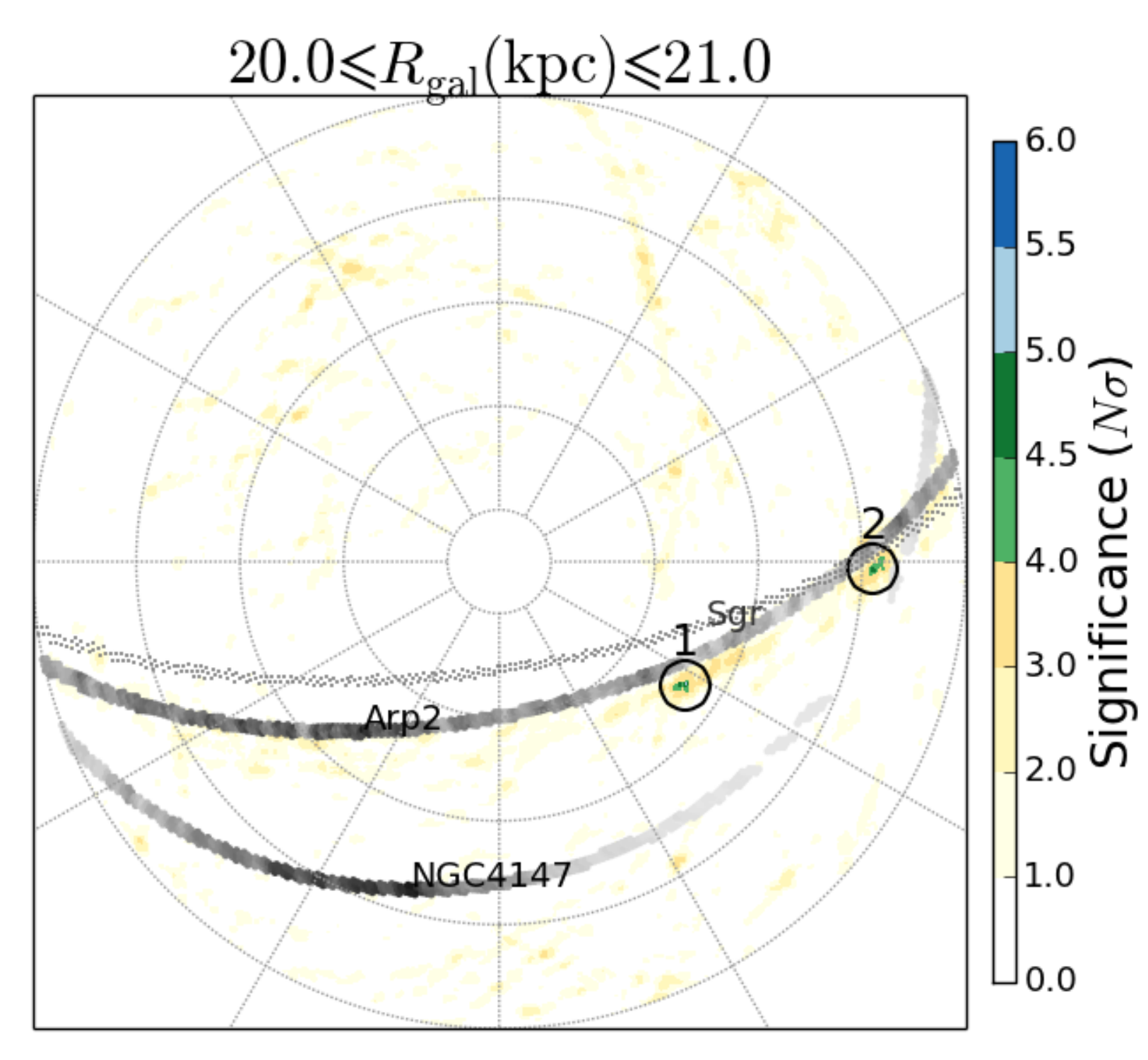}
\includegraphics[width=1.1\columnwidth]{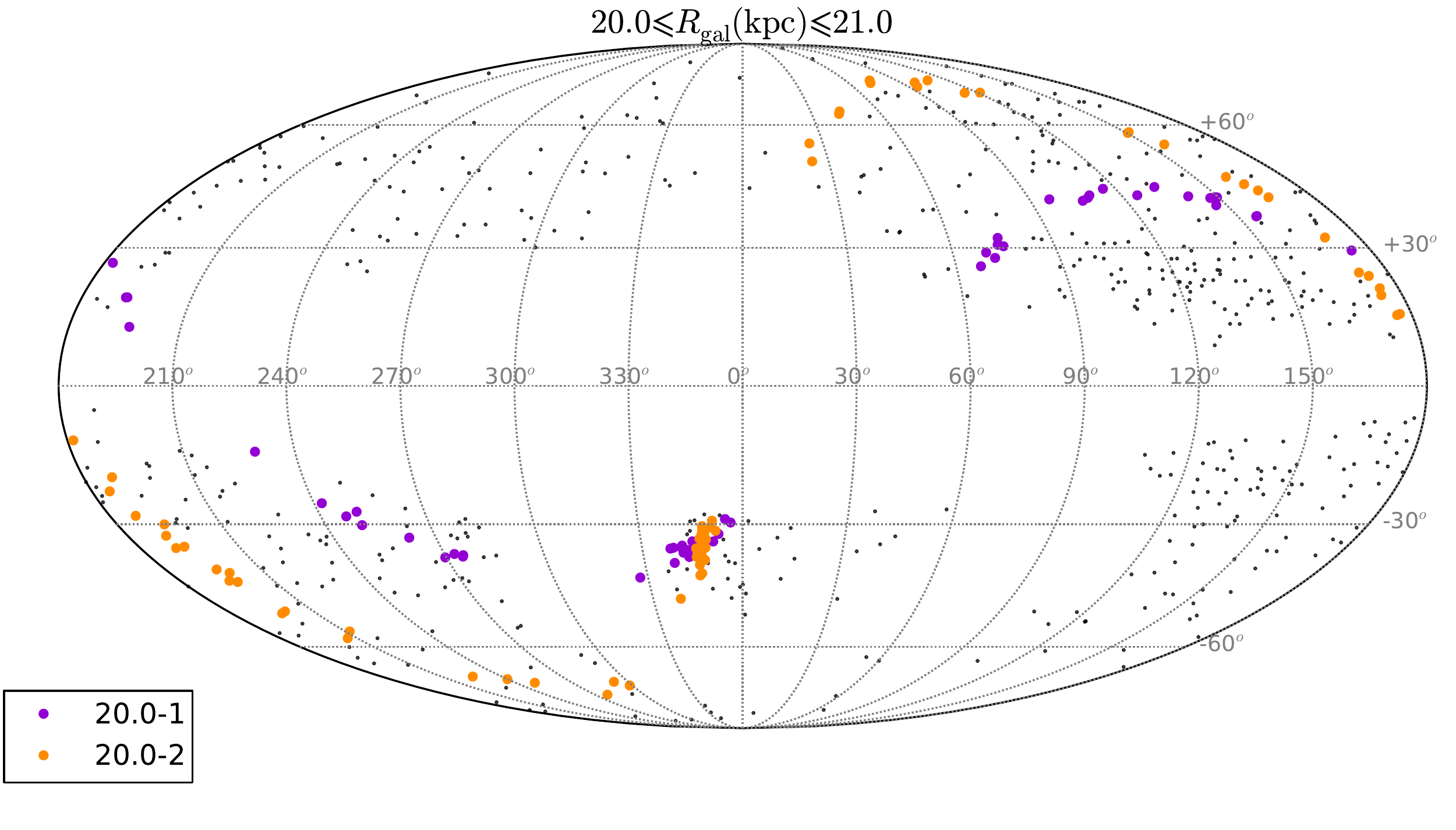}   
\includegraphics[width=0.8\columnwidth]{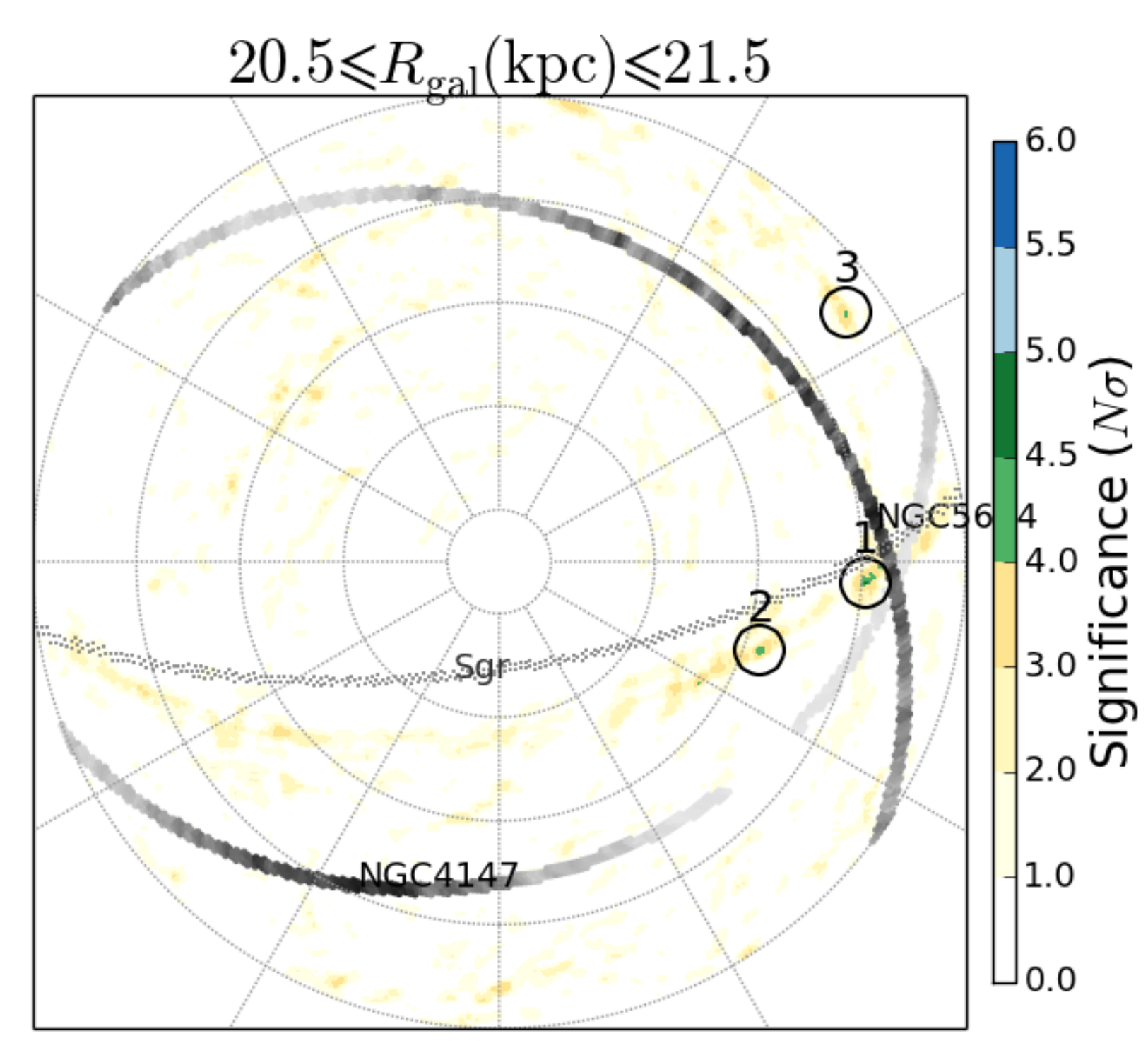}  
\includegraphics[width=1.1\columnwidth]{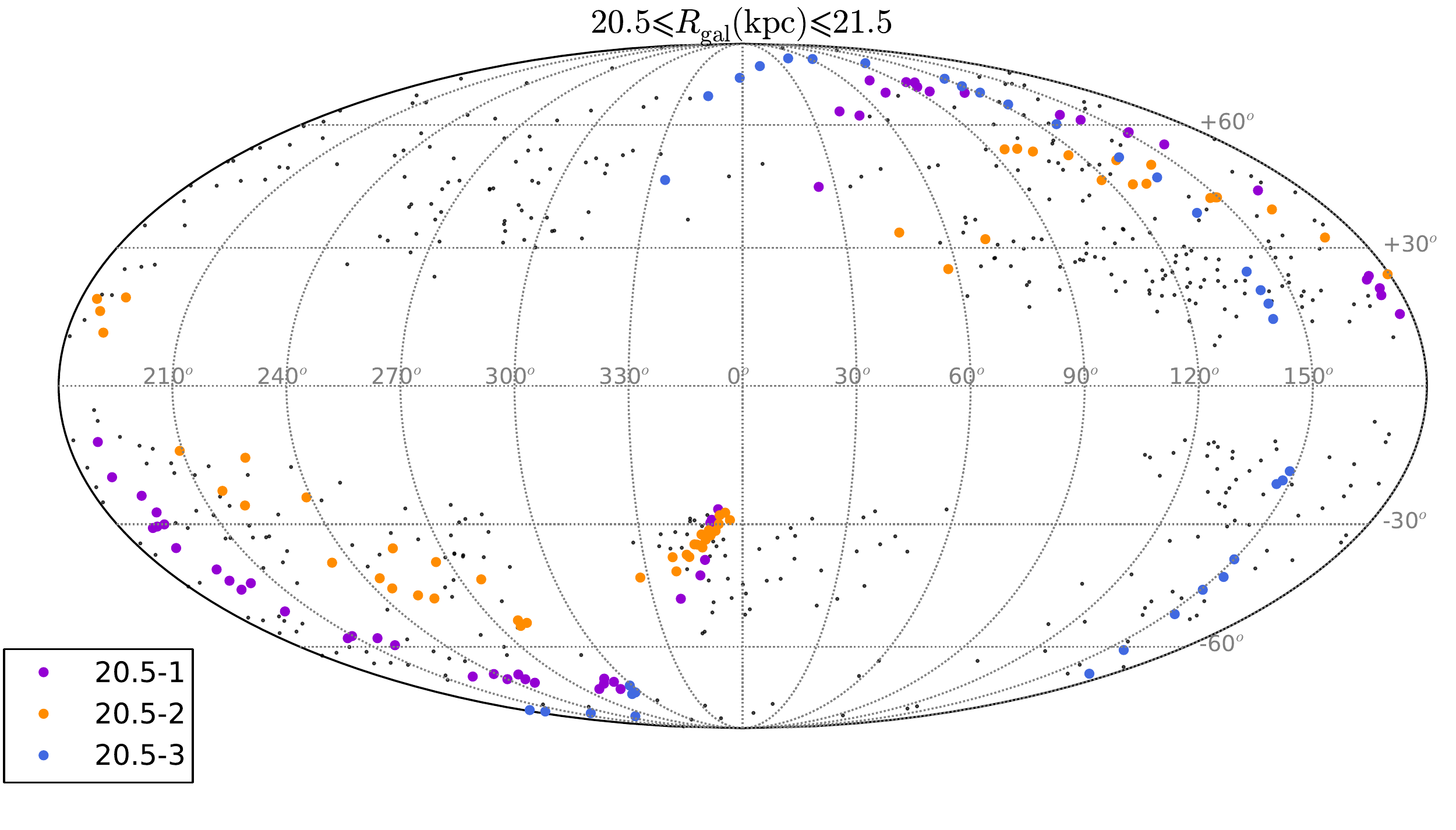} 
\end{center}
\contcaption{ -- \emph{Left:} Combined unsharp-masked \gc+\ngc~PCMs.  \emph{Right:} \emph{Galactocentric} coordinates map. }
\end{figure*}

\begin{figure*}
\begin{center}
\includegraphics[width=0.8\columnwidth]{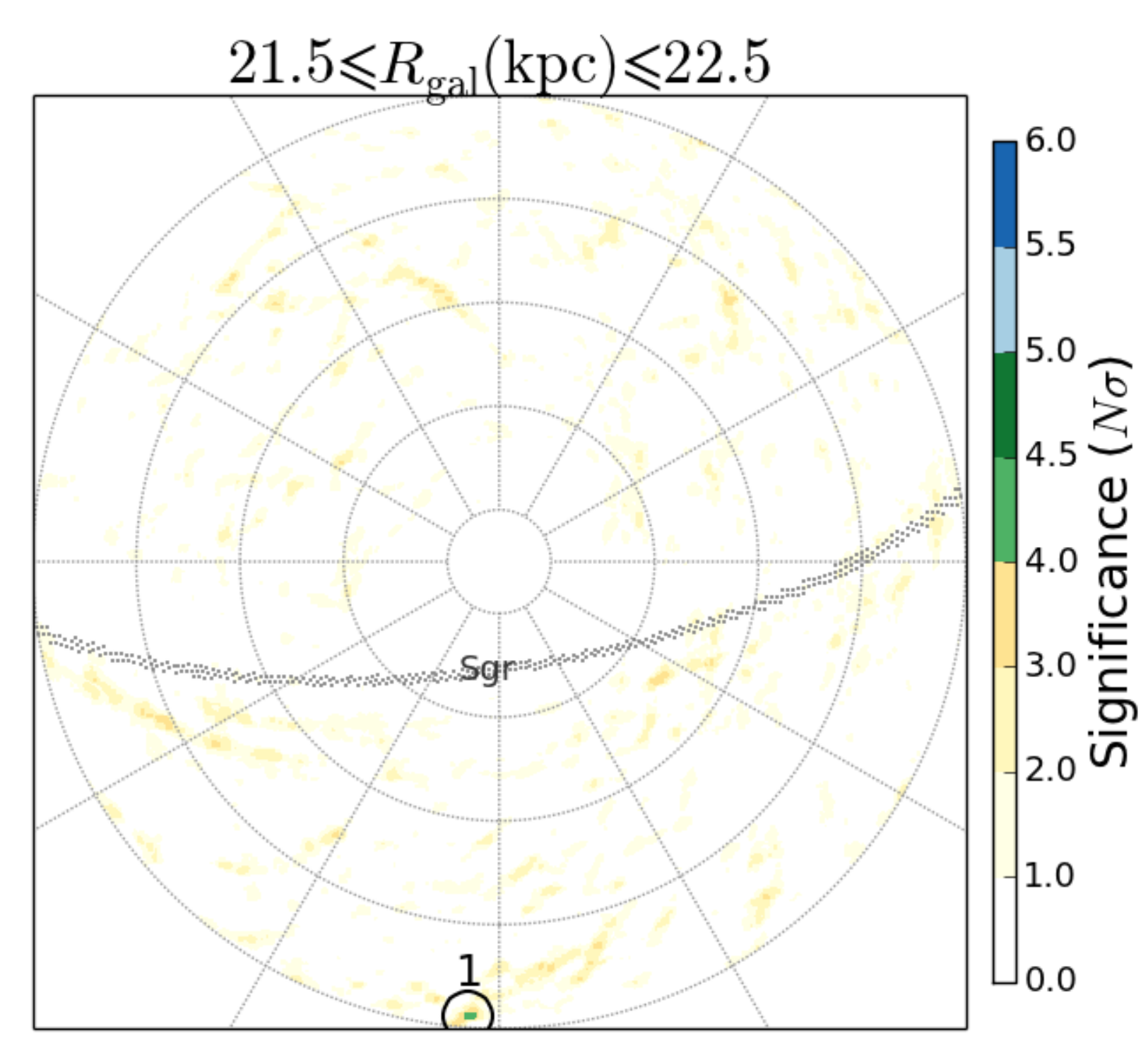} 
\includegraphics[width=1.1\columnwidth]{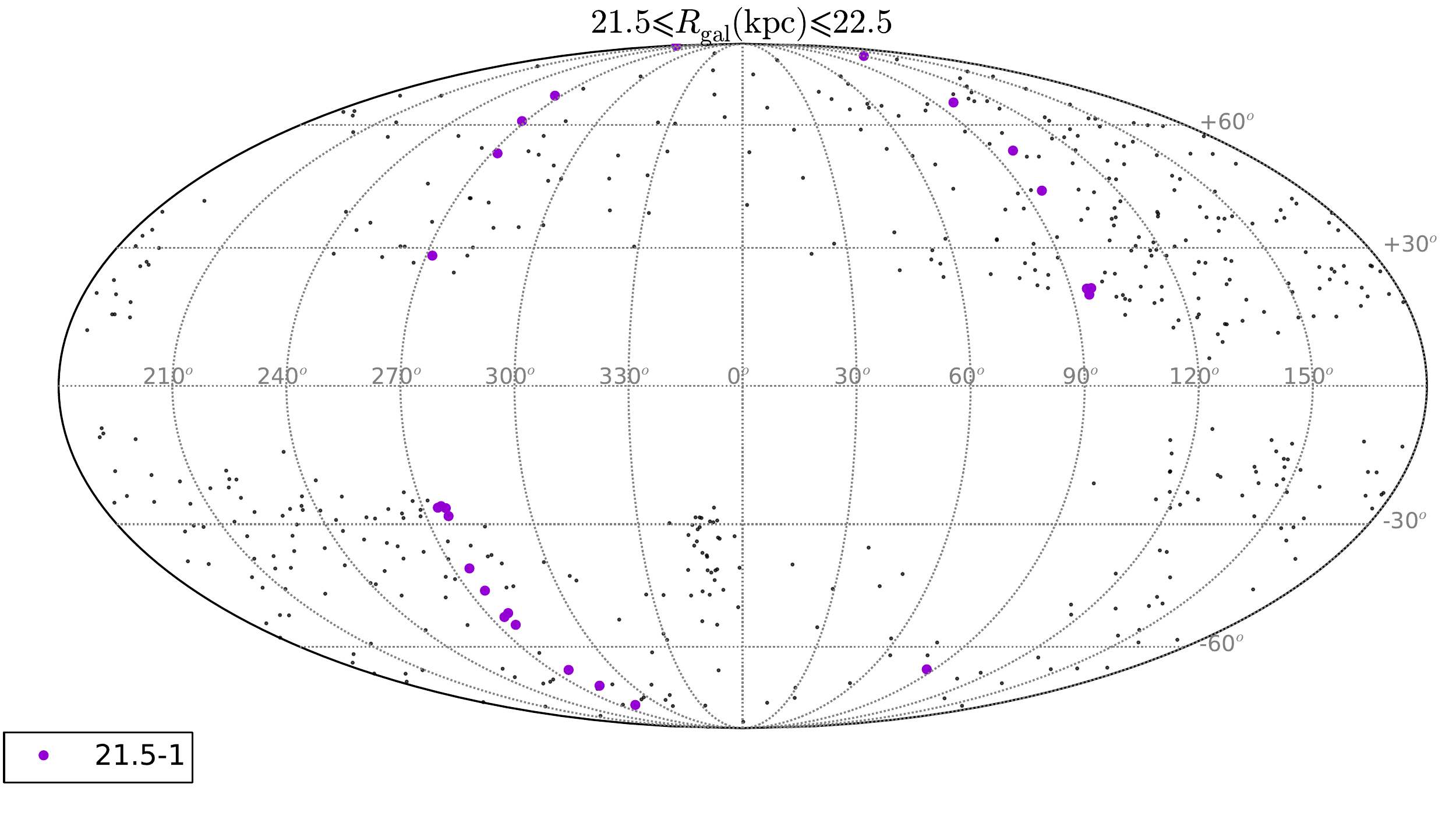}   
\includegraphics[width=0.8\columnwidth]{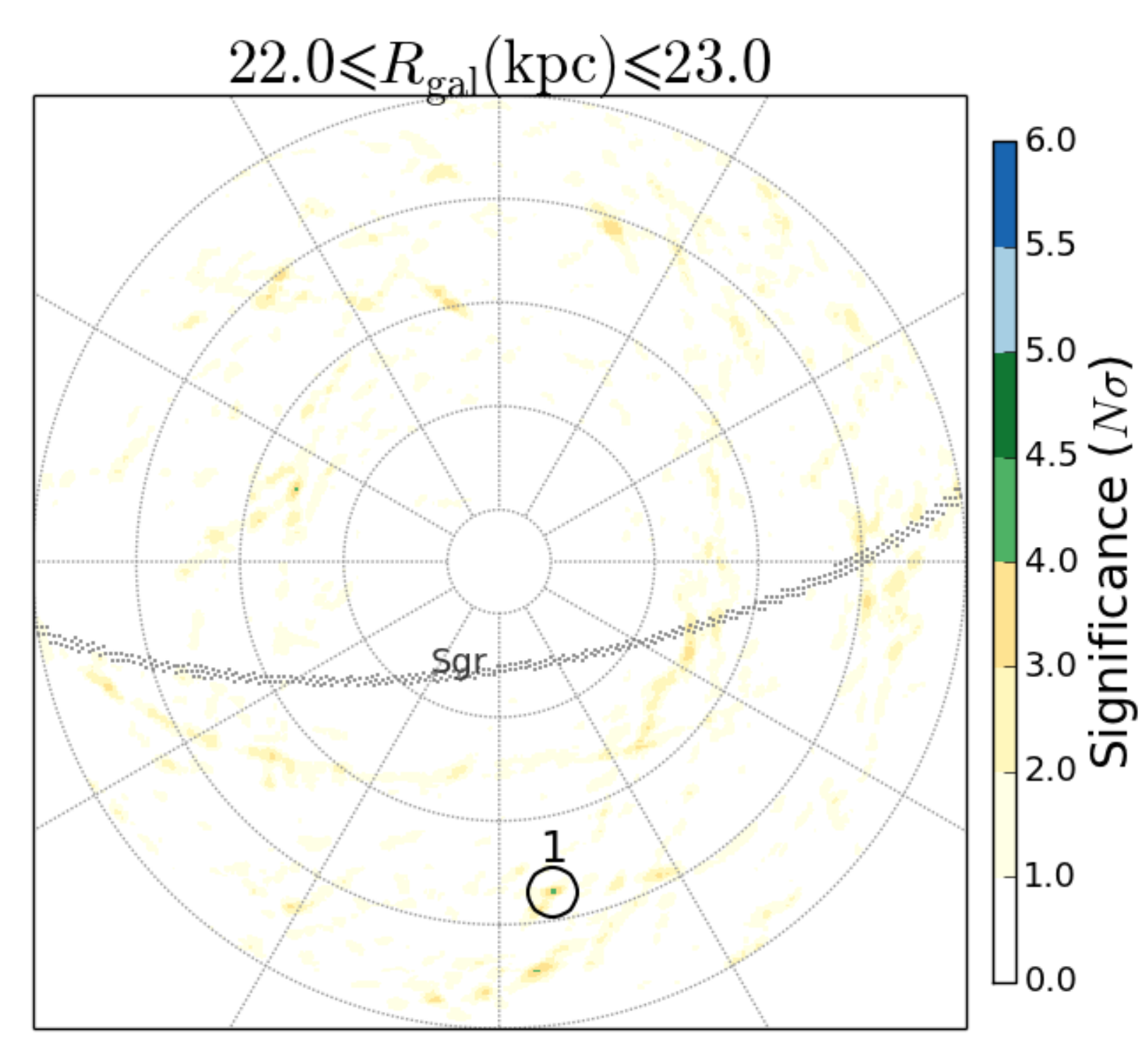} 
\includegraphics[width=1.1\columnwidth]{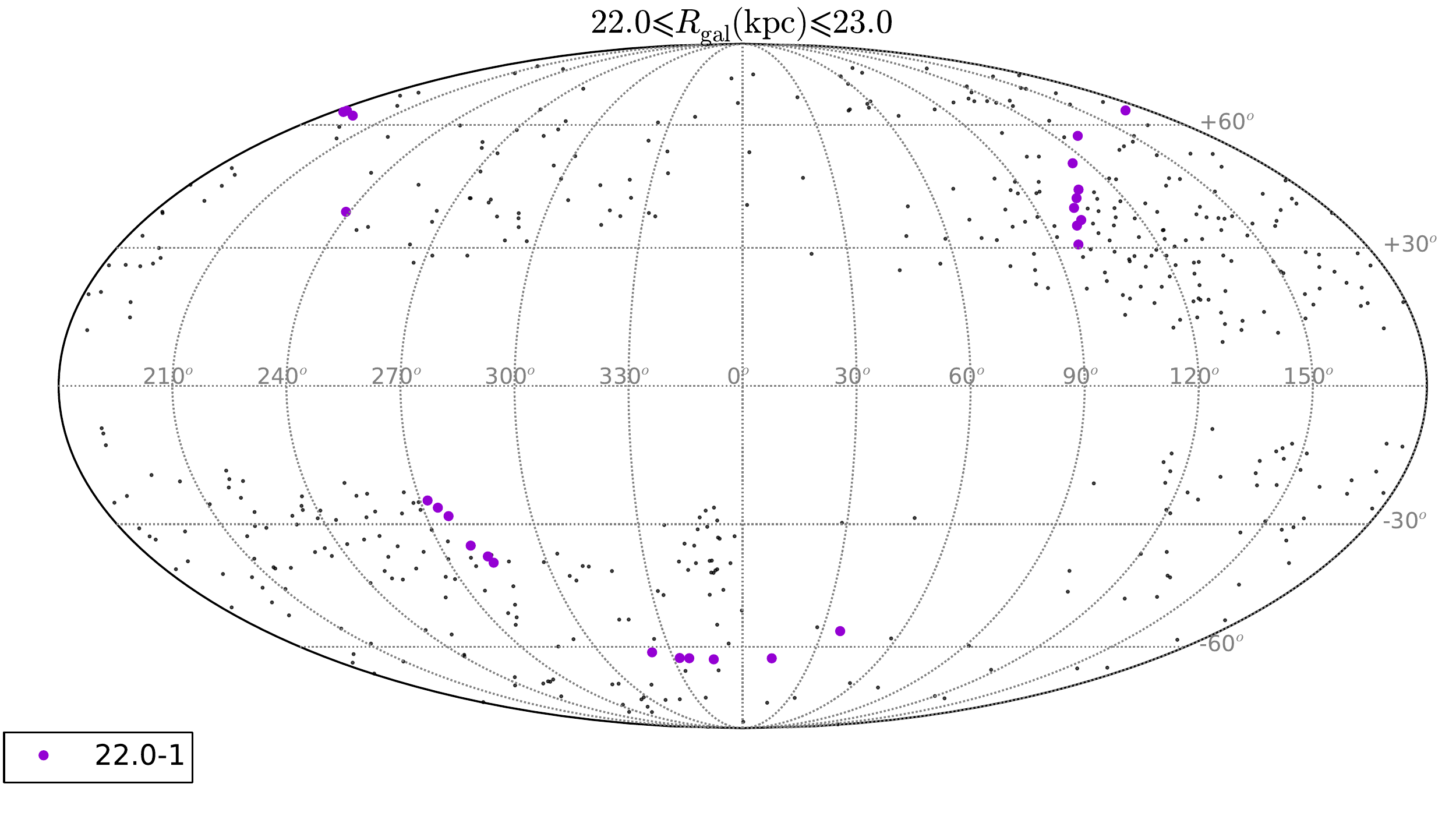}  
\includegraphics[width=0.8\columnwidth]{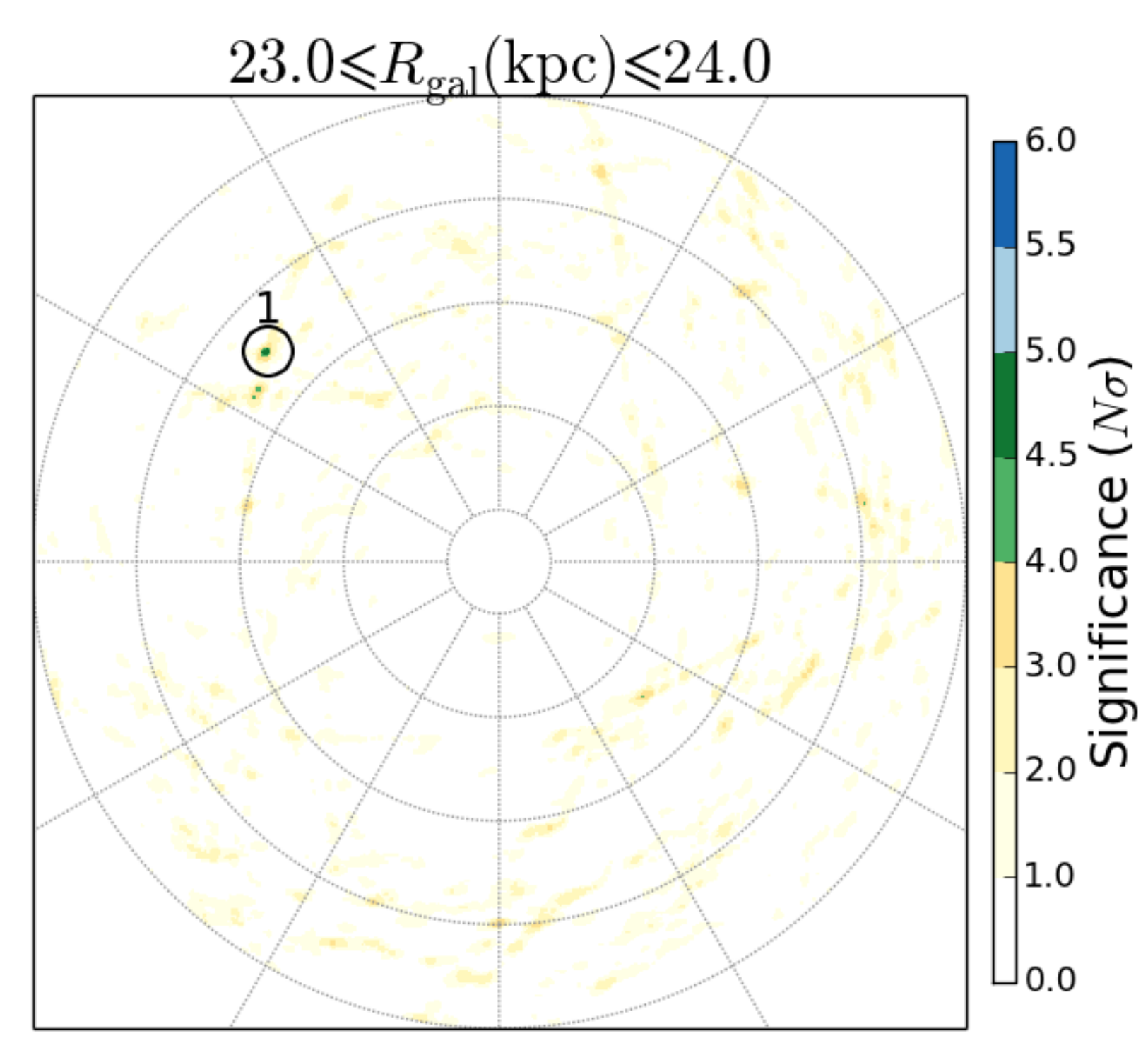}
\includegraphics[width=1.1\columnwidth]{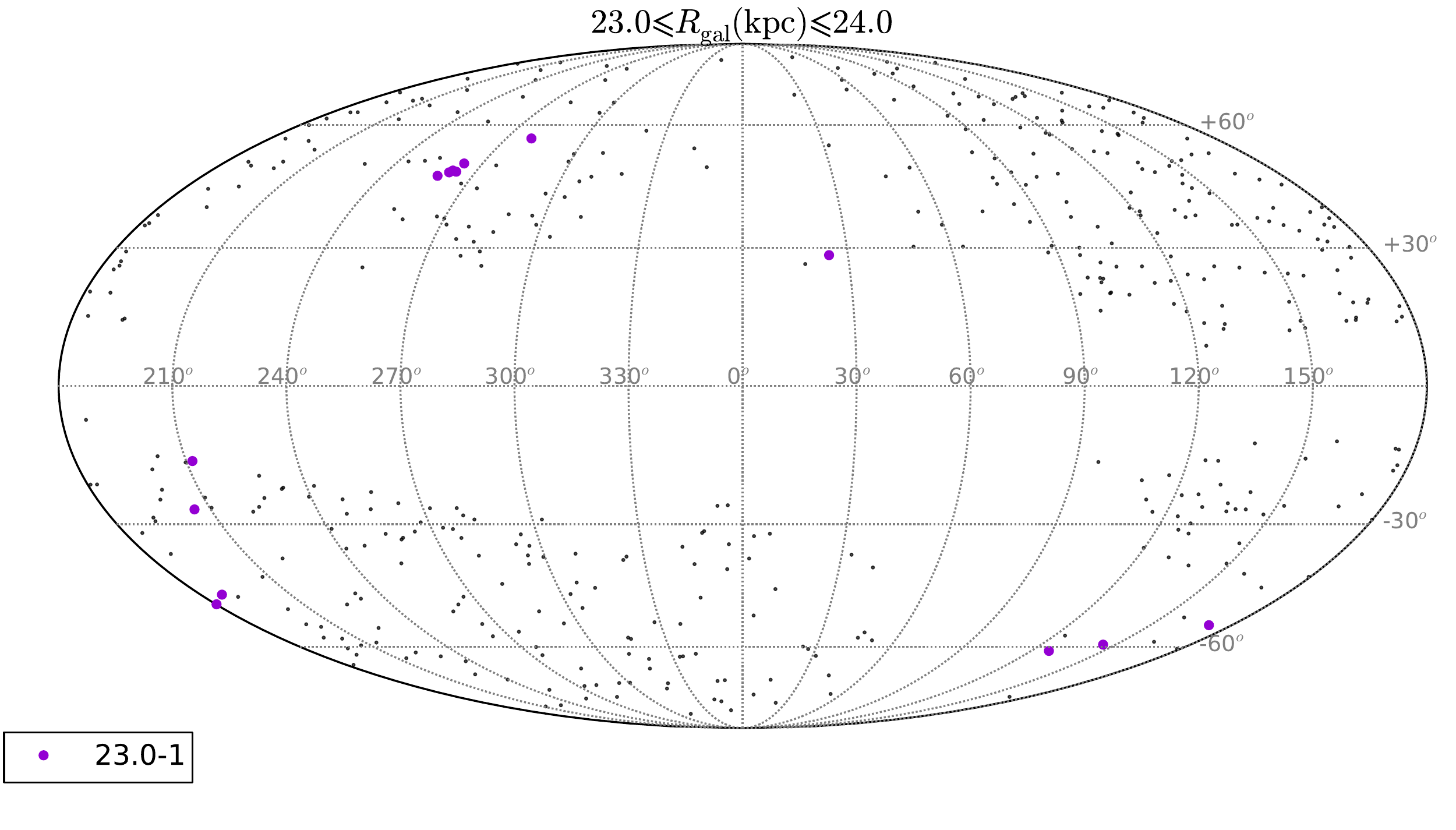}   
\includegraphics[width=0.8\columnwidth]{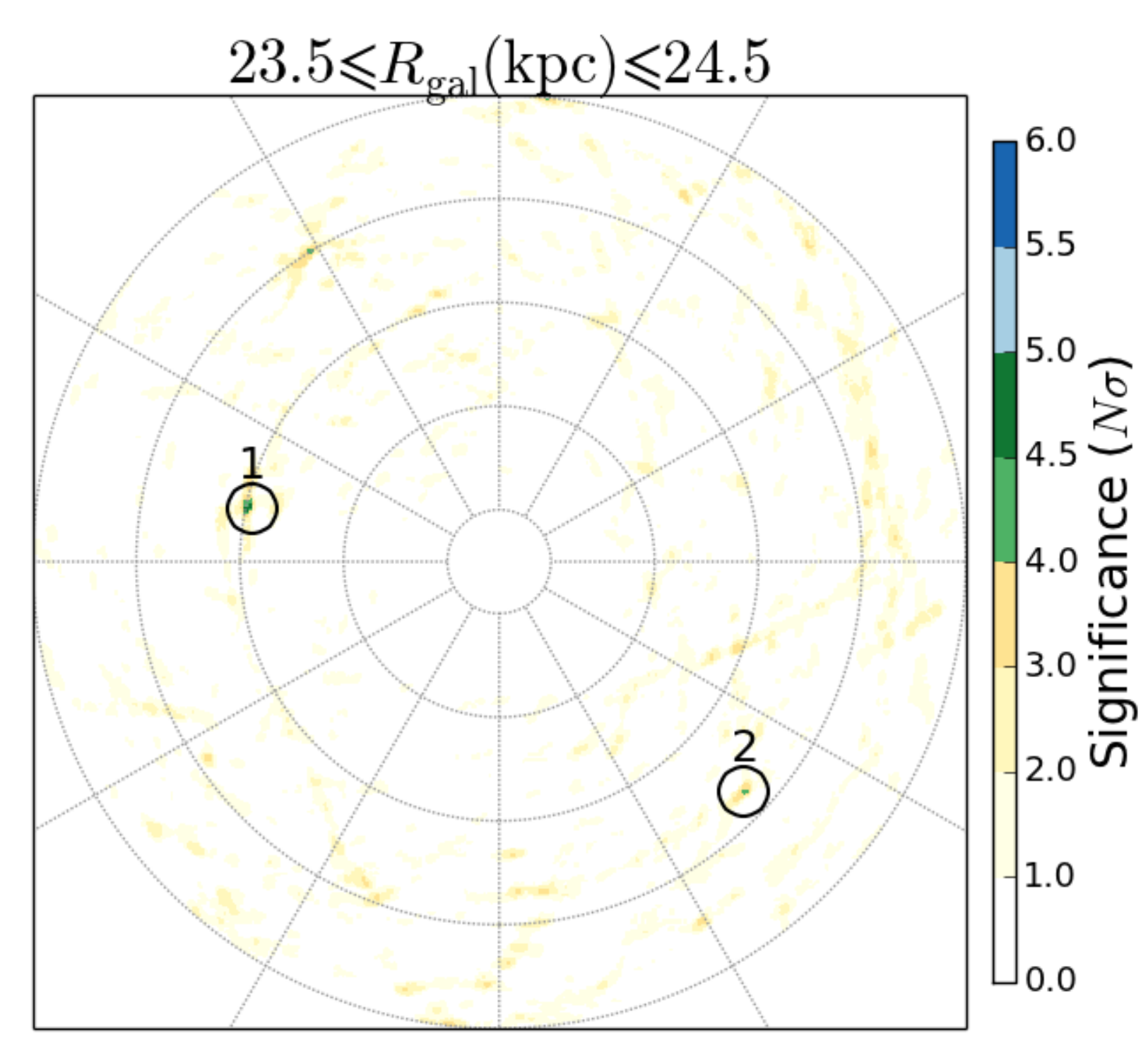}  
\includegraphics[width=1.1\columnwidth]{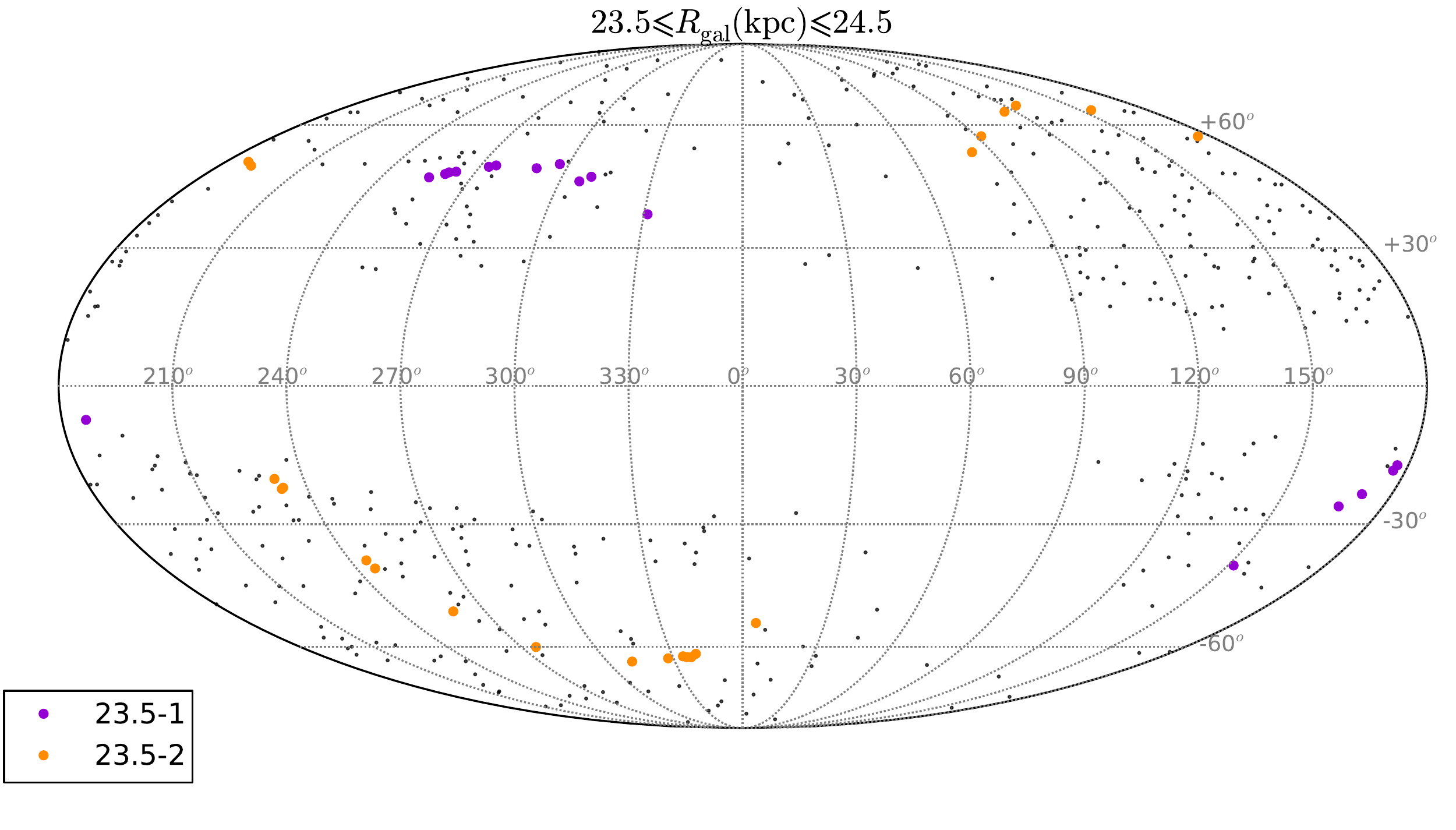} 
\end{center}
\contcaption{ -- \emph{Left:} Combined unsharp-masked \gc+\ngc~PCMs.  \emph{Right:} \emph{Galactocentric} coordinates map. }
\end{figure*}

\begin{figure*}
\begin{center}
\includegraphics[width=0.8\columnwidth]{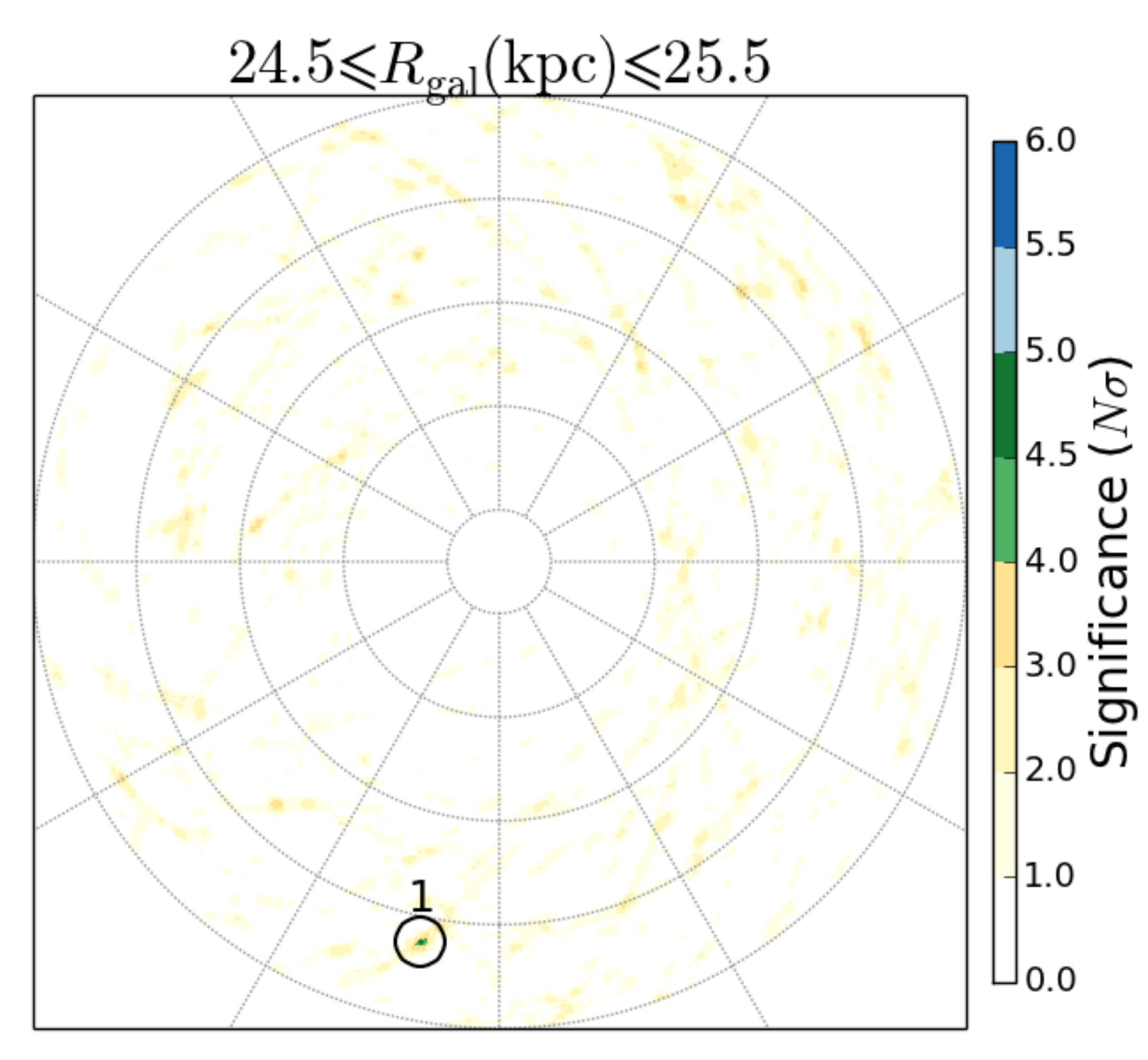} 
\includegraphics[width=1.1\columnwidth]{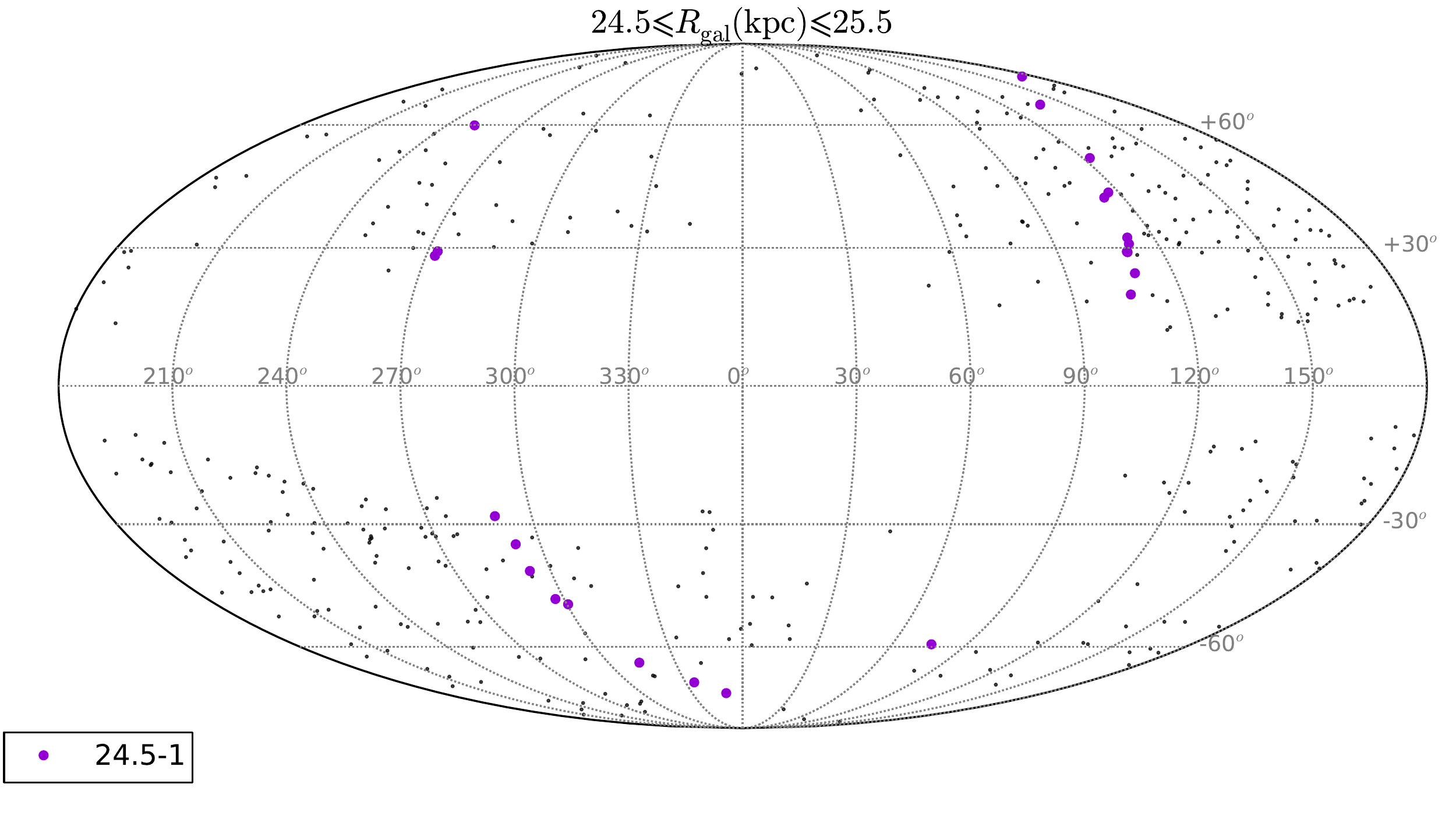}   
\includegraphics[width=0.8\columnwidth]{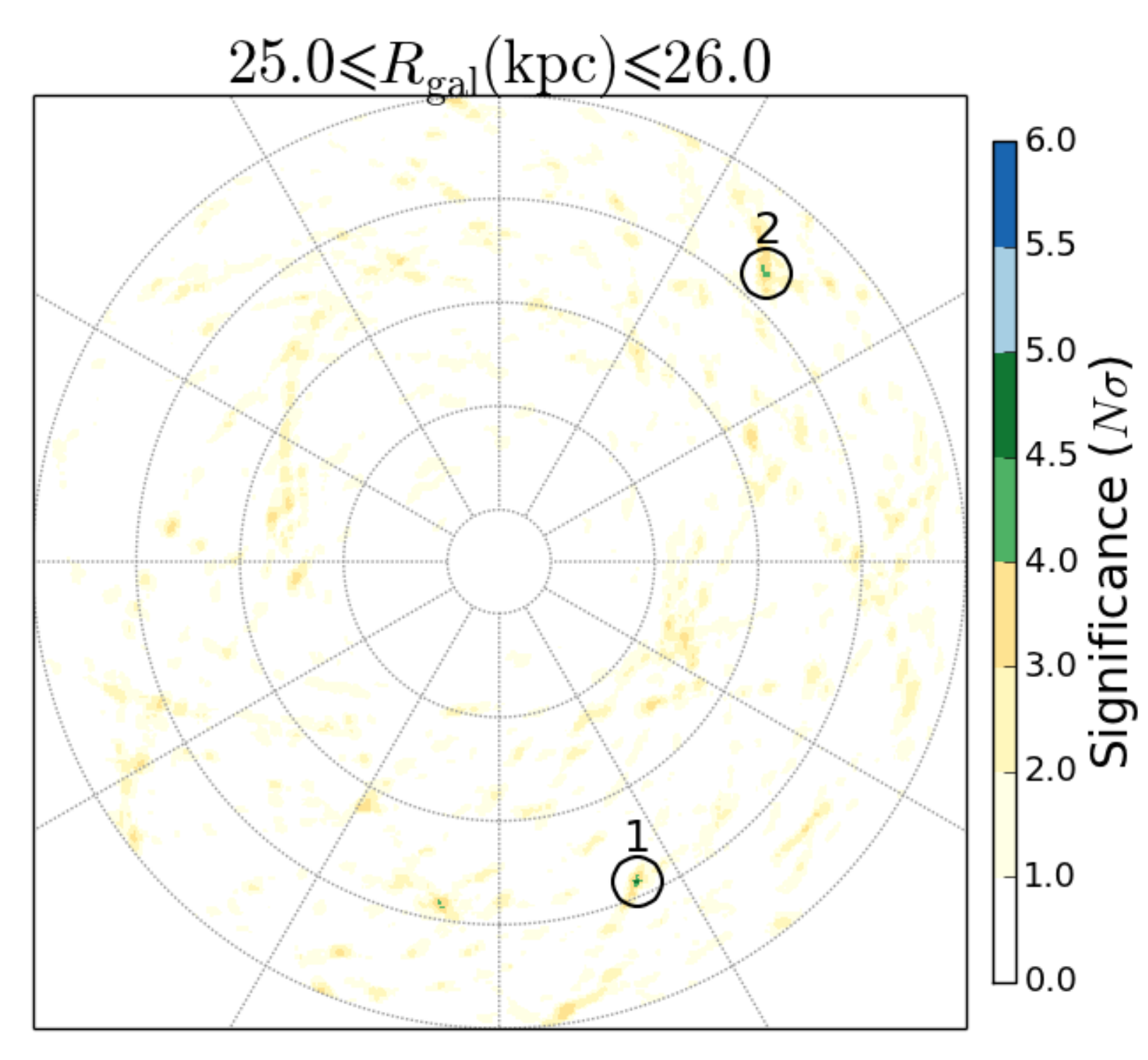} 
\includegraphics[width=1.1\columnwidth]{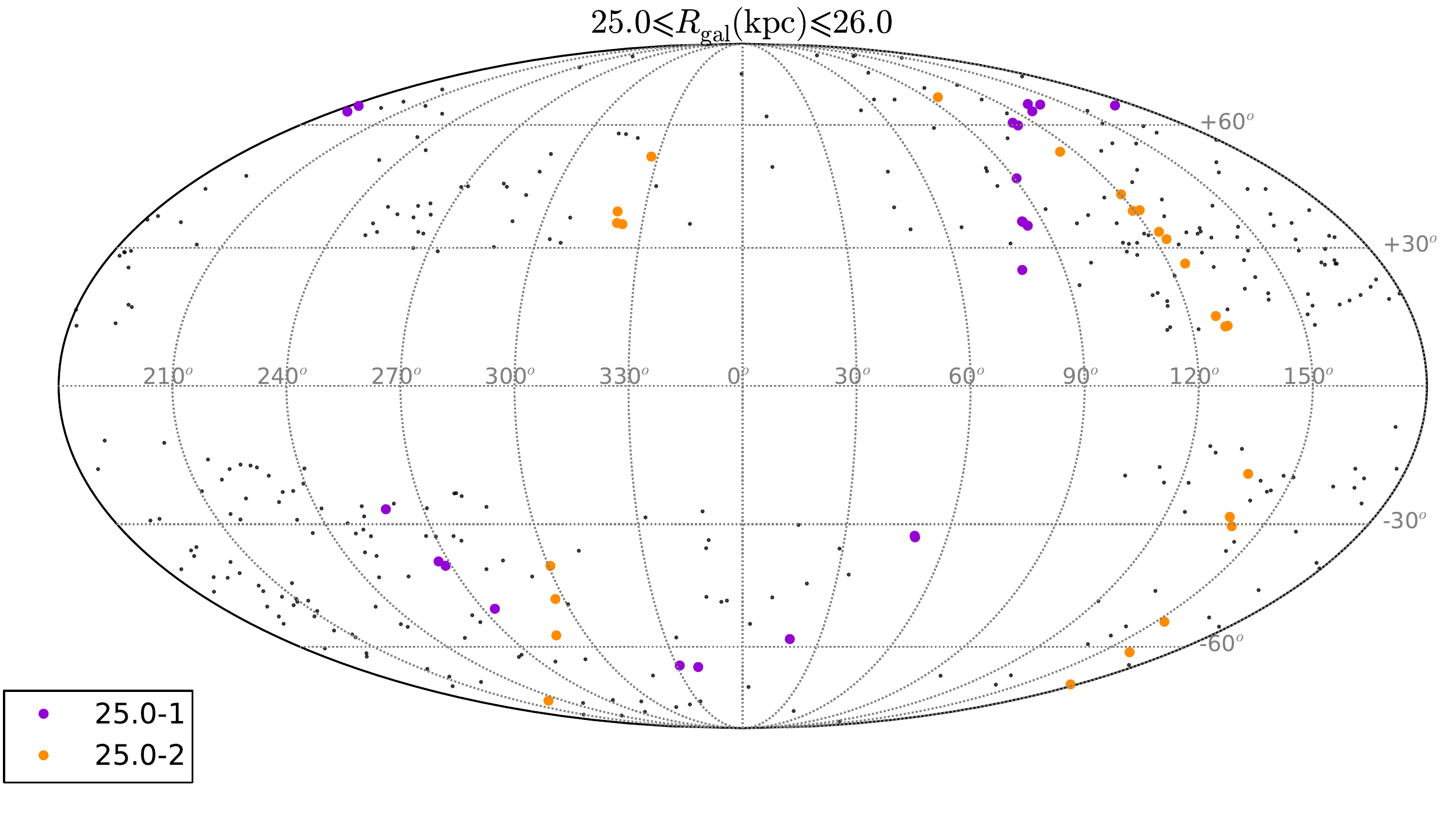}  
\end{center}
\contcaption{ -- \emph{Left:} Combined unsharp-masked \gc+\ngc~PCMs.  \emph{Right:} \emph{Galactocentric} coordinates map. }
\end{figure*}

\begin{table*}
\caption{Detection and geometric properties of the stream candidates detected in \gc+\ngc~PCMs.}\label{t:stream_detections}
\tabcolsep=0.15cm
\begin{tabular}{lccrrcrrrrrrl}
\hline
  \multicolumn{1}{c}{ID} &
  \multicolumn{1}{c}{Pole $(\phi,\theta)$} &
    \multicolumn{1}{c}{Pole $(l,b)$} &
  \multicolumn{2}{c}{$\Rgal$~(kpc)} &
  \multicolumn{1}{c}{$(l_\circ,b_\circ)$} &
  \multicolumn{2}{c}{$\Rhel$~(kpc)} &
  \multicolumn{2}{c}{$\Delta\Theta$} &
  \multicolumn{2}{c}{N$\sigma$-Significance} &
  \multicolumn{1}{l}{Comment} \\
  \multicolumn{1}{c}{} &
  \multicolumn{1}{c}{$(^\circ)$} &
  \multicolumn{1}{c}{$(^\circ)$} &
    \multicolumn{1}{c}{mean} &
  \multicolumn{1}{c}{st.dev.} &
  \multicolumn{1}{c}{($^\circ$)} &
    \multicolumn{1}{c}{min} &
  \multicolumn{1}{c}{max} &
  \multicolumn{1}{c}{($^\circ$)} &
  \multicolumn{1}{c}{(pc)} &
  \multicolumn{1}{c}{$1\degr(0\fdg5)$} &
  \multicolumn{1}{c}{Bts} &
         \multicolumn{1}{l}{} \\
\hline
 \multicolumn{13}{c}{High-confidence Candidates} \\
\hline
11.0-1& (104.5 ,32.46)&   (115.80,43.50)& 11.05 &    0.33  &  (288.98,46.21 )& 6.35   &  12.89  &   1.63  &  313  &  5.1(4.7) &  4.3 & Corvus\\  
20.0-1& (56.38 ,46.98)&    (55.26,56.50)& 20.43 &    0.26  &  (7.75  ,-23.94)& 13.25  &  28.35  &   1.8   &  641  &  5.2(5.1) &  4.7 & Arp2/PS1-C \\  
\hline
\multicolumn{13}{c}{Tentative Candidates}  \\
\hline
04.5-1& (196.08,42.64)&  ( 48.94,32.00) & 5.38  &    0.31  &  (328.83,-20.9 )& 6.0    &  8.78   &   1.68  &  158  &  4.9(5.2) &  3.5 & \\  
08.0-1& (38.31 ,17.41)&  ( 59.46,34.50) & 8.64  &    0.26  &  (352.46,-29.64)& 6.85   &  15.42  &   1.9   &  286  &  4.7(6.2) &  3.7 & \\  
08.0-2& (106.51,1.59 )&  ( 95.50, 0.00) & 8.96  &    0.3   &  (4.45  ,-39.97)& 2.6    &  13.78  &   1.45  &  226  &  4.4(5.3) &  3.5 & \\  
09.0-1& (42.99 ,33.73)&  ( 56.87, 6.50) & 9.78  &    0.31  &  (322.69,26.85 )& 5.91   &  16.77  &   1.94  &  330  &  5.0(4.8) &  3.8 & \\  
15.0-1& (359.05,36.6 )&  ( 22.72,14.50) & 15.54 &    0.34  &  (286.6 ,21.15 )& 12.8   &  21.72  &   1.88  &  511  &  4.6      &  3.3 & \\  
17.0-1& (135.58,18.42)&  (309.51, 0.50) & 17.29 &    0.28  &  (261.26,19.67 )& 12.52  &  21.46  &   1.83  &  552  &  4.4(5.6) &  3.9 & not-Pal5\\  
17.5-2& (225.46,23.67)&  (209.89, 3.50) & 18.05 &    0.2   &  (75.32 ,42.76 )& 12.5   &  18.15  &   1.87  &  590  &  4.7      &  3.5 & Hermus? \\  
17.5-3& (220.72,48.79)&  (221.17,29.50) & 18.01 &    0.26  &  (251.24,-55.26)& 11.41  &  22.25  &   1.18  &  371  &  4.2(6.5) &  3.5 & NGC1261? \\  
22.0-1& (9.23  ,25.58)&  (357.09,41.00) & 22.06 &    0.3   &  (285.28,38.03 )& 19.76  &  27.57  &   1.81  &  696  &  4.3(5.5) &  3.6 & \\  
23.5-1& (257.8 ,41.11)&  (258.43,43.00) & 23.91 &    0.19  &  (66.24 ,45.15 )& 15.34  &  26.05  &   1.46  &  611  &  4.8(6.7) &  3.9 & Hyllus?\\  
23.5-2& (46.77 ,25.44)&  ( 44.12,37.50) & 24.18 &    0.31  &  (14.69 ,-48.64)& 24.07  &  29.02  &   1.69  &  712  &  4.4      &  3.2 & \\  
24.5-1& (348.31,15.17)&  (  2.00, 0.00) & 24.94 &    0.3   &  (272.77,28.92 )& 23.29  &  29.43  &   1.69  &  734  &  4.9      &  3.4 & \\  
\hline
\multicolumn{13}{c}{Possible Artefacts}  \\
\hline
04.5-2& (38.02 ,29.35)&  (240.09,23.00) & 4.95  &    0.23  &  (337.72,16.98 )& 8.0    &  11.82  &   1.74  &  150  &  4.4      &  3.0 & \\  
07.0-1& (94.86 ,40.23)&  ( 90.08,40.50) & 7.61  &    0.29  &  (335.65,26.65 )& 1.19   &  14.31  &   1.96  &  260  &  4.7      &  3.2 & \\  
07.0-2& (347.69,4.56 )&  (210.49,34.00) & 7.34  &    0.26  &  (334.57,38.45 )& 9.82   &  12.38  &   1.72  &  220  &  4.4(5.1) &  3.7 & not-Pal11? \\  
18.5-1& (30.19 ,0.51 )&  (192.93,17.50) & 19.16 &    0.26  &  (86.49 ,41.98 )& 16.83  &  24.2   &   1.64  &  550  &  4.2      &  3.2 & \\  
20.5-3& (125.68,7.61 )&  (128.35,22.00) & 21.01 &    0.25  &  (23.65 ,-62.65)& 14.69  &  25.47  &   1.89  &  692  &  4.2      &  3.2 & \\  
21.5-1& (356.09,2.25 )&  (338.56,14.00) & 21.74 &    0.25  &  (35.71 ,-59.58)& 22.43  &  24.67  &   1.84  &  700  &  4.5      &  3.1 & \\  
25.0-1& (23.41 ,22.93)&  ( 31.23, 7.00) & 25.23 &    0.24  &  (280.83,69.63 )& 23.2   &  31.18  &   1.7   &  749  &  4.7      &  3.5 & \\  
25.0-2& (137.18,14.05)&  (154.84,20.00) & 25.5  &    0.22  &  (259.98,38.09 )& 20.56  &  29.81  &   1.76  &  782  &  4.3      &  3.4 & \\  
\hline
\multicolumn{13}{c}{Repeated Detections}  \\
\hline
17.5-1& (128.97,11.14)& (150.48,23.50)& 17.76 &    0.29  &  (285.04,57.96 )& 12.2   &  18.72  &   1.81  &  561  &  5.3      &  3.6 & same as 17.0-1\\  
19.5-1& (63.96 ,45.02)& ( 65.10,35.50)& 20.27 &    0.27  &  (8.42  ,-22.76)& 12.1   &  27.8   &   1.75  &  619  &  4.5(5.3) &  4.6 & same as 20.0-1\\  
20.5-2& (71.06 ,37.03)& ( 81.22,32.50)& 21.1  &    0.23  &  (7.46  ,-24.16)& 27.79  &  28.89  &   1.71  &  631  &  4.4(5.4) &  3.7 & same as 20.0-1\\  
23.0-1& (227.9 ,29.72)& (183.47,22.50)& 23.33 &    0.32  &  (67.03 ,46.51 )& 16.88  &  25.52  &   1.75  &  713  &  5.0      &  3.3 & same as 23.5-1\\  
\hline
\multicolumn{13}{c}{Unambiguous Detections of Known Streams}  \\
\hline
19.5-2& (95.94 ,9.9  )& ( 95.52, 6.00)& 20.43 &    0.25  &  (8.46  ,-24.26)& 11.65  &  27.87  &   1.86  &  663  &  4.2(4.6) &  3.7 & Sgr \\  
20.0-2& (88.89 ,17.96)& ( 89.65,17.50)& 20.74 &    0.26  &  (7.93  ,-24.91)& 12.17  &  28.47  &   1.84  &  667  &  4.7(5.2) &  4.2 & Sgr \\  
20.5-1& (86.8  ,19.4 )& ( 88.58,18.50)& 20.95 &    0.3   &  (4.85  ,-19.2 )& 12.54  &  29.22  &   1.83  &  668  &  4.7(5.6) &  4.4 & Sgr \\  

\hline
\end{tabular}
\end{table*}

\begin{table*}
\caption{Properties of the stream candidates detected in \gc+\ngc~PCMs.}\label{t:stream_rrprops}
\centering
\tabcolsep=0.13cm
\begin{tabular}{lcrcrrrrrcccl}
\hline
  \multicolumn{1}{c}{ID} &
  \multicolumn{1}{c}{$N_{ab}^{all}$} &
  \multicolumn{1}{c}{Purity} &
  \multicolumn{1}{c}{$N_{ab}^{exp}$} &
  \multicolumn{3}{c}{Oo. fractions} &
  \multicolumn{2}{c}{Oosterhoff} &
  \multicolumn{1}{c}{$M_V$} &
    \multicolumn{1}{c}{$M_V$} &
  \multicolumn{1}{c}{$L_V$ (P84)} &
   \multicolumn{1}{l}{Comment} \\
 \multicolumn{1}{c}{} &
  \multicolumn{1}{c}{} &
  \multicolumn{1}{c}{} &
  \multicolumn{1}{c}{} &
  \multicolumn{1}{c}{I} &
  \multicolumn{1}{c}{Int} &
  \multicolumn{1}{c}{II} &
  \multicolumn{1}{c}{Type} &
  \multicolumn{1}{c}{Prob.} &
    \multicolumn{1}{c}{Mode} &
  \multicolumn{1}{c}{($-\Delta$,$+\Delta$)} &
  \multicolumn{1}{c}{($\times10^5L_\odot$)} &
    \multicolumn{1}{l}{} \\
\hline
 \multicolumn{13}{c}{High-confidence Candidates} \\
\hline 11.0-1 & 40 & 0.48 & 19 & 0.68 & 0.17 & 0.15 &    .. & 0.13 & -7.5 & (-1.51,1.50) & $<$8.3 & Corvus  \\ 
20.0-1 & 32 & 0.53 & 17 & 0.84 & 0.09 & 0.06 &     I & 0.05 & -7.3 & (-1.46,1.47) & $<$6.7 & Arp2/PS1-C \\ 
\hline
\multicolumn{13}{c}{Tentative Candidates}  \\
\hline
  04.5-1 & 23 & 0.56 & 13 & 0.61 & 0.26 & 0.13 &    .. & 0.06 & -6.7 & (-1.35,1.32) & $<$3.3 &  \\ 
  08.0-1 & 34 & 0.54 & 18 & 0.76 & 0.06 & 0.18 &    .. & 0.07 & -7.4 & (-1.47,1.50) & $<$6.7 &  \\ 
  08.0-2 & 27 & 0.53 & 14 & 0.85 & 0.11 & 0.04 &    .. & 0.06 & -6.9 & (-1.38,1.37) & $<$5.0 &  \\ 
  09.0-1 & 42 & 0.51 & 22 & 0.76 & 0.14 & 0.10 &    .. & 0.07 & -7.8 & (-1.55,1.54) &$<$10.0 & \\ 
  15.0-1 & 42 & 0.49 & 21 & 0.81 & 0.10 & 0.10 &     I & 0.05 & -7.7 & (-1.53,1.54) &$<$10.0 & \\ 
  17.0-1 & 36 & 0.47 & 17 & 0.75 & 0.19 & 0.06 &    .. & 0.07 & -7.3 & (-1.46,1.47) & $<$6.7 &  not-Pal5\\ 
  17.5-2 & 23 & 0.61 & 14 & 0.61 & 0.26 & 0.13 &    .. & 0.06 & -6.9 & (-1.38,1.37) & $<$5.0 &  Hermus? \\ 
  17.5-3 & 18 & 0.57 & 10 & 0.78 & 0.11 & 0.11 &    .. & 0.10 & -6.2 & (-1.18,1.16) & $<$1.7 &  NGC1261?\\ 
  22.0-1 & 25 & 0.54 & 13 & 0.88 & 0.04 & 0.08 &     I & 0.05 & -6.7 & (-1.35,1.32) & $<$3.3 &  \\ 
  23.5-1 & 13 & 0.68 &  9 & 0.85 & 0.15 & 0.00 &    .. & 0.10 & -6.0 & (-1.20,1.00) & $<$1.7 &  Hyllus?\\ 
  23.5-2 & 22 & 0.56 & 12 & 0.86 & 0.14 & 0.00 &    .. & 0.06 & -6.6 & (-1.26,1.30) & $<$3.3 &  \\ 
  24.5-1 & 21 & 0.56 & 12 & 0.81 & 0.14 & 0.05 &    .. & 0.08 & -6.6 & (-1.26,1.30) & $<$3.3 &  \\ 

\hline
\end{tabular}
\end{table*}

\begin{table*}
\caption{RRLSs associated to each of the stream candidates. (This table is published in its entirety as Supporting Information with the electronic version of the article. A portion is shown here for guidance regarding its form and content).}
\label{t:cand_rrls_list}
\tabcolsep=0.10cm
\begin{tabular}{|lrrrrrrrrrrrl}
\hline 
  \multicolumn{1}{c}{ID} &
  \multicolumn{1}{c}{CSS-SSS-ID} &
  \multicolumn{1}{c}{RAJ2000} &
  \multicolumn{1}{c}{DEJ2000} &
  \multicolumn{1}{c}{Rhel} &
  \multicolumn{1}{c}{Rgal} &
  \multicolumn{1}{c}{Period} &
  \multicolumn{1}{c}{AmpV} &
  \multicolumn{1}{c}{FeH} &
    \multicolumn{1}{c}{flg} &
  \multicolumn{1}{c}{pmRA} &
  \multicolumn{1}{c}{pmDE} &
  \multicolumn{1}{l}{HSOY-ID} \\
  \multicolumn{1}{c}{} &
  \multicolumn{1}{c}{} &
  \multicolumn{1}{c}{($\degr$)} &
  \multicolumn{1}{c}{($\degr$)} &
  \multicolumn{1}{c}{(kpc)} &
  \multicolumn{1}{c}{(kpc)} &
  \multicolumn{1}{c}{(d)} &
  \multicolumn{1}{c}{(mag)} &
  \multicolumn{1}{c}{(dex)} &
    \multicolumn{1}{c}{} &
  \multicolumn{1}{c}{(mas/yr)} &
  \multicolumn{1}{c}{(mas/yr)} &
  \multicolumn{1}{l}{} \\  
\hline
04.5-1 & J165107.7-185500 & 252.78216 & -18.91671 & 12.09 & 4.56 & 0.60154 & 0.64 & -1.34 & 0 & -8.1 $\pm$ 2.2 & -6.3 $\pm$ 2.2 & 4.8426375499386829E18\\
  04.5-1 & J203208.8-251433 & 308.03653 & -25.24252 & 6.01  & 5.18 & 0.57042 & 0.78 & -1.38 & 0 &  1.8 $\pm$ 2.1 & -0.5 $\pm$ 2.1 & 5.0354848055937546E18\\
  04.5-1 & J204034.3-251227 & 310.14282 & -25.20737 & 6.00  & 5.37 & 0.54851 & 0.77 & -0.96 & 0 & -1.6 $\pm$ 2.1 & -5.0 $\pm$ 2.1 & 5.0362798660131287E18\\
  04.5-1 & J204750.0-271645 & 311.95815 & -27.27923 & 6.11  & 5.47 & 0.66466 & 0.46 & -1.19 & 0 &  1.3 $\pm$ 2.7 & -7.4 $\pm$ 2.7 & 5.0313484201172142E18\\
  04.5-1 & J202002.4-292359 & 305.00997 & -29.39977 & 6.08  & 4.80 & 0.50794 & 1.12 & -1.16 & 0 &  2.8 $\pm$ 2.1 & -7.4 $\pm$ 2.1 & 5.014405122247722E18\\

\hline
\end{tabular}
\end{table*}

\section{Comparison with known streams and halo substructure}\label{s:comparison_with_known_streams}

\subsection{Milky Way Streams Library and Python Package}

\citet{Grillmair2016} have made a recent compilation to summarise basic average properties of known, well-established streams and clouds in the MW (their Tables~4.1 and 4.2). The data they provide in their Table~4.1 to illustrate the extent of each stream, as the authors point out, refers only to the coordinate range occupied by each stream in the celestial sphere. Although useful as a first approximation, and reasonably well suited to describe the extent of diffuse cloud-like structures, more specific footprint data is warranted for streams in order to make a fair comparison of the literature to any newly identified substructure.

With this motivation in mind, we have built the Python Package \textsc{galstreams}, publicly available 
at \href{https://github.com/cmateu/galstreams}{https://github.com/cmateu/galstreams}. The package contains a MW Streams Library with standardised information about known Galactic streams and clouds and providing a series of utility classes and methods to define, manipulate and plot the footprint data of all streams registered in the library; with flexible ways to define a stream, making the library easy to expand.
The MW Streams Library is based on the \citet{Grillmair2016} review and expanded to include new streams published up to June 2017. Table~\ref{t:streams_lib} summarises the streams and clouds included in the library by default. 

The main object classes in \textsc{galstreams} are as follows: the \textsc{MWstreams} object handles the entire library as a whole, with each stream/cloud being represented as a \textsc{Footprint} object, which handles the set of coordinates that represent a given stream.  

The \textsc{Footprint} object represents each stream's celestial footprint as a collection of points, i.e. a set of sky coordinate arrays. So for each stream the \textsc{Footprint} object holds as attributes, the coordinates in all pre-defined systems: equatorial, heliocentric galactic, spherical Galactocentric, and cartesian helio and Galactocentric coordinates. The main feature of the \textsc{Footprint} class is that it allows instantiating or creating any stream in one of four different ways, defined in the \textsc{gcutils} library, by giving one of the following:

\begin{itemize}
\item start and end point coordinates
\item orbital pole coordinates and, optionally, the stream's center, length and width
\item a coordinate range (more suitable for clouds)
\item a list of individual coordinates 
\end{itemize}

For any of this four options the input coordinates can be provided in the equatorial or (heliocentric) Galactic reference frames
and heliocentric distance information is optional. Using the available information, the \textsc{Footprint} class computes all needed (and possible) coordinate transformations in order for the \textsc{Footprint} object to have the coordinates in all pre-defined systems mentioned above.  

It is this overall flexibility that allows the library to be easily extended as new streams are reported or recovered from the literature. It also allows the user to quickly create a \textsc{Footprint} object for a stream of interest, without needing to include it in the library.

The \textsc{galstreams} GitHub repository\footnote{\url{https://github.com/cmateu/galstreams}} includes a detailed description of the \textsc{galstreams} package capabilities, so we refer the reader there for more details. Also, for non-Python users, individual files are provided containing the footprints of all streams and clouds in the library at this URL\footnote{\url{https://github.com/cmateu/galstreams/tree/master/footprints}}. In the remainder of this section, Figures~\ref{f:hc_cands_lit}, \ref{f:hc_cands_comparison_sss} and \ref{f:tent_cands_lit} will showcase the footprints for the streams and clouds stored in the MW Streams Library, summarised in Table~\ref{t:streams_lib}. Currently the library includes spatial information, i.e. celestial coordinates and distances where available, but it can be easily extended to include further information such as radial velocities, proper motions, metallicity and elemental abundances.

\begin{table*}
\caption{Streams and Clouds in the MW Streams Library included in the \textsc{galstreams} Python Package.}\label{t:streams_lib}
\centering
\begin{tabular}{llllll}
\hline
  \multicolumn{1}{l}{Name} &
  \multicolumn{1}{l}{Reference} &
    \multicolumn{1}{l}{Name} &
  \multicolumn{1}{l}{Reference} &
  \multicolumn{1}{l}{Name} &
    \multicolumn{1}{l}{Reference} \\
\hline Alpheus   &  \citet{Grillmair2013}  &  Monoceros     & \citet{Grillmair2016}   & PS1-C     &  \citet{Bernard2016}   \\
 Acheron   &  \citet{Grillmair2009}  &  Molonglo      & \citet{Grillmair2017b}  & PS1-D     &  \citet{Bernard2016}   \\
 ACS       &  \citet{Grillmair2006}  &  Murrumbidgee  & \citet{Grillmair2017b}  & PS1-E     &  \citet{Bernard2016}   \\ 
 ATLAS     &  \citet{Koposov2014}    &  NGC5466       &  \citet{Grillmair2006}  & Sangarius &  \citet{Grillmair2017} \\  
 Cetus     &  \citet{Newberg2009}    &  Ophiucus      &  \citet{Bernard2014}    & Scamander &  \citet{Grillmair2017} \\  
 Cocytos   &  \citet{Grillmair2009}  &  Orphan        &  \citet{Newberg2010}    & Styx      &  \citet{Grillmair2009} \\  
 GD-1      &  \citet{Grillmair2006}  &  Orinoco       &  \citet{Grillmair2017b} & Tri-And   &  \citet{Grillmair2016} \\  
 EBS       &  \citet{Grillmair2016}  &  Pal 5         &  \citet{Grillmair2006}  & Tri-And2  &  \citet{Grillmair2016} \\  
 Eridanus  &  \citet{Myeong2017}     &  Pal 15        &  \citet{Myeong2017}     & Tri/Pis   &  \citet{Bonaca2012}    \\  
 Hermus    &  \citet{Grillmair2014}  &  PAndAS        &  \citet{Grillmair2016}  & VOD/VSS   &  \citet{Grillmair2016} \\  
 Her-Aq    &  \citet{Grillmair2016}  &  Phoenix       &  \citet{Balbinot2016}   & WG1       &  \citet{Agnello2017}   \\  
 Hyllus    &  \citet{Grillmair2014}  &  PiscesOv      &  \citet{Grillmair2016}  & WG2       &  \citet{Agnello2017}   \\  
 Kwando    &  \citet{Grillmair2017b} &  PS1-A         &  \citet{Bernard2016}    & WG3       &  \citet{Agnello2017}   \\  
 Lethe     &  \citet{Grillmair2009}  &  PS1-B         &  \citet{Bernard2016}    & WG4       &  \citet{Agnello2017}   \\

\hline
\end{tabular}
\end{table*}

\subsection{High-confidence Candidates}

\begin{figure*}
\begin{center}
 \includegraphics[width=0.99\textwidth]{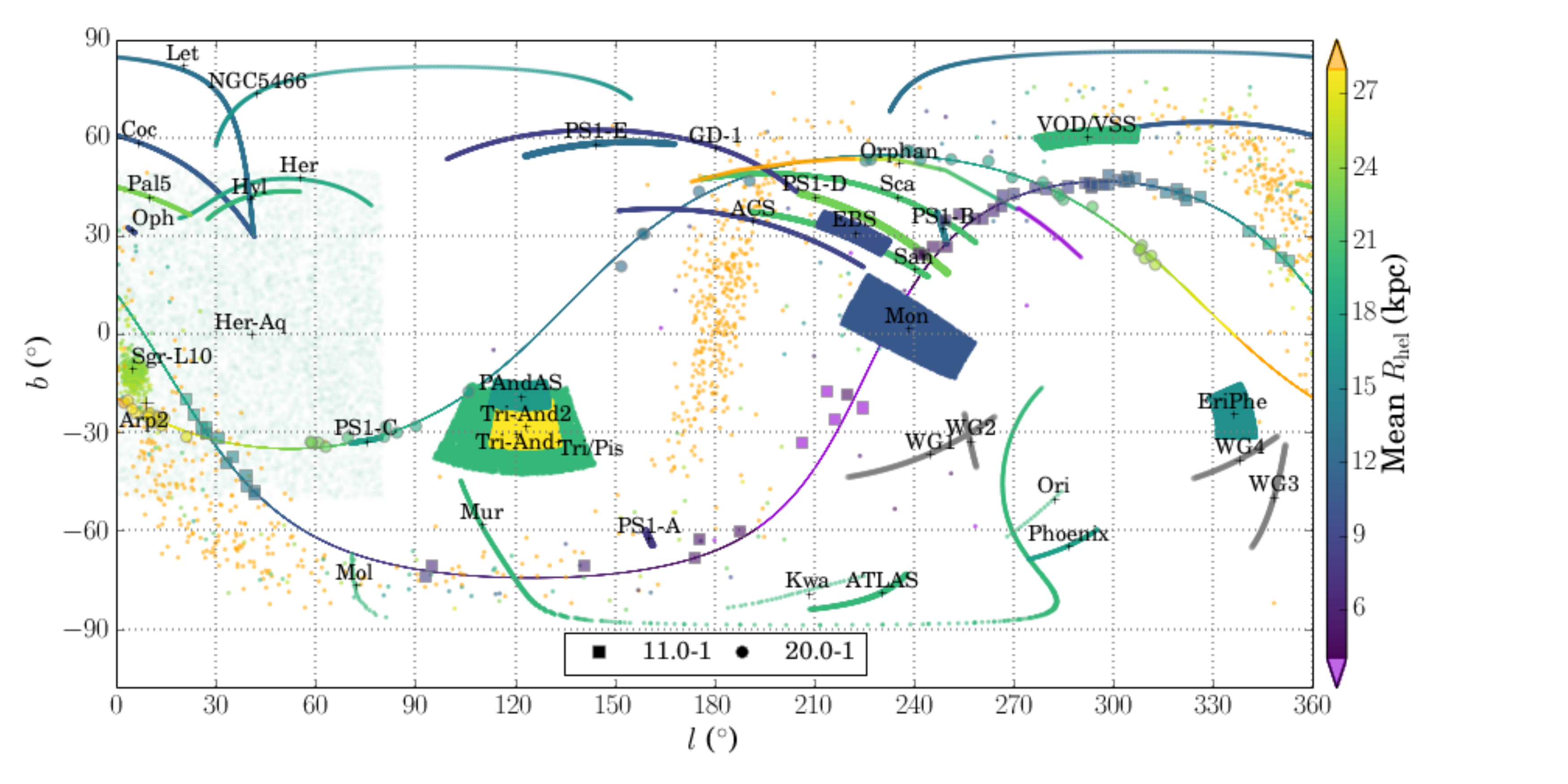} 
 \caption{Galactic (heliocentric) latitude versus longitude map for the high-confidence candidates. The RRLS belonging to each candidate are indicated with different symbols, as shown in the legend and the colour scale is proportional to the heliocentric distance. Each of our candidate's great-circle is shown with a solid line passing through the RRLSs. The streams and clouds from the MW Streams Library that have any overlap in the figure's heliocentric distance range are shown with coloured dots and labelled in the plot. The corresponding references are summarised in Table~\ref{t:streams_lib}.}
\label{f:hc_cands_lit}
\end{center}
\end{figure*}

Figure~\ref{f:hc_cands_lit} shows the sky distribution of the high-confidence candidates relative to known streams and other halo substructure in the MW Streams Library, in a (heliocentric) latitude versus longitude plot, with a colour scale proportional to the heliocentric distance $\Rhel$. The symbols represent the RRLS associated to each candidate: squares for candidate 11.0-1 and circles for candidate 20.0-1. The solid line that goes along each candidate's RRLS is the great circle defined by the candidate's pole given in Table~\ref{t:stream_detections}. Note that, because this plot is heliocentric, each candidate spans a range of several kpcs in $\Rhel$ as the colour scale shows even though, by construction of our search strategy, it's \emph{Galactocentric} radial extent is $<$1~kpc.

\begin{figure*}
\begin{center}
 \includegraphics[width=0.99\textwidth]{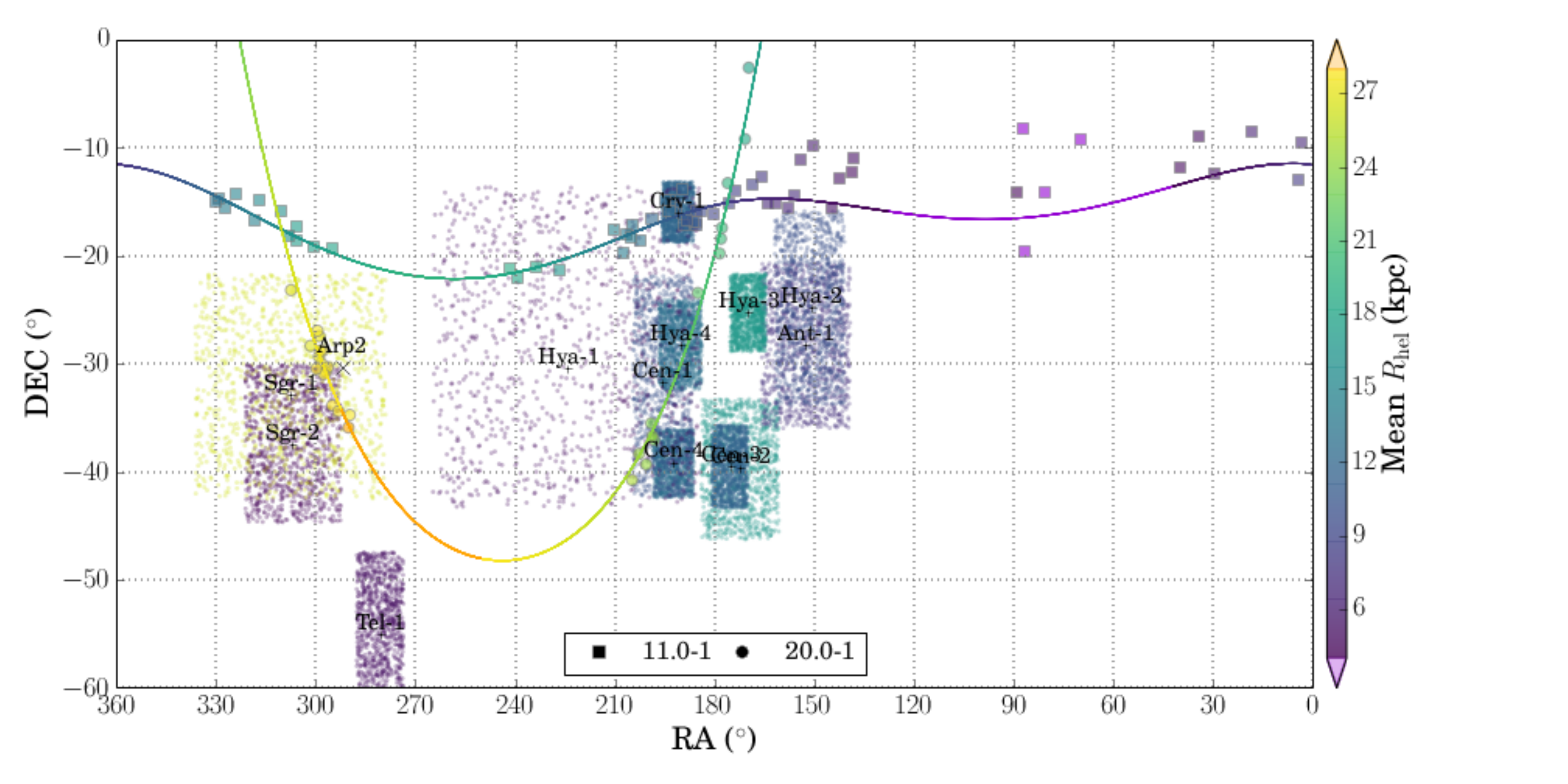} 
 \caption{Equatorial coordinates map for the high-confidence candidates and significant overdensities found by \citet{Torrealba2015} in the SSS survey. The RRLS belonging to each candidate are indicated with the same symbols as in Figure~\ref{f:hc_cands_lit}, as shown in the legend. The colour scale is proportional to the heliocentric distance. Each of our candidate's great-circle is shown with a solid line passing through the RRLSs. }
\label{f:hc_cands_comparison_sss}
\end{center}
\end{figure*}

\citet{Torrealba2015} has also searched for overdensities in the southern SSS part of the Catalina survey, included in our Catalina+HSOY RRLS catalogue (Sec.~\ref{s:catalina_hsoy}). In their Table~3 they report the central equatorial coordinates and extent of the 26 overdensities they identified, shown also in their Figure~14. In Figure~\ref{f:hc_cands_comparison_sss} we compare the location of our high-confidence candidates to their 12 reported overdensities with a significance $\geqslant3\sigma$, in an RA-DEC map similar to Figure~\ref{f:hc_cands_lit}, with a colour scale proportional to heliocentric distance. This comparison is approximate since we use the data in \citeauthor{Torrealba2015}'s Table 3 to represent their overdensities as rectangles in RA-DEC and their Figure 14 shows them to be more general polygonal areas. Nevertheless, even with this simplification, the figure still serves our purpose to check for approximate spatial coincidences between our candidates and their overdensities.

\subsubsection{Candidate 11.0-1 -- Corvus Stream}\label{s:cand11-1}

Candidate 11.0-1 is the one with the highest bootstrap significance. The PCM in Figure~\ref{f:pcms_detections} shows it is not related to any known GC or dwarf galaxy. 

The densest part of this candidate, which causes the peak in the PCM, is located at $240\degr \lesssim l \lesssim 330\degr$. In particular, as Figure~\ref{f:hc_cands_lit} shows, the region around $240\degr \lesssim l \lesssim 270\degr$ is quite crowded with known substructures: the PS1-B, PS1-D, Sangarius, Scamander and Orphan streams cross that region, but are most likely independent of our stream candidate as they cross the 11.0-1 great circle almost orthogonally. They are also much more distant than our candidate: in this region, the 11.0-1 RRLSs are located at $\Rhel\sim7$~kpc, while these streams are at least twice as distant, with PS1-B being the closest at $\sim14.5^{+3.7}_{-3.0}$~kpc \citep{Bernard2016} and the rest having distances $>$18~kpc. The southern part of  the 11.0-1 candidate  at $l\lesssim 40\degr$ lies close to the edge of the Sgr tail,  but in this area our candidate's RRLS lie at $\Rhel\sim18$~kpc and the Sgr tail is much more distant ($>28$ kpc), so it is not a likely contaminant. 

Out of the total 40 \rrab~stars associated to this detection, we expect $\sim19$ \rrab~to be real stream stars, given that its purity is estimated at $0.48$. The fraction of RRLS of different Oosterhoff types are fairly consistent with those of the typical halo field (see Sec.~\ref{s:cand_rrls_props}), so no information can be inferred at this point about the possible Oosterhoff type of this candidate.  

In Figure~\ref{f:hc_cands_comparison_sss} we compare in an RA-DEC plot the spatial distribution of our high-confidence candidates and SSS overdensities from \citet{Torrealba2015}, where we have also included Crv~1 (2.99$\sigma$), although it is just below the $3\sigma$ threshold. This overdensity lies right along our 11.0-1 candidate and at the same distance ($\sim$12~kpc) as our candidate's RRLS  in that region.  Nevertheless, after removing the RRLS in the Crv~1 region the 11.0-1 detection is still recovered at the same pole, with a $4.2\sigma$ significance. This confirms 11.0-1 is an independent detection and traces a structure larger than the reported extent of Crv~1, although without kinematics it is not possible to estimate its real extent.
The central coordinates $(l,b)=(300\fdg4,+46\fdg84)$ reported for Crv~1 by \citet{Torrealba2015} are also fairly close to the densest part of our candidate, $(l_\circ,b_\circ)=(288\fdg98,+46\fdg21)$, at an angular distance $<7\degr$, supporting that the RRLS in that area are the best follow-up candidates to characterise the stream and to look for a potential progenitor.

\citet{Duffau2016} and \citet{Navarrete2016} suggest Crv~1 might be a possible southern extension of the Virgo Stellar Stream. However, the orbital plane shown in Figure~\ref{f:hc_cands_lit} for 11.0-1 with respect to the VSS does not seem to support this. Gaia proper motions combined with radial velocities from the spectroscopic survey being conducted by these authors will be decisive to clarify whether the two are related.

In summary, the properties of the 11.0-1 detection suggest the Crv~1 overdensity is more extended than originally reported and is a stream-like overdensity. We propose naming it the \emph{Corvus stream} to keep \citeauthor{Torrealba2015}'s designation and following the usual convention for streams. 

\subsubsection{Candidate 20.0-1 -- Arp2/PS1-C}\label{s:cand20-1}

Figure~\ref{f:pcms_detections} shows our multiple detections of this candidate (19.5-2, 20.0-1 and 20.5-2) lie right on top of the Arp 2 GC's signature in the PCM, suggesting a possible association between the two. At first this is just a hint. The proper motion errors for Arp 2 are quite high, which is why its PCM signature shows up as a full great circle, as it is only constrained by the cluster's position. Nevertheless, in this candidate, the largest concentration of RRLS \emph{is} found very near the center of Arp 2, at $(l_\circ,b_\circ)=(7\fdg8,-23\fdg9)$ (see Table~\ref{t:stream_detections}), as Figure~\ref{f:hc_cands_lit} also illustrates. The fact that there are two more detections of this candidate (19.5-1 and 20.5-2) with very similar poles differing only by $\sim5\degr$, reinforces  that it is unlikely this candidate would be a random excess due to outlying Arp 2 RRLS, in which case it would be more natural to expect pole count excesses at random poles along the Arp 2 great circle. Also, since in our pre-processing of the RRLS catalogue all the stars inside the tidal radius of each GC were removed (see Sec.~\ref{s:removing_globulars}), \emph{the RRLS excess is likely due to tails and not to the GC's bound core itself}.

It also seems unlikely that this candidate's detection is due to the Sgr stream. The \citet{Law2010} Sgr spherical model (shown in the figures) predicts the tails in this region to be much more distant ($>$30~kpc) and if the peak detected in the PCM were just a spurious peak induced by left-over RRLS just outside the Sgr core, there is no reason why the detection should coincide so well with the Arp 2 great circle in the PCM \emph{and do so over three radial distance bins}. In the event that Sgr RRLS were causing a random excess, it is unlikely that a significant PCM peak would be detected almost in the same position over three radial bins (two of which, 19.5 and 20.5, have no overlapping stars and, so, are entirely independent), differing by only a few degrees and always along the Arp 2 great circle.

Figure~\ref{f:hc_cands_lit} shows this candidate might also be associated with the PS1-C stream \citep{Bernard2016}.  The PS1-C stream coincides exactly with the 20.0-1 great circle in the plane of the sky and the distance is very similar:  \citet{Bernard2016} reports a heliocentric distance of $17.4^{+3.5}_{-3.6}$~kpc for PS1-C and the RRLS in this area have a distance around $\sim$19~kpc, which are in agreement within the errors. We also removed the RRLS around the PS1-C stream (17 RRLS with $19.5<\Rgal\mathrm{(kpc)}<21.5$) and found the peak is still detected in the PCM at the same pole, with only a slightly decreased significance ($4.8\sigma$ compared to the initial $5.2\sigma$). This confirms that 20.0-1 is an independent detection since the PCM excess is not dominated by PS1-C RRLS, but rather by the RRLS closer to Arp 2. 

\citet{Bernard2016} suggests PS1-C might be related to the Balbinot~1 stellar cluster \citep{Balbinot2013}, which lies very near ($(l,b)=(75\fdg2,-32\fdg6)$) the reported center for PS1-C (shown with a cross in Fig.~\ref{f:hc_cands_lit}). However this association does not seem likely at first since, according to \citeauthor{Balbinot2013}'s distance estimate ($\Rhel$=31.9~kpc), the cluster is located at $\Rgal$=31.2~kpc, much more distant than our stream candidate.

The Orphan stream also partly coincides with the 20.0-1 candidate in the $l$-$b$ plane at $l\in[180\degr-210\degr]$, but it is much more distant lying at $\gtrsim30$ kpc compared to a mean distance $\sim$14~kpc of the candidate's RRLS  in this area. 
This implies the Orphan stream is not the cause of the 20.0-1 PCM excess, but it is interesting to note that the two orbital planes --and, therefore, their poles-- are relatively similar, which suggests the possibility that they might be related.

The RRLS content of 20.0-1 has a fraction of $0.84$ Oo type I stars, compared to the expectation of $0.71$, which marginally favours an Oosterhoff type I classification, as does detection 20.5-1, with a very similar Oo I fraction. However, detection 19.5-1 more clearly favours an Oo Int type, with an observed fraction of 0.27 Oosterhoff Intermediate stars, almost twice the expected fraction of 0.14. 
According to \citet{Catelan2009}, Arp 2 is one of the few existing Oo Intermediate GCs (only 4 out of their 41 clusters are classified as Oo Int) and it is considered to be associated to the Sgr dwarf galaxy, also Oosterhoff Intermediate. Although we have argued that the 20.0-1 candidate's PCM detection is unlikely to be due to Sgr entirely, some Sgr contamination is inevitably expected due to our current lack of kinematic information. Therefore, the evidence regarding the Oo type of this candidate is still inconclusive, but worth being analysed in a future study.

Figure~\ref{f:hc_cands_comparison_sss} shows the 20.0-1 candidate also overlaps with Sgr~1, a large overdensity spanning over 80 sq. deg. according to the RA/DEC extent reported by \citeauthor{Torrealba2015}, which they claim ``is almost certainly a part of the Sgr stream".  We have two arguments to believe 20.0-1 is a distinct substructure within the Sgr~1 overdensity: the angular scale of Sgr~1 is much larger, with a typical width of $\sim$20$\degr$ compared to $1\fdg8$ for 20.0-1; and we have argued above why, despite the inevitable contamination from Sgr, this candidate is more likely to be caused by a tidal tail from GC Arp~2. 

Thus, we take the 20.0-1 candidate as a \emph{new detection}, that appears to be a thin stream \emph{within} the candidate Sgr~1 overdensity, it is potentially related to the Arp~2 cluster and even possibly the PS1-C stream.  

The differences and partial coincidences with the \citet{Torrealba2015} results are to be expected since our RRLS catalogue contains the SSS catalogue in its entirety, but our search methods are very different. \citeauthor{Torrealba2015} searched for overdensities by comparing the local estimated density of RRLS in SSS to the density expected from a halo-only model, a method more suited to search for wide extended or cloud-like overdensities; our method, on the other hand, is specifically designed to search for planar substructure. 

\subsection{Tentative Candidates}

\begin{figure*}
\begin{center}
 \includegraphics[width=0.9\textwidth]{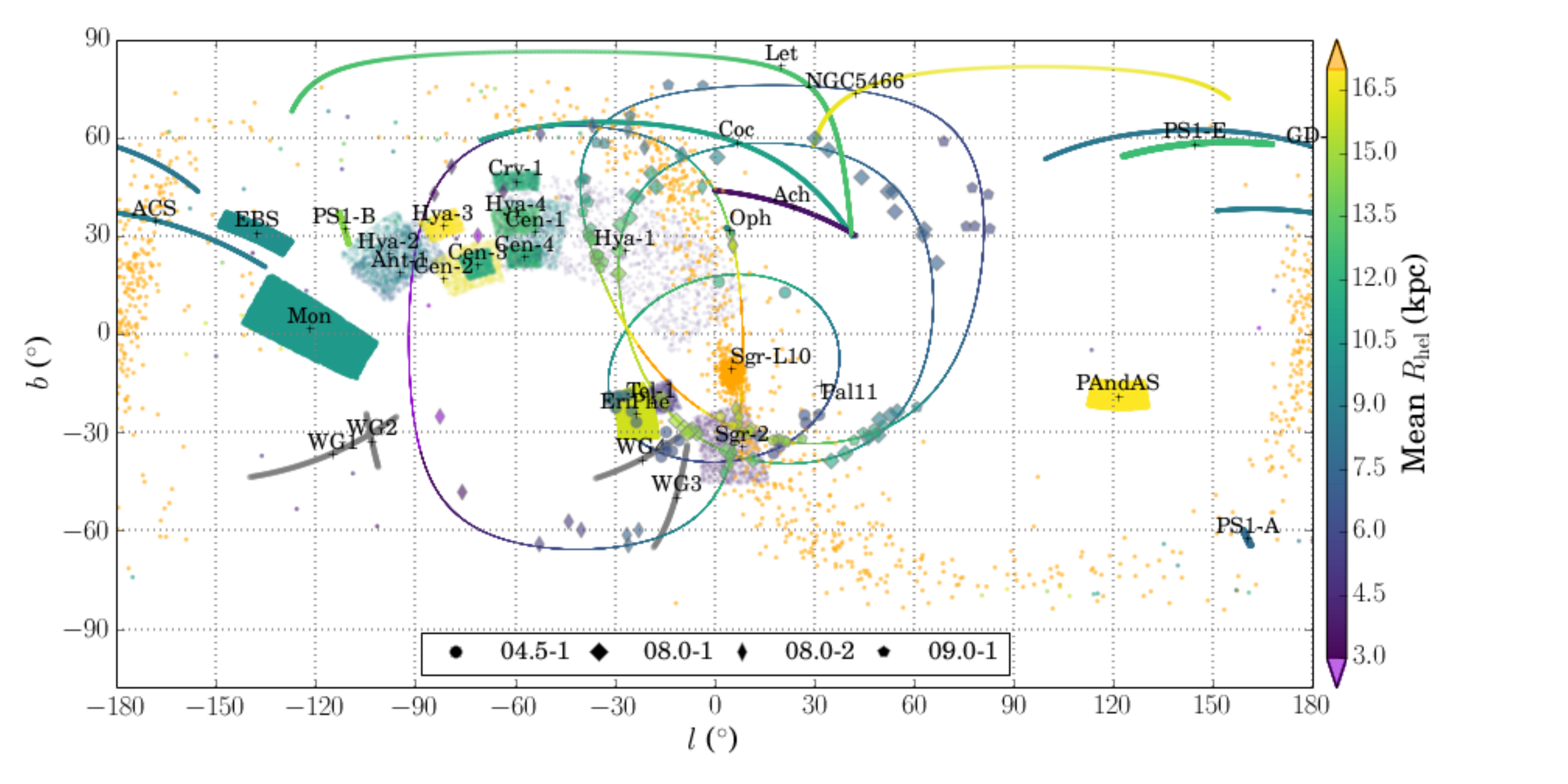} 
 \includegraphics[width=0.9\textwidth]{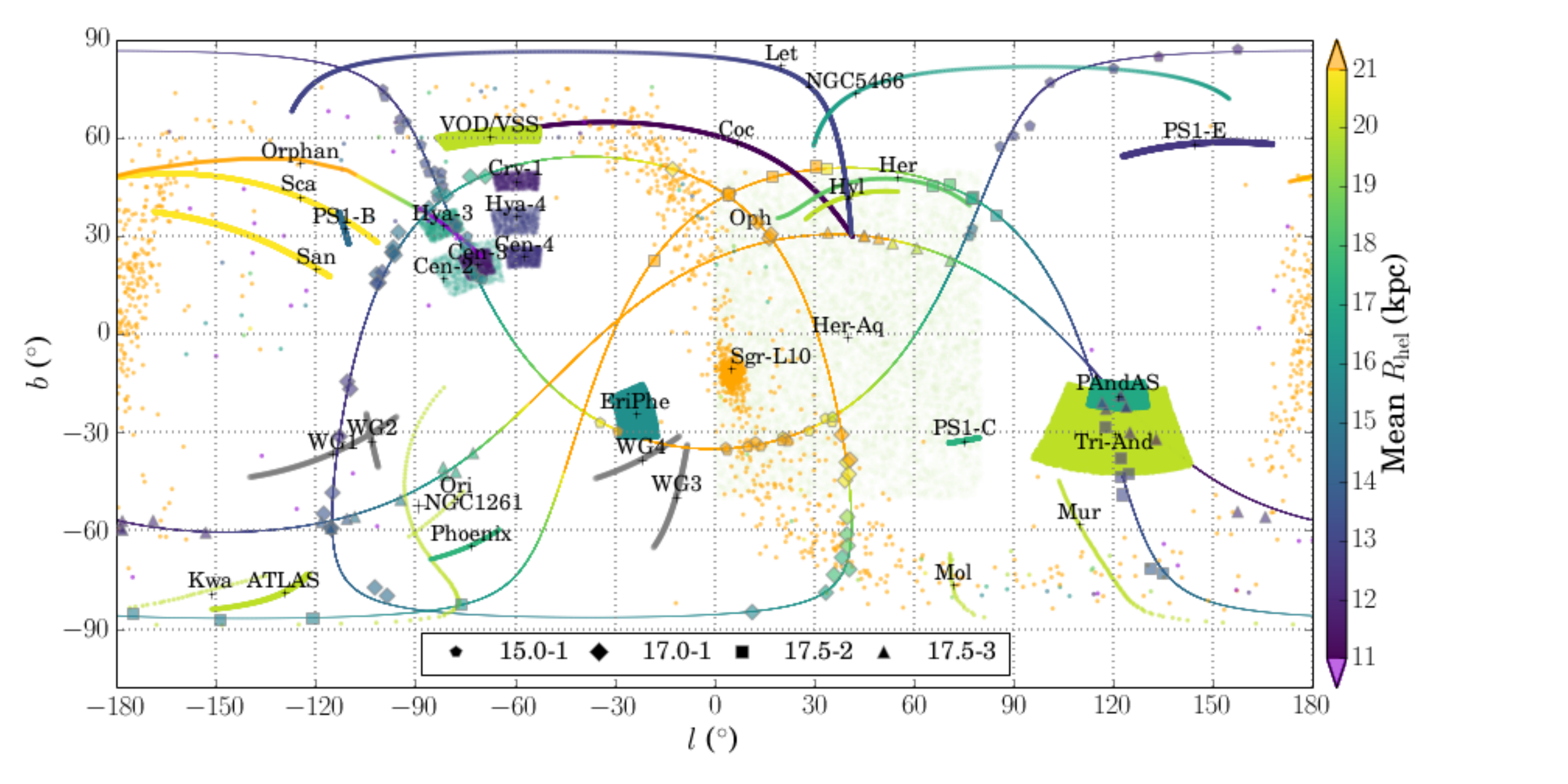} 
  \includegraphics[width=0.9\textwidth]{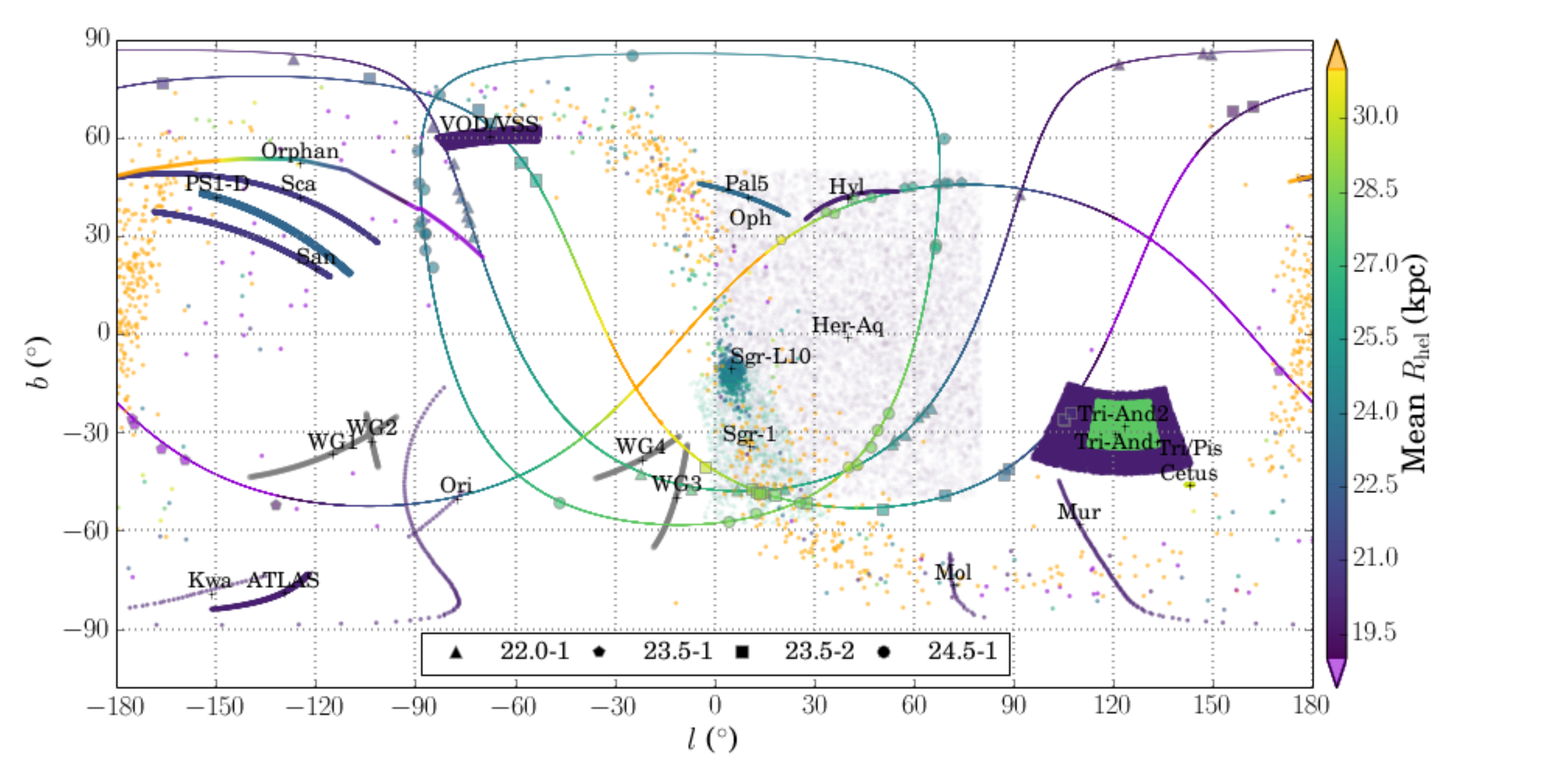} 
 \caption{Galactic (heliocentric) latitude versus longitude map for the tentative candidates. The three panels show candidates ordered by their Galactocentric distance: 4--10 kpc \emph{top}, 15--21 kpc \emph{middle} and 21--26 kpc \emph{bottom}.  As in Figure~\ref{f:hc_cands_lit}, the colour scale is proportional to the heliocentric distance and the RRLS belonging to each candidate are indicated with different symbols, as shown in the legend. Each of our candidate's great-circle is shown with a solid line passing through the RRLSs. The streams and clouds from the MW Streams Library that have any overlap in each plot's heliocentric distance range are shown with coloured dots and labelled in the plot. The corresponding references are summarised in Table~\ref{t:streams_lib}.}
\label{f:tent_cands_lit}
\end{center}
\end{figure*}

Figure~\ref{f:tent_cands_lit} shows the spatial distribution of our tentative candidates in an $l$-$b$ map, similarly to Figure~\ref{f:hc_cands_lit}, with a colour scale proportional to heliocentric distance. The three panels span different heliocentric distance ranges chosen appropriately to show case the candidates in three galactocentric distance ranges: $\Rgal\in[4,10]$ in the \emph{top} panel; $\Rgal\in[15,18]$ in the \emph{middle} panel; and $\Rgal\in[22,25]$ in the \emph{bottom} panel. As in Fig.~\ref{f:hc_cands_lit}, each candidate's great circle is shown with a solid line. Known MW streams and  \citet{Torrealba2015} SSS overdensities are also shown.

In what follows we will discuss the tentative candidates that may be associated to known streams or GCs.

\subsubsection{Candidate 17.0-1 - not Pal 5}\label{s:cand_notpal5}

The PCM in Figure~\ref{f:pcms_detections} shows candidate 17.0-1's  peak lies $\sim40\degr$ away from the Pal 5 orbital pole in the same great circle,  well outside the signature expected for Pal 5 accounting for 3$\sigma$ proper motion errors, which  suggests the two are probably unrelated. This is also illustrated in the middle panel of Figure~\ref{f:tent_cands_lit}, where candidate 17.0-1 is shown with thick diamond symbols ($\diamond$). 17.0-1's great circle passes close to Pal 5 at a very different angle, and also the densest concentration of RRLS along the great circle is found at $(l_\circ,b_\circ)=(261.3,19.7)$, quite far from Pal 5, over $90\degr$ away in longitude. Therefore, this candidate is not likely to be associated to Pal~5. On the other hand, the main overdensity in  the 17.0-1 candidate is relatively close, and at the same distance $\Rhel\sim17$ kpc, as the Hya~3 overdensity from \citet{Torrealba2015}, suggesting a possible connection between the two. 

\subsubsection{Candidate 17.5-2 - Hermus?}\label{s:cand_hermus}

Candidate 17.5-2 is not related to any known GC according to the PCM in Figure~\ref{f:pcms_detections}. As the middle panel of Figure~\ref{f:tent_cands_lit} shows (square symbols) it passes close to the Hermus stream \citep{Grillmair2014} in the sky. The densest part of 17.5-2, at $(l_\circ,b_\circ)=(75\fdg3,42\fdg8)$ (see Table~\ref{t:stream_detections}), is right at the end of the Hermus stream as reported by \citet{Grillmair2014}, and at the same distance $\sim18$~kpc. 

\citet{Grillmair2016} cites a possible metallicity for this stream around $\FeH$$\sim-2.3$ and mentions that efforts are in progress to associate RRLS or other tracers, such as blue horizontal branch stars, to this stream. Our's would then be \emph{the first identification of RRLS potentially associated to the Hermus stream}, which will have to be confirmed with kinematic data. This will also help us confirm which of this candidate's RRLS in the rest of the great circle are a true coherent overdensity and whether what we're seeing could be a bifurcation of the Hermus stream. 

\subsubsection{Candidate 17.5-3 - NGC1261?}\label{s:cand_ngc1261}

The PCM of Figure~\ref{f:pcms_detections} shows candidate 17.5-3's pole coincides with GC NGC1261's great circle, which hints at a possible association between the two. This is just a hint at first since the proper motions for this cluster have very large uncertainties ($>100\%$ in RA) and do not constrain the orbital plane well, which is why the PCM signature is a full great circle rather than a less extended feature (see Sec.~\ref{s:xgc3}).

There is little evidence regarding the RRLS content of this candidate either. According to \citet{Catelan2009} this is a Young-Halo GC, Oosterhoff type I. Its metallicity is $\FeH=-1.35$ and it has $13$ known \rrab~stars. This is consistent with the $78\%$ of OoI RRLS in this candidate, but this fraction is also fairly close to the mean expected one from field RRLS. Hence, a possible association of 17.5-3 with NGC1261 remains tentative. 

\subsubsection{Candidate 23.5-1 - Hyllus?}\label{s:cand_hyllus}

Candidate 23.5-1 is shown with pentagon symbols in the bottom panel of Figure~\ref{f:tent_cands_lit}, which shows it passes very close in the sky to the Hyllus stream \citep{Grillmair2014}. 
\citet{Grillmair2016} cite a \emph{mean} heliocentric distance of $\sim$20 kpc for Hyllus (which we use in the figure), but \citet{Grillmair2014} report a distant gradient, estimating distances of $18.5\pm3$ and $23\pm3$ kpc for the northern and southern ends respectively (in equatorial coordinates). The 23.5-1 RRLS in its vicinity span heliocentric distances from 28 to 25 kpc at the $l\sim30\degr$ and $l\sim60\degr$ ends which correspond to Hyllus's (equatorial) northern and southern ends respectively. At the southern end the distances are well in agreement with \citeauthor{Grillmair2014}'s estimate of $23\pm3$ kpc, within the errors, with our RRLS having a mean distance of 25~kpc at $l\in[60\degr,80\degr]$. This is also where the densest part of the 23.5-1 candidate is located, $(l_\circ,b_\circ)=(66\degr,+45\degr)$ (see Table~\ref{t:stream_detections}). 

Note also that the bootstrap significance for this candidate is $3.9\sigma$, so it was not classified as high-confidence but it is just below the selected threshold of $4\sigma$; thus, even though classified as tentative, this seems to be a fairly good detection.  Therefore, it seems likely that \emph{ the 23.5-1 candidate is the first independent detection of the Hyllus stream made with RRLSs} and that \emph{we may have identified a southern extension of the stream going from $(l,b)\sim(56\degr,44\degr)$ to $(76\degr,47\degr)$}.

\section{Comparison with predictions and known globular cluster tidal tails}\label{s:comparison_with_predictions}

Three GCs have tidal tails detected in SDSS or Pan-STARRS, according to \citet{Balbinot2017}: Pal~5 \citep{Odenkirchen2003}, NGC~5897 \citep[Price-Whelan priv. comm. cited by][]{Balbinot2017} and NGC~5466 \citep{Belokurov2006b,Grillmair2006c,Fellhauer2007}. 
The first two have a very scarce population of \rrab~stars: The Pal 5 cluster itself has 5 known RRLS, all of type \typec~\citep{Clement2001,SawyerHogg1973}, and  \citet{Vivas2001,Vivas2006}  have identified 2 type~\typeab~RRLS (out of a total of 6) in the cluster tails; and NGC~5897 has 3 type~\typeab~out of a total of 11 RRLS \citep{Clement2001,Clement2010}. Again, it is natural that with such a low number of type~\typeab~RRLS these clusters' tails could be not recovered by our analysis. NGC~5466, on the other hand, has 13 known \rrab~stars \citep{Catelan2009} at $\Rgal=16.9$~kpc, well within the distance range probed by our study, and \citet{Grillmair2006} reports a $1\fdg4$ width for this cluster's tidal tail. We do not find any pole count excess in the PCMs at NGC~5466's radial distance. However, as we have cautioned, we do not expect our stream candidate sample to be complete due to the lack of kinematic data.

\citet{Balbinot2017} present a recent study of the formation of tidal tails in GCs, in which they take into account the collisional dynamical effects, such as mass segregation, produced by internal evolution of the cluster. They find that low mass stars are preferentially ejected at the early stages of a cluster's disruption and higher mass stars are only ejected as the cluster comes close to complete dissolution. This effect naturally produces an observational bias, making tidal tails of fully dissolved GCs more easily observable than those of clusters that still retain a bound core. \citeauthor{Balbinot2017} offer this as an explanation of why tidal tails have been found only around very few GCs and predict that new surveys will preferentially find `orphan' or progenitor-less GC streams. This prediction supports our findings that only 2 out of our 14 candidate streams could (possibly) be associated with known GCs. These are 17.5-3 and 20.0-1 that might be related to NGC~1261 and Arp~2, respectively (see Secs.~\ref{s:cand_ngc1261} and \ref{s:cand_ngc1261}).

\citet{Balbinot2017} also offer general predictions for which known GCs are most likely to have detectable tidal tails, based on the cluster's position and velocity data. Their candidates: Pal~1 (15.9~kpc), Pal~7 (3.6~kpc) and M56 (NGC~6779, 9.5~kpc) are in the low-latitude exclusion zone $|b|\leqslant20\degr$ of the CSS+SSS surveys, while Whiting~1 (49.5~kpc) is outside the distance range probed in our analysis ($\Rgal<25$~kpc). This leaves us with their GC candidates AM4 ($\Rgal=24.8$~kpc), NGC~288 (11.4~kpc) and M92 (NGC~6341, 9.5~kpc), the latter two being their best candidates in terms of optimal detectability.

For a proper comparison with our results, we must also consider each cluster's stellar population and known RRLS content, since our survey is based on RRLS of type~\typeab~alone, which are not present in some GCs and are scarce in several.  According to \citet{Catelan2009}, NGC~288 is a classical second-parameter cluster, with a very blue horizontal branch and only one reported \rrab~star (and one \rrc) according to \citet{ArellanoFerro2013}. AM~4 is an extremely faint cluster \citep[$M_V=-1.8$,][]{Carraro2009} with almost no discernible red giant branch \citep{Hamren2013}. It harbours no variable stars according to \citet{Clement2001}, \citet{Clement2010} and references therein. Therefore, neither NGC~288's nor AM~4's tails would be detectable with an RRLS sample.

Thus, out of the candidates proposed by \citet{Balbinot2017}, only M92 could be detected with our RRLS sample. This cluster has 11 known \rrab~stars \citep{Catelan2009} and is well within the distance range of our study ($\Rgal=9.5$~kpc). Although its tidal tails, if present, should be detectable with our current sample, in our PCMs we do not find any evidence of a pole count excess that could be associated with this cluster. There is one candidate in the same distance bin as M92 (9.0-1). However, Figure~\ref{f:pcms_detections} shows that its pole is tens of degrees away from the pole-count signature expected for this cluster. We stress, however, that given our lack of kinematic information (required to reduce background contamination), \emph{we cannot rule out the possibility that M92 has tidal tails.}

\section{Conclusions}\label{s:conclusions}

We have applied the GC3 stream-finding method to RRLSs in the Catalina survey that is sensitive to nearby streams over the Galactocentric distance range range $4 < D/{\rm kpc} < 26$. Our key results are as follows:

\begin{itemize}
\item We detect two high-confidence ($>4\sigma$) new RRLS stream candidates:

-- {\it Candidate 11.0-1} includes the recently discovered Crv~1 RRLS overdensity \citep{Torrealba2015}, but is a larger structure. We call this the `Corvus stream'.

-- {\it Candidate 20.0-1} appears to trace tidal tails around the Arp~2 GC that might be connected to the PS1-C stream \citep{Bernard2016}. Proper motions and/or radial velocities are needed to test this scenario. This candidate also spatially coincides with the Sgr~1 overdensity found by \citet{Torrealba2015}, but is much thinner ($\sim1\fdg8$) than Sgr-1 ($\sim$20$\degr$, see Sec.~\ref{s:cand20-1}). We call this the `Arp2/PS1-C stream'.

\item We detect of 12 tentative RRLS stream candidates ($>3.5\sigma$). Out of these, three are of particular interest: candidates 17.5-2 and 23.5-1 could be possible extensions of the Hermus and Hyllus streams respectively, and candidate 17.5-3 could be associated with GC NGC1261.

\item Our high confidence stream candidates are expected to host $\sim$17-19 \rrab, accounting for MW halo contaminants. This number of RRLS implies an absolute magnitude $M_V\sim-7.4\pm1.5$ for the underlying population, which translates into an upper bound of $\sim7\times10^5\Lsun$ for the total luminosity. For the low confidence candidates, the number of expected \rrab \ stars ranges from 9 to 
$\sim$20 and inferred absolute magnitudes from $M_V\sim-6$ to $-7.8$. These are summarised in Table~\ref{t:stream_rrprops}.

\item We do not find any candidate stream around M92, the only cluster out of the tidal tail candidates proposed by \citet{Balbinot2017} that could be detectable with our RRLS sample. However, due to background contamination from the Milky Way stellar halo, we are not able to rule out tidal tails around M92. For this, kinematic data are required.

\item Of our 14 stream candidates, only two -- 17.5-3 (NGC 1261) and 20.0-1 (Arp2) -- are potentially associated with known GCs. This supports the idea that, due to mass segregation, tidal tails around GCs only become detectable close to full dissolution, leading to a high fraction of orphan GC streams \citep{Balbinot2017}.

\item Our detections are likely a lower bound on the total number of dissolving GCs in the inner Galaxy. Many GCs have few RRLS, while only the brightest streams are visible over the Galactic RRLS background. A more complete census will be possible with the inclusion of velocity data.

\end{itemize}

We make all of our data public, and provide the Python Package \textsc{galstreams}\footnote{\href{https://github.com/cmateu/galstreams}{https://github.com/cmateu/galstreams}.} which stores footprint information for all currently known Galactic streams and clouds, with utility classes and methods to define, manipulate and plot these data. This library is extensible so that more detailed information on each stream can be added, where publicly available. We will keep the database updated with new streams and structures as they are found.

\section*{Acknowledgements}

CM is indebted to Luis Aguilar, Cesca Figueras, Merc\`e Romero-G\'omez and B\'arbara Pichardo for their unwavering support. CM, DK, JIR also thanks Luis Aguilar and B\'arbara Pichardo for organising the Mexico Gaia Meeting 2016, where this work began, and in which they strived to provide an enjoyable environment that would foster open and relaxed discussions and collaborations between the participants. CM acknowledges support from the ICC University of Barcelona Maria de Maeztu visiting academic grants. CM is grateful for the hospitality of the University of Surrey, MSSL, and the organisers of the IV Gaia Challenge Workshop, where part of this research was carried out and warmly thanks Mark Gieles for an interesting discussion about the formation of tails in globular clusters. JIR would like to acknowledge support from STFC consolidated grant (CG) ST/M000990/1 and the MERAC foundation. DK acknowledges support from STFC CG (ST/N000811/1).

\bibliographystyle{mnras}

\label{lastpage}

\end{document}